\documentclass[twocolumn,twocolappendix]{aastex631}

\definecolor{ForestGreen}{RGB}{34,139,34}

\usepackage{graphicx}	
\usepackage{amsmath}	
\usepackage{multirow}

\newcommand{\kms}{\mathrm{km\ s^{-1}}}
\newcommand{\msun}{\ensuremath{M_\odot}}
\newcommand{\mej}{\ensuremath{M_{\rm ej}}}
\newcommand{\mlrp}{\ensuremath{M_{\rm lrp}}}
\newcommand{\mni}{\ensuremath{M_{\rm Ni}}}
\newcommand{\Xla}{\ensuremath{X_{\rm La}}}
\newcommand{\iso}[2]{\ensuremath{^{#2}\mathrm{#1}}}
\newcommand{\rp}{\emph{r}-process}
\newcommand{\gw}{GW190521}

\shorttitle{Super-kilonovae from massive collapsars}
\shortauthors{Siegel et al.}
\graphicspath{{./}{figures/}}

\begin{document}

\title[Super-Kilonovae from Massive Collapsars]{``Super-Kilonovae'' from Massive Collapsars as Signatures of Black-Hole Birth \\in the Pair-instability Mass Gap}

\author[0000-0001-6374-6465]{Daniel M.~Siegel}
\affiliation{Perimeter Institute for Theoretical Physics, Waterloo, Ontario, Canada, N2L 2Y5}
\affiliation{Department of Physics, University of Guelph, Guelph, Ontario, Canada, N1G 2W1}

\author[0000-0002-8685-5477]{Aman Agarwal}
\affiliation{Perimeter Institute for Theoretical Physics, Waterloo, Ontario, Canada, N2L 2Y5}
\affiliation{Department of Physics, University of Guelph, Guelph, Ontario, Canada, N1G 2W1}

\author[0000-0003-3340-4784]{Jennifer Barnes}
\affiliation{Columbia Astrophysics Laboratory, Columbia University, New York, New York 10027, USA}
\affiliation{Kavli Institute for Theoretical Physics, University of California, Santa Barbara, CA 93106, USA}

\author[0000-0002-4670-7509]{Brian D.~Metzger}
\affiliation{Columbia Astrophysics Laboratory, Columbia University, New York, New York 10027, USA}
\affiliation{Center for Computational Astrophysics, Flatiron Institute, New York, NY 10010, USA}

\author[0000-0002-6718-9472]{Mathieu Renzo}
\affiliation{Columbia Astrophysics Laboratory, Columbia University, New York, New York 10027, USA}
\affiliation{Center for Computational Astrophysics, Flatiron Institute, New York, NY 10010, USA}

\author[0000-0002-5814-4061]{V.~Ashley Villar}
\affiliation{Columbia Astrophysics Laboratory, Columbia University, New York, New York 10027, USA}
\affiliation{Center for Computational Astrophysics, Flatiron Institute, New York, NY 10010, USA}
\affiliation{Department of Astronomy \& Astrophysics, The Pennsylvania State University, University Park, PA 16802, USA}
\affiliation{Institute for Computational \& Data Sciences, The Pennsylvania State University, University Park, PA, USA}
\affiliation{Institute for Gravitation and the Cosmos, The Pennsylvania State University, University Park, PA 16802, USA
}



\begin{abstract}

The core collapse of rapidly rotating massive $\sim 10 M_{\odot}$ stars (``collapsars''), and resulting formation of hyper-accreting black holes, are a leading model for the central engines of long-duration gamma-ray bursts (GRB) and promising sources of $r$-process nucleosynthesis.  Here, we explore the signatures of collapsars from progenitors with extremely massive helium cores $\gtrsim 130M_{\odot}$ above the pair-instability mass gap.  While rapid collapse to a black hole likely precludes a prompt explosion in these systems, we demonstrate that disk outflows can generate a large quantity (up to $\gtrsim 50M_{\odot}$) of ejecta, comprised of $\gtrsim 5-10M_{\odot}$ in $r$-process elements and $\sim 0.1-1M_{\odot}$ of $^{56}$Ni, expanding at velocities $\sim 0.1$\,c.  Radioactive heating of the disk-wind ejecta powers an optical/infrared transient, with a characteristic luminosity $\sim 10^{42}$ erg s$^{-1}$ and spectral peak in the near-infrared (due to the high optical/UV opacities of lanthanide elements) similar to kilonovae from neutron star mergers, but with longer durations $\gtrsim$ 1 month.  These ``super-kilonovae'' (superKNe) herald the birth of massive black holes $\gtrsim 60M_{\odot}$, which---as a result of disk wind mass-loss---can populate the pair-instability mass gap ``from above'' and could potentially create the binary components of \gw{}.  SuperKNe could be discovered via wide-field surveys such as those planned with the {\it Roman Space Telescope} or via late-time infrared follow-up observations of extremely energetic GRBs.  Gravitational waves of frequency $\sim 0.1-50$ Hz from non-axisymmetric instabilities in self-gravitating massive collapsar disks are potentially detectable by proposed third-generation intermediate and high-frequency observatories
at distances up to hundreds of Mpc; in contrast to the ``chirp'' from binary mergers, the collapsar gravitational-wave signal decreases in frequency as the disk radius grows (``sad trombone'').

\end{abstract}



\section{Introduction} 
\label{sec:intro}

The astrophysical locations which give rise to the synthesis of heavy nuclei via the rapid capture of neutrons onto lighter seed nuclei (the $r$-process;
\citealt{Burbidge+57,cameron_nuclear_1957}) remains a topic of active debate (see \citealt{Horowitz+19,Cowan+21,siegel_21} for recent reviews).  Several lines of
evidence, ranging from measurements of radioactive isotopes on the sea
floor (e.g.,~\citealt{Wallner+15,Hotokezaka+15}) to the abundances of
metal-poor stars formed in the smallest dwarf galaxies
(e.g.,~\citealt{Ji+16,Tsujimoto+17}), suggest that the dominant site of the
$r$-process is much rarer than ordinary core collapse supernovae (SNe), both in the early history of our Galaxy and today.  The most promising contenders are the mergers of neutron star binaries (e.g.,~\citealt{Lattimer&Schramm74,Symbalisty&Schramm82}) or rare channels of core collapse SNe, such as those which give birth to a rapidly spinning magnetar \citep{Thompson+04,Metzger+07,Winteler+12,Nishimura+15} or a hyper-accreting black hole (``collapsar''; e.g.,~\citealt{Surman+08,siegel_collapsars_2019}; see also \citealt{Grichener&Soker19}).  Perhaps not coincidentally, the same two types of events$-$neutron star mergers and collapsars$-$are the leading models for the central engines of gamma-ray bursts (GRB) of the short- and long-duration classes, respectively (e.g., \citealt{Woosley&Bloom06,berger_short-duration_2014}).

The radioactive decay of $r$-process elements in the ejecta of a
neutron star mergers power a short-lived optical/infrared transient known as a kilonova \citep{Li&Paczynski98,Metzger+10,Barnes&Kasen13}. However, the large quantity of $r$-process ejecta $\gtrsim 0.02-0.06M_{\odot}$ inferred from the kilonova to accompany GW170817, as well as the
relatively low inferred outflow velocity $\sim 0.1$ c of the bulk of
this material (e.g., \citealt{cowperthwaite2017electromagnetic, drout2017light, Villar+17}), do not agree with
predictions from numerical relativity for the mass ejected during the
early dynamical phase of the merger (see
\citealt{Metzger19,Siegel19,Margutti&Chornock20,Nakar20} for reviews).
Instead, the dominant ejecta source in
GW170817, and likely in the majority of neutron star
mergers, are delayed outflows from the accretion disk which forms
around the black hole (BH) or neutron star remnant (e.g.,
\citealt{Metzger+08,Fernandez&Metzger13,Just+15,siegel_three-dimensional_2017,siegel_three-dimensional_2018,Fujibayashi+18}).
General relativistic magnetohydrodynamical (GRMHD) simulations of the
long-term evolution of post-merger disk find that up to
$\sim 30\%$ of its original mass is unbound in outflows with average
velocities $\sim 0.1$ c
\citep{siegel_three-dimensional_2017,Fujibayashi+18,Fernandez+19,Christie+19,fujibayashi_mass_2020},
broadly consistent with the kilonova observed from GW170817.

As emphasized by \citet*{siegel_collapsars_2019}, similar accretion disk outflows to those generated in neutron star mergers occur also in collapsars (see also \citealt{MacFadyen&Woosley99,Janiuk+04,Surman+06,Miller+20,just_neutrino_2021}).  Unlike in the merger case, the collapsing stellar material feeding the disk is composed of roughly equal numbers of protons and neutrons (electron fraction $Y_e \simeq 0.5$).  However, for mass accretion rates above a critical threshold value ($\gtrsim 10^{-1}-10^{-3}M_{\odot}$ s$^{-1}$, which depends on the effective viscosity and BH mass; \citealt{chen_neutrino-cooled_2007,metzger_conditions_2008}), the inner regions of the disk are electron degenerate and act to ``self-neutronize'' via electron captures on protons (e.g.,~\citealt{Beloborodov03}), thus maintaining a low electron fraction $Y_e \approx 0.1$ in a regulated process \citep{siegel_three-dimensional_2017}.  As a result, the collapsar disk outflows, which feed on this neutron-rich reservoir, can themselves possess a sufficiently high neutron concentration to enable an $r$-process, throughout much of the epoch over which the GRB jet is being powered.  However, the details of the synthesized composition$-$particularly the partitioning between light and heavy $r$-process elements$-$are sensitive to the impact of neutrino absorption processes on the electron fraction of the outflowing material (\citealt{Surman+06,Miller+20,Li&Siegel21}).

In comparison to neutron star mergers, collapsars hold several
complementary advantages as $r$-process sources \citep{siegel_collapsars_2019}.  Firstly, as a
result of being generated promptly from very massive stars and
empirically found to occur in small dwarf galaxies at
low metallicity (e.g.,~\citealt{Fruchter+06}), collapsars naturally explain the $r$-process enrichment in ultra-faint dwarf
galaxies such as Reticulum II (e.g., \citealt{Ji+16}) and metal-poor stars in the Galactic halo (e.g., \citealt{Brauer+21}).  Furthermore,
if the gamma-ray luminosity of GRBs scales with BH accretion
rate in the same way in mergers as in collapsars, then from the relative rate and gamma-ray fluence distributions of long- versus short-duration GRBs, one is led to conclude that the total mass accreted through collapsar disks over cosmic time (and hence the integrated amount of disk wind ejecta) could exceed that in neutron star mergers \citep{siegel_collapsars_2019}.  Arguments based on the chemical evolution history of the early Milky Way galaxy have been made both in favor of rare SNe/collapsars \citep{Cote+19,siegel_collapsars_2019,vandeVoort+20,Yamazaki+21,Brauer+21} and mergers \citep{Shen+15,Duggan+18,Macias&RamirezRuiz19,Bartos&Marka19,Holmbeck+19,Tarumi+21} as sources of early $r$-process.

While the kilonova from GW170817 provided ample evidence that neutron
star mergers can execute an $r$-process, the same signature has
not yet been seen from the SNe observed in coincidence with long GRBs.
However, this fact is not necessarily constraining yet, insofar that
$r$-process material is easier to hide in the collapsar case.  In
particular, a prompt and powerful supernova explosion may be required
to explain the large masses of $^{56}$Ni inferred from GRB supernova
light curves (e.g.,~\citealt{Cano16,barnes_grb_2018}; however, see
\citealt{zenati_nuclear_2020}).  By contrast, the
$r$-process-generating disk outflows occur over longer times, up to
tens of seconds or more after collapse, commensurate with the observed
duration of the long GRB.  Unless efficiently mixed to the highest
velocities, the $r$-process elements (and any associated photometric
or spectroscopic signatures) are therefore buried behind several solar
masses of ``ordinary'' supernova ejecta (dominated by
$\alpha$-elements such as oxygen).

Nevertheless, if present in the inner ejecta layers, $r$-process elements could manifest as a late-time infrared signal \citep{siegel_collapsars_2019} arising from the high opacity of heavy $r$-process nuclei \citep{Kasen+13,Tanaka&Hotokezaka13}.  This signal is challenging to detect given the typically large distances to GRB SNe and the late times required (at which point the emission is faint).  The detection prospects will improve with the advent of the {\it James Webb Space Telescope} ({\it JWST}), particularly if the nebular spectra of lanthanide-rich material also peaks in the infrared \citep{Hotokezaka+21}.

Collapsars of the type observed so far as SNe may also only represent a subset of accretion-powered core collapse events.  The progenitor stars which give rise to long GRBs are typically believed to possess ZAMS masses $\lesssim 40M_{\odot}$ with helium cores at death of $\lesssim 10M_{\odot}$ (e.g., \citealt{Woosley&Heger06}).  Upon collapse of their iron cores, these events first go through a rapidly rotating proto-neutron star phase \citep{Dessart+08}, in which a millisecond magnetar is formed \citep{Thompson+93,Raynaud+20}.  The strong and collimated outflow from such a magnetar during the first seconds after its birth \citep{Thompson+04,Metzger+07}, before it accretes sufficient matter to collapse into a BH, may play an important role in shock heating and in unbinding much of the outer layers of the star and generating the required large $^{56}$Ni masses (e.g., \citealt{Shankar+21}).

On the other hand, the number of long GRBs with detected SNe only number around a dozen, and the majority of these are associated with the volumetrically more common but physically distinct ``low luminosity'' class of GRBs (e.g., \citealt{Liang+07}).  It thus remains unclear whether the more energetic, classical long GRBs always occur in coincidence with $^{56}$Ni-powered SNe.  Indeed, luminous SNe have been ruled out for a few nominally long GRBs (e.g.,~\citealt{Fynbo+06,Gehrels+06}), though the nature of these events (e.g., whether they are actually short GRBs masquerading as collapsars) remains unclear (e.g., \citealt{Zhang+07}).

Within this context, we consider in this paper the fate of initially much more massive stars, those with ZAMS masses $M_{\rm ZAMS} \gtrsim 260M_{\odot}$ which are predicted to evolve
helium cores by the time of core collapse above the pair-instability
(PI) gap $\gtrsim 130M_{\odot}$ (e.g., \citealt{woosley:02}, \citealt{woosley:17},
\citealt{renzo:20csm}, \citealt{farmer:20}, \citealt{woosley:21}).  If
the initial mass function (IMF) is an indication, such stars are
potentially much rarer than the ordinarily considered collapsar
progenitors with $M_{\rm ZAMS} \lesssim 40M_{\odot}$.  On the other
hand, if such stars are rapidly spinning \citep[e.g.,][]{marchant:19,
  marchant:20}---possibly because of continuous gas accretion
throughout their life \citep[e.g.][]{Jermyn+21, dittmann:21}---and if
these form collapsar-like disks upon collapse in proportion to their
(much higher) helium core masses, their resulting yield of $r$-process ejecta in disk winds could be substantially greater.

Another key difference is that a prompt explosion (e.g., as attributed to
a proto-magnetar above, or fallback accretion; \citealt{powell:21}) is more challenging to obtain for these very massive stars.  This is because (1) the nominal timescale for BH formation is much faster, within $\lesssim 0.1$ s, due to large
masses exceeding~$2M_{\odot}$ and high compactness of their iron cores
\citep[e.g.,][]{renzo:20csm}; (2) their larger $\gtrsim 10^{53}$ erg
gravitational binding energies relative to those of lower mass helium
cores $\lesssim 10^{52}$ erg exceed the rotational energies of even maximally spinning neutron stars.  As a result of the assuredly failed initial explosion of post-PI cores, these systems are unlikely to eject a large quantity of prompt, shock-synthesized $^{56}$Ni and unprocessed stellar material (however, see \citealt{Fujibayashi+21}).  Instead, the bulk of the ejecta will arise over longer timescales from disk outflows, which--scaling up from low-mass collapsars---could amount to $\gtrsim 10M_{\odot}$ of $r$-process and Fe-group elements (including $^{56}$Ni).

Rather than the usual picture of GRB SNe, the type of collapse
transient above the PI-gap we envision is in some ways more akin to a scaled-up neutron star merger.  At risk of committing etymological heresy, we therefore refer to these massive collapsar transient events as ``super-kilonovae'' (superKNe).  As we shall discuss, if superKNe exist, their long durations and red colors may render them potentially identifiable through either follow-up infrared observations of long GRBs (e.g., with {\it JWST}), or blindly in surveys with the {\it Vera Rubin Observatory} (\citealt{Tyson+02}) or the {\it Nancy Grace Roman Space Telescope} ({\it Roman}; \citealt{Spergel+15}).

The gravitational wave observatory LIGO/Virgo detected a binary BH merger, \gw{}, for which both binary components
of masses $\sim\!85M_{\odot}$ and $\sim\!66M_{\odot}$, respectively
\citep{Abbott+20_190521}, were inside the nominal PI mass
gap.\footnote{However, see \cite{fishbach:20, Nitz&Capano21}, who interpret \gw{}~as a merger between one BH below the PI gap and one above.}  Tentative evidence suggests an effective high BH spin
of the progenitor binary, albeit with the spin axis misaligned with the orbital momentum
axis (however, see \citealt{Mandel&Fragos20,Nitz&Capano21}).  These
unusual properties have motivated a number of theoretical studies
proposing new ways to populate the PI mass gap, such as through
dynamical stellar mergers \citep[e.g.][]{dicarlo:19, dicarlo:20, renzo:20merger}, hierarchical black hole mergers in dense environments (e.g.,
\citealt{Antonini&Rasio16,Yang+19,Tagawa+21,Gerosa&Fishbach21}), modifying stellar
physics at low metallicity \citep[e.g.][]{farrell:21, vink:21}, or through external gas accretion (e.g., \citealt{Safarzadeh&Haiman20}).  As we shall describe, if both
the BHs acquired their low expected masses and high spin as a
result of inefficient disk accretion, superKN events of the type envisioned here provide a novel single-star channel for filling the PI mass gap ``from above''.

This paper is organized as follows.  Sec.~\ref{sec:outflow_model} presents a semi-analytic model for the collapse of rotating massive stars, their accretion disks and disk wind ejecta, and resulting heavy element nucleosynthesis, which builds on earlier work in \citet{siegel_collapsars_2019}.  Calibrating the model such that collapsars generate BH accretion events consistent with the observed properties of long GRB jets, we then apply the model to more massive $\gtrsim 130M_{\odot}$ progenitors above the PI mass gap.  Using our results for the disk wind ejecta, in Sec.~\ref{sec:light_curve_models} we calculate the light curves and spectra of their superKN emission by means of Monte Carlo radiative transfer simulations.  Sec.~\ref{sec:discovery} explores the prospects for discovering superKNe by future optical/infrared surveys or in follow-up observations of long GRBs. Section~\ref{sec:implications} discusses several implications of our findings, including gravitational-wave emission from self-gravitating phases of the collapsar disk evolution; the astrophysical origin of \gw{}; and the luminous radio and optical emission that results from the superKN ejecta interacting with surrounding gas.  Sec.~\ref{sec:conclusions} summarizes our results.

\section{Disk Outflow Model}
\label{sec:outflow_model}

\subsection{Stellar models}
\label{sec:stellar_models}

To model the pre-collapse structure of the superKN progenitors, we
employ the \texttt{MESA} stellar evolution models of
\cite{renzo:20csm}, publicly available at
\url{https://zenodo.org/record/3406357}.  The simulations start from
naked helium cores of metallicity $Z=0.001$ which are then
self-consistently evolved from helium core ignition, through possible
(pulsational) pair-instability (PPI), to the onset of core-collapse
(defined as when the radial in-fall velocity exceeds $1000\,\kms$).
We label the input stellar models according to their initial helium core
mass, e.g., model \texttt{200.25} corresponds to
$M_{\rm He,init} = 200.25M_{\odot}$, and focus on models ``above'' the
PI gap, which do not experience pair-instability driven pulses.

The models are computed using a 22-isotope nuclear reaction network,
which is sufficient to capture the bulk of the energy generation
throughout the stellar evolution, but cannot accurately capture the
weak-interactions in the innermost core
\citep[e.g.,][]{farmer:16}. However, the deepest layers of the core
promptly fall into the newly formed BH (see Sec.~\ref{sec:fallback})
and hence do not contribute to the accretion disk and its outflows.

These models were evolved without rotation, which we instead
artificially impose at the point of core collapse (see
Sec.~\ref{sec:fallback}). The main effect of rotation during the
pre-core-collapse evolution is mixing at the core-envelope interface,
which leads to more massive helium cores for a given initial mass. In
the extreme case of chemically homogeneous evolution
\citep{maeder:00}, the entire star may become a helium core. This
will impact how many stars develop core masses reaching into the
PI/pulsational PI regime or beyond, and thus the predicted population
statistics. However, because \cite{renzo:20csm} only simulate the helium
core, this does not affect our present study.  Rotation can also
enhance the wind mass-loss rate \citep[e.g.,][]{langer:98}, and
increase the radius in the outer layers at the rotational equator by up to 50\%, which is neglected in the progenitors we use.
Finally, by adding centrifugal support to the core, rotation can
modestly increase the PI/pulsational PI mass range
\citep[e.g.,][]{glatzel:85}.\footnote{For example, using a setup
  similar to \cite{renzo:20csm}, \cite{marchant:20} study the impact
  of an initial rotation frequency $\omega/\omega_\mathrm{crit}=0.9$,
  where
  $\omega_\mathrm{crit} \equiv \sqrt{(1-L_{\star}/L_\mathrm{Edd})GM_{\star}/R_{\star}^3}$
  and $L_{\star}/L_{\rm Edd}$ is the stellar luminosity in units of
  the Eddington luminosity. They found a $\sim{}4\%$ ($\sim{}15\%$)
  increase in the maximum BH mass below the PI mass gap assuming
  angular momentum is transported by a Spruit-Tayler dynamo (assuming
  no angular momentum transport).  The stronger angular momentum
  coupling found by \citet{fuller:19} would likely result in an even
  more modest effect.}

For sufficiently large initial core masses $M_\mathrm{He, init}\gtrsim 200\,M_\odot$, the final mass at collapse would nominally produce a BH above the PI mass gap (neglecting subsequent mass-loss in accretion disk outflows, as explored in the present study).  This arises because the gravitational energy released by the PI-driven collapse acts to photo-disintegrate nuclei produced in the thermonuclear explosion, instead of generating outwards bulk motion \citep[e.g.,][]{bond:84}.  Since these models do not experience pulses of mass loss, their pre-collapse total mass is determined by the assumed wind mass-loss prescription: the minimum final helium core mass above the PI mass gap for our \texttt{MESA} models is $M_\mathrm{He, fin}\gtrsim 125\,M_\odot$.
\cite{renzo:20csm} estimated the corresponding final BH mass (again,
neglecting post-collapse disk outflows) as the total baryonic mass
with binding energy $<10^{48}$ ergs, which effectively corresponds to
the total final mass within a few $0.1\,M_\odot$
\citep[e.g.,][]{farmer:19, renzo:20csm}.

A key ingredient in modeling fallback accretion is the radial density profile of the star at collapse (Sec.~\ref{sec:fallback}).  Despite their large masses, helium stars above the PI gap remain compact throughout their lives, never expanding as a result of
PI pulses.  Their typical radii $R_{\star} \approx 10\,R_\odot$ 
are similar to the normally considered Wolf-Rayet progenitors of GRBs (e.g.,
\citealt{Woosley&Heger06}; Fig.~\ref{fig:stellar_rotation_profile}). If stars in this mass range reach core collapse with their hydrogen envelope intact (e.g., for sufficiently low metallicity, as in population III stars) their radii could be considerably larger; however, no red supergiants of this mass
have yet been observed.

We also employ the $15\,M_\odot$ and $20\,M_\odot$ single (hydrogen-rich) star models from \cite{heger_presupernova_2000} to test our collapsar model on more canonical long-GRB progenitors (see Appendix \ref{app:collapsars} for a discussion of results). These are computed starting from a
surface equatorial velocity of $200\,\mathrm{km\ s^{-1}}$ at ZAMS and assume that mean molecular weight gradients do not impede rotational mixing ($f_\mu=0$, ``weak molecular weight barriers''). They are labeled \texttt{E15} and
\texttt{E20} respectively, and are publicly available at \href{https://2sn.org/stellarevolution/rotation/}{https://2sn.org/stellarevolution/rotation/}.

\subsection{Collapsar Model}
\label{sec:fallback}

The masses and composition of the superKNe ejecta are computed by modeling
the collapse of a progenitor star. Depending on the stellar angular momentum
profile, the collapse and fallback of envelope material leads to the formation of an accretion disk, which gives rise to massive neutron-rich disk outflows
\citep{siegel_collapsars_2019,Miller+20,just_neutrino_2021}. Although
rotation profiles of massive stars at the time of core collapse, in
particular of those above the PI mass gap considered here, are
highly uncertain (e.g.,
\citealt{heger_presupernova_2000,ma_angular_2019, marchant:20}), the specific angular momentum $j_z$ generally increases with stellar radius. In-falling stellar material thus circularizes at increasingly larger radii from the BH with time.

\begin{figure}
\centering
\includegraphics[width=0.99\linewidth]{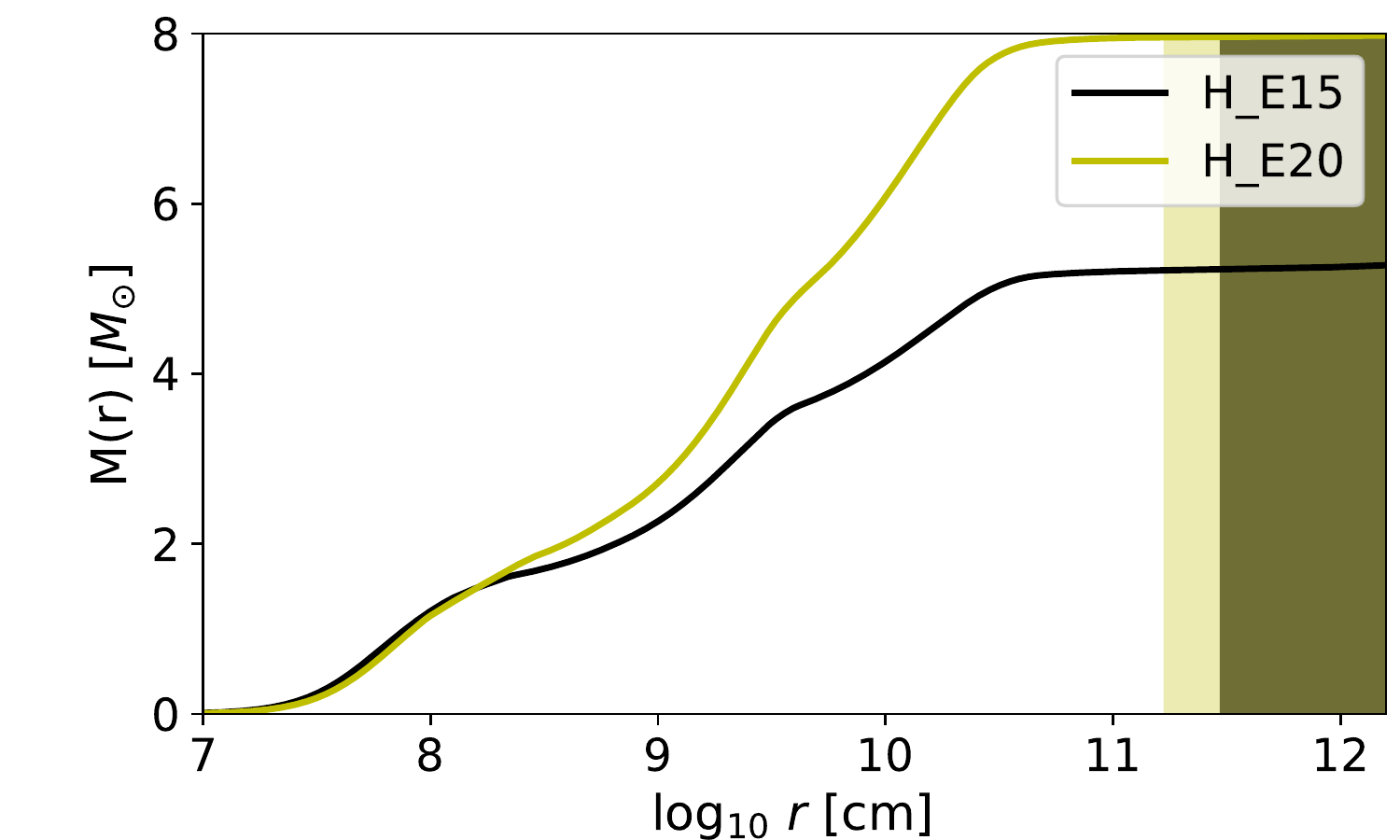}
\includegraphics[width=0.99\linewidth]{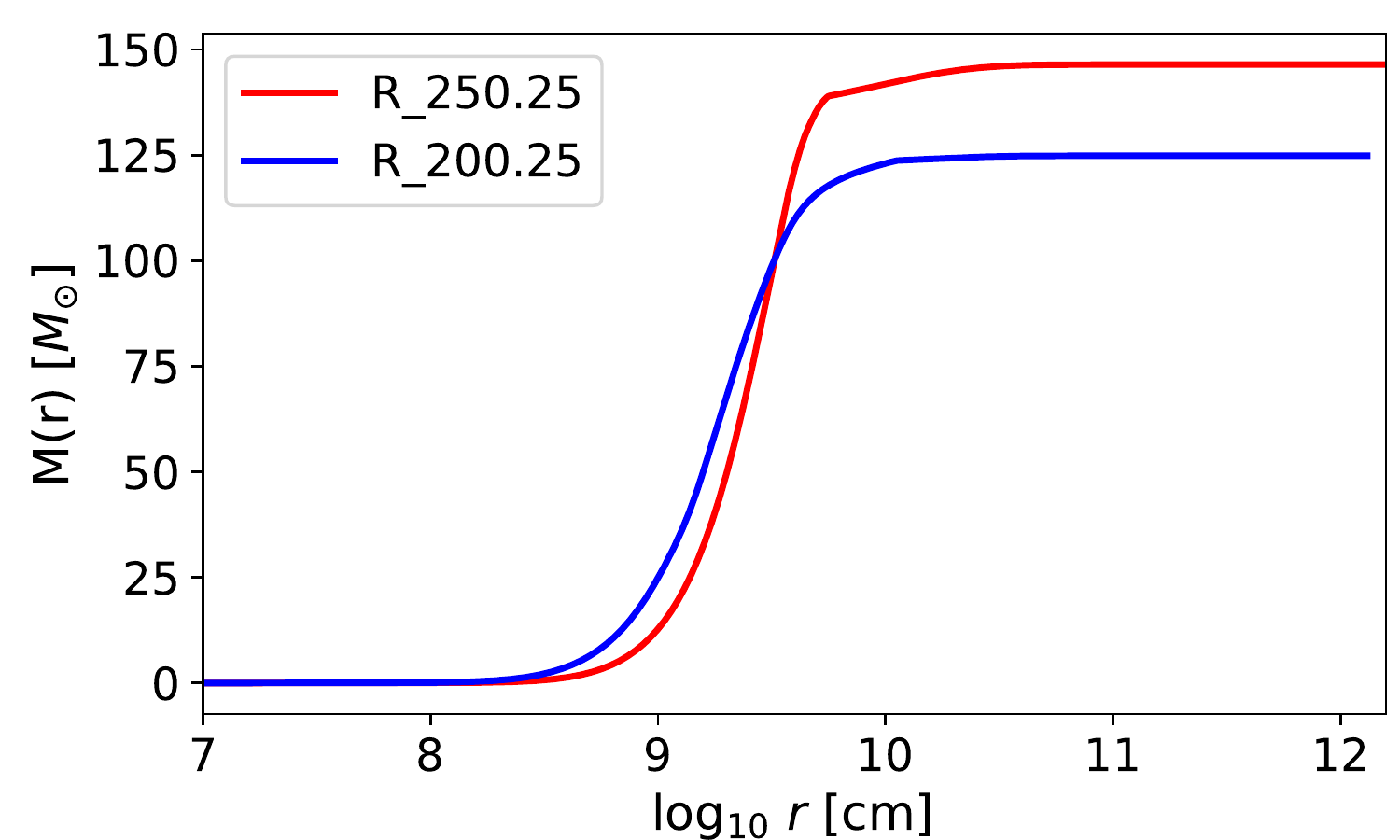}
\includegraphics[width=0.99\linewidth]{
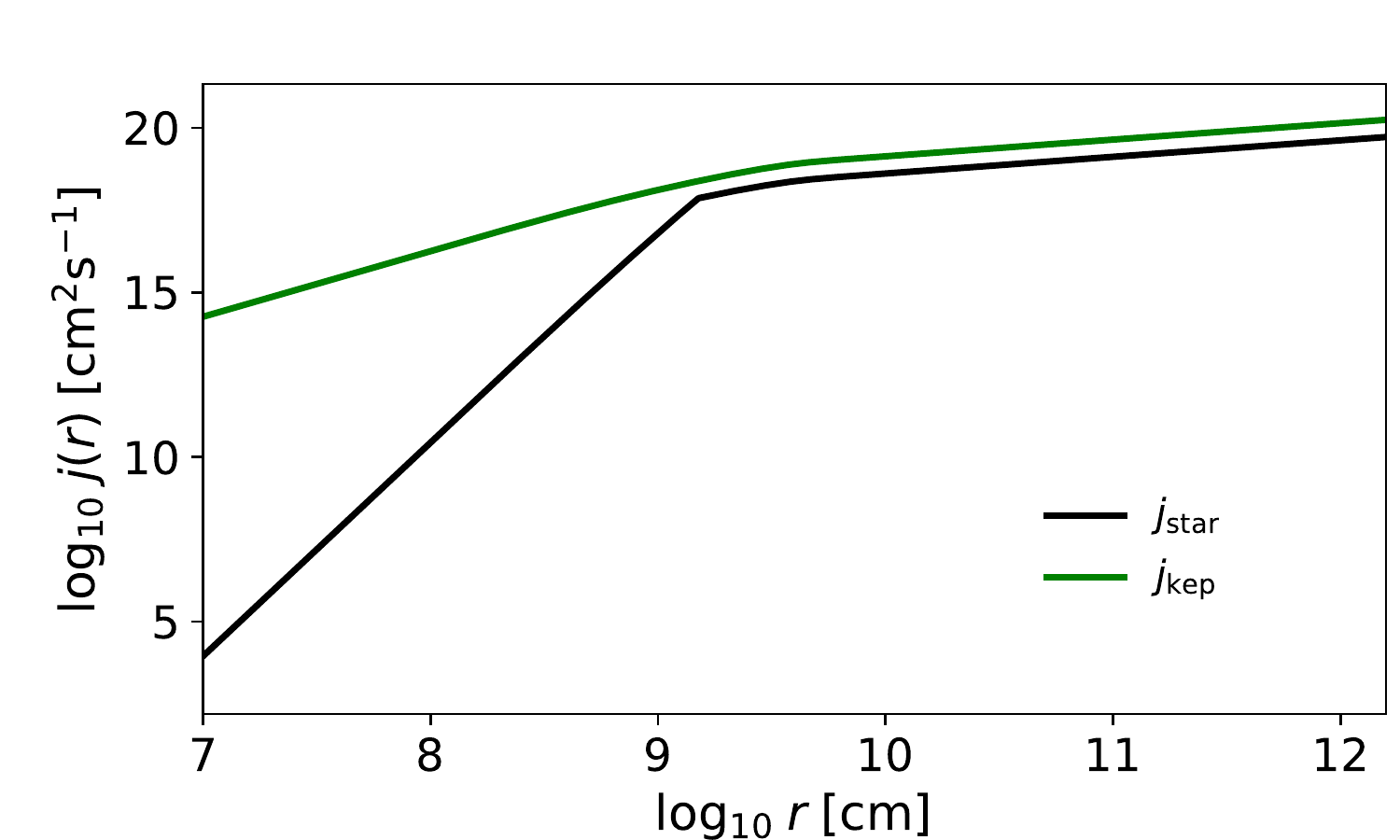}
\caption{Properties of stellar models at the onset of collapse,
  showing the enclosed mass as a function of stellar radius (top: models \texttt{E15} and \texttt{E20} of \citealt{heger_presupernova_2000}; center: models \texttt{200.0} and \texttt{250.25} of \citealt{renzo:20csm}), and an example of the imposed specific angular momentum profile
  for model \texttt{250.25} with $p=4.5$,
  $r_{\rm b}=1.5\times 10^{9}$\,cm, and $f_{\rm K}=0.3$
  (cf.~Eq.~\ref{eq:j_profile}) compared to the corresponding Keplerian
  profile (green solid line; bottom). The light (dark) shaded region in the top panel represents the hydrogen envelopes of the \texttt{E20} (\texttt{E15}) models. Such envelopes are absent in the models of \citet{renzo:20csm}.
}
\label{fig:stellar_rotation_profile}
\end{figure}

We endow the stellar models with mass $M_\star$ and radius $R_\star = R_{\rm He,fin}$ at the time of core collapse (Sec.~\ref{sec:stellar_models}) with an angular momentum profile that assumes rigid rotation on spherical shells, with angular velocity $\Omega(r,\theta)=\Omega(r)$.  This results in
\begin{equation}
    j_z(r,\theta) = j(r) \sin^2(\theta),
\end{equation}
where $r,\theta$ are the radial and polar angle coordinates, respectively.  We adopt a general parametrized angular momentum profile of the form
\begin{equation}
	j(r) = \left\{ \begin{array}{cc}
		f_{\rm K} j_{\rm K}(r)\left(\frac{r}{r_{\rm b}}\right)^{p}, & r < r_{\rm b} \\
		f_{\rm K} j_{\rm K}(r), & r_{\rm b} \le r \le R_\star
		\end{array} \right., \label{eq:j_profile}
\end{equation}
where $r_{\rm b}$, $p$, and $f_{\rm K}$ are free parameters. This
corresponds to a low-density `envelope', composed primarily of helium
in the models considered here, rotating at a fraction $f_{\rm K} < 1$ of the local Keplerian angular momentum $j_{\rm K} = \sqrt{G M_{\rm enc}(r)r}$, where $M_{\rm enc}(r)$ is the mass enclosed interior to radius $r$, and an inner `core', in which rotation is suppressed by a power-law with index $p$ relative to the fraction of local break-up rotation adopted for the envelope. Although the parameter values ($r_{\rm b}$, $p$, $f_{\rm K}$) are uncertain, as we discuss below, they can be ``calibrated'' to produce the timescales and energetics of the disk accretion consistent with the observed properties of long GRB jets.  Figure \ref{fig:stellar_rotation_profile} illustrates the parametrized rotation profile for model \texttt{250.25}.

Assuming an axisymmetric rotating star, we discretize the progenitor
stellar model into $(n_r, n_\theta)$ mass elements, logarithmically
spaced in stellar radius $r$, and uniformly spaced in $\cos\theta$. The angular resolution is chosen sufficiently high (typically $n_\theta = 1001$) that the accuracy in numerically computing global quantities by integration (total mass, total fall-back mass, etc.) is dominated by the finite radial resolution of the stellar progenitor models. Defining $t=0$ as the onset of core-collapse, a given stellar layer at radius $r$ will start to collapse onto the centre upon the sound travel time $t_{\rm s}(r)=\int_0^r c_s^{-1}(r){\rm d}r$ from the centre to $r$. Due to its finite angular momentum, a given fluid element will do so on an eccentric trajectory and circularize on the equatorial plane at time (cf.~\citealt{kumar_mass_2008})
\begin{eqnarray}
	t_{\rm circ}(r,\theta) &=& t_{\rm s}(r)\label{eq:t_circ}\\ 
	&+& \frac{(1+e)^{-3/2}}{\Omega_{\rm K}(r)}\left[\cos^{-1}(-e) + e(1-e^2)^{1/2}\right]\nonumber
\end{eqnarray}
and radius
\begin{equation}
	r_{\rm circ}(r,\theta) = \left(\frac{8}{3\pi}\right)^2 r (1-e), \label{eq:r_circ}
\end{equation}
where $e(r,\theta) = 1- [\Omega^2(r)/\Omega^2_{\rm K}(r)]\sin^2\theta$ is the eccentricity of the trajectory and $\Omega_{\rm K} = (G M_{\rm enc}(r) /r^3)^{1/2}$ the Keplerian angular velocity.

The innermost parts of the stellar core may not possess sufficient angular momentum to circularize in an accretion disk and, instead, directly collapse into a BH. We define the initial BH as a `seed BH' formed by the innermost stellar layers up to radius $r_{\bullet,0}$ with enclosed mass $M_{\bullet,0}=0.5\,M_\odot$, a safe assumption for all stellar models considered here. This seed BH has dimensionless spin parameter
\begin{equation}
  a_{\bullet,0} = \frac{c J_{\rm enc}(r_{\bullet,0})}{GM^2_{\rm enc}(r_{\bullet,0})},
\end{equation}
where $J_{\rm enc}(r)$ is the enclosed angular momentum, $G$ is the gravitational constant, and $c$ is the speed of light. The corresponding innermost stable circular orbit (ISCO) is given by \citep{bardeen_rotating_1972}
\begin{eqnarray}
  r_\mathrm{ISCO}(M_{\bullet},a_{\bullet}) &=& \frac{GM_{\bullet}}{c^2}\\
  & \times &\left\{ 3 + z_2 - \left[ (3-z_1)(3+z_1+2z_2)\right]^{1/2}\right\}, \nonumber
  \label{eq:r_isco}
\end{eqnarray}
where
\begin{eqnarray}
  z_1 &=& 1 + (1-a_{\bullet}^2)^{1/3}\left[ (1+a_{\bullet})^{1/3} + (1-a_{\bullet})^{1/3}\right], \label{eq:z1}\\
  z_2 &=& (3a_{\bullet}^2 + z_1^2)^{1/2}. \label{eq:z2}
\end{eqnarray}

Upon initial BH formation, we follow the collapse of the outer stellar layers according to Eqs.~\eqref{eq:t_circ} and \eqref{eq:r_circ} and distinguish between mass elements that circularize outside the BH to form a disk ($r_{\rm circ}(r(t),\theta) > r_{\rm ISCO} (t)$), giving rise to a `disk feeding rate' $\dot{m}_{\rm fb, disk}$, and those that directly fall into the BH without accreting through a disk ($r_{\rm circ}(r(t),\theta) \le r_{\rm ISCO} (t)$), giving rise to a direct fallback rate onto the BH $\dot{m}_{\rm fb, \bullet}$. Here, $r(t)$ refers to the radius of a stellar element at polar angle $\theta$ which circularizes at time $t$ in the equatorial plane. We denote the associated rates of angular momentum supplied to the disk and the BH by $\dot{J}_{\rm fb, disk}$ and $\dot{J}_{\rm fb, \bullet}$, respectively.

We follow the evolution of the BH, disk, and ejecta properties solving
the following equations:
\begin{eqnarray}
	\frac{{\rm d}M_{\rm \bullet}}{{\rm d} t} &=& \dot{m}_{\rm fb, \bullet} + \dot{m}_{\rm acc}, \label{eq:evol_eqn_1}\\
	\frac{{\rm d}J_{\rm \bullet}}{{\rm d} t} &=& \dot{J}_{\rm fb, \bullet} + \dot{m}_{\rm acc} j_{\rm ISCO},\\
	\frac{{\rm d}M_{\rm disk}}{{\rm d} t} &=& \dot{m}_{\rm fb, disk} - \dot{m}_{\rm acc} - \dot{m}_{\rm wind}, \\
	\frac{{\rm d}J_{\rm disk}}{{\rm d} t} &=& \dot{J}_{\rm fb, disk} - \dot{m}_{\rm acc} j_{\rm ISCO} - \dot{J}_{\rm wind},\\
	\frac{{\rm d}M_{\rm ejecta}}{{\rm d} t} &=& \dot{m}_{\rm wind}, \\
	\frac{{\rm d}J_{\rm ejecta}}{{\rm d} t} &=& \dot{J}_{\rm wind}. \label{eq:evol_eqn_6}
\end{eqnarray}
Here,
\begin{eqnarray}
	j_{\rm ISCO} &=& (G M_{\rm \bullet}r_{\rm ISCO})^{1/2} \times\\
	&& \mskip-40mu \frac{r^2_{\mathrm{ISCO}}- a_{\rm \bullet}r_g(r_{\mathrm{ISCO}}r_g/2)^{1/2}+a_{\rm \bullet}^2 r_g^2/4}{r_{\mathrm{ISCO}}\left[r^2_{\mathrm{ISCO}}-3r_{\mathrm{ISCO}} r_g/2+ a_{\rm \bullet} r_g(r_{\mathrm{ISCO}}r_g/2)^{1/2}\right]^{1/2}} \nonumber
\end{eqnarray}
is the specific angular momentum of a fluid element at the ISCO of the BH with mass $M_{\rm \bullet}$, spin $a_{\rm \bullet}= c J_{\rm \bullet}/G M_{\rm \bullet}^2$, and gravitational radius $r_g = 2G M_{\rm \bullet}/c^2$ \citep{bardeen_rotating_1972}. Mass is accreted onto the BH at a rate
\begin{equation}
	\dot{m}_{\rm acc} = f_{\rm acc} \frac{M_{\rm disk}}{t_{\rm visc}}, \label{eq:accretion_rate}
\end{equation}
where
\begin{equation}
	t_{\rm visc} = \alpha^{-1} \Omega_{\rm K, disk}^{-1} h_{\rm z, disk}^{-2} \label{eq:t_visc}
\end{equation}
is the viscous timescale of the disk, with $\alpha$ being the standard dimensionless disk viscosity \citep{shakura_black_1973}, 
\begin{equation}
    \Omega_{\rm K, disk}(t) = (G M_{\rm \bullet}/r_{\rm disk}^{3})^{1/2} \label{eq:Omega_disk}
\end{equation}
is the Keplerian angular velocity of the disk, and $h_{\rm z, disk}$ its scale height (we take $h_{\rm z, disk}\approx 0.5$ as a fiducial value). The disk radius $r_{\rm disk}(t)$ is defined by the current disk mass and angular momentum,
\begin{equation}
	j_{\rm disk} \equiv (G M_{\rm \bullet} r_{\rm disk})^{1/2} = \frac{J_{\rm disk}}{M_{\rm disk}}. \label{eq:r_disk}
\end{equation}
The disk accretion flow gives rise to powerful outflows with mass-loss at a rate
\begin{equation}
	\dot{m}_{\rm wind} = (1-f_{\rm acc}) \frac{M_{\rm disk}}{t_{\rm visc}}, \label{eq:wind_outflow_rate}
\end{equation}
and associated angular momentum loss rate
\begin{equation}
	\dot{J}_{\rm wind} = \dot{m}_{\rm wind} j_{\rm disk}. \label{eq:angular_momentum_wind}
\end{equation}

Neutrinos cool the disk effectively above the critical ``ignition'' accretion rate for weak interactions \citep{chen_neutrino-cooled_2007,metzger_conditions_2008,siegel_collapsars_2019,de_igniting_2020}, which is approximately given by (see Appendix \ref{sec:ignition})
\begin{equation}
	\dot{M}_{\rm ign}\approx 2\times 10^{-3} M_\odot {\rm s}^{-1}  
	 \left(\frac{\alpha}{0.02}\right)^{5/3}
	 \left(\frac{M_{\rm \bullet}}{3 M_\odot}\right)^{4/3}. \label{eq:Mdot_ign}
\end{equation}
Motivated by the findings of GRMHD simulations of neutrino-cooled accretion flows \citep{siegel_three-dimensional_2018,Fernandez+19,siegel_collapsars_2019,de_igniting_2020}, we assume that for high accretion rates $>\!\dot{M}_{\rm ign}$ a fraction $1 - f_{\rm acc} \approx 0.3$ of the disk mass is unbound in outflows.  This fraction is assumed to increase to $\approx 0.6$ below $\dot{M}_{\rm ign}$, under the assumption that inefficient cooling will result in excess heating and outflow production (e.g., \citealt{Blandford&Begelman99,de_igniting_2020}).  Similarly, we assume that enhanced outflow production occurs also at very high accretion rates, for which neutrinos become effectively trapped in the optically thick accretion disk and are advected into the BH before radiating.  This threshold ``trapping'' accretion rate is given by (see Appendix \ref{sec:trapping})
\begin{equation}
	\dot{M}_{\nu, {\rm trap}}\approx 1 M_\odot {\rm s}^{-1}  
    \left(\frac{\alpha}{0.02}\right)^{1/3}\left(\frac{M_{\rm \bullet}}{3M_{\odot}}\right)^{4/3}. \label{eq:Mdot_trap}
\end{equation}
Insofar as $\dot{M}_{\nu, {\rm trap}}$ scales in the same way with the (growing) BH mass as $\dot{M}_{\rm ign}$, we find this trapped regime is of little practical importance in our models.  In summary, the accretion efficiency is given by
\begin{equation}
	f_{\rm acc} = \left\{ \begin{array}{cc}
		0.4, & \dot{M}_{\nu,{\rm trap}} \le \frac{M_{\rm disk}}{t_{\rm visc}} \\
		0.7, & \dot{M}_{\rm ign} < \frac{M_{\rm disk}}{t_{\rm visc}} < \dot{M}_{\nu,{\rm trap}}  \\
		0.4, & \frac{M_{\rm disk}}{t_{\rm visc}} \le \dot{M}_{\rm ign}
	\end{array}\right. . \label{eq:accretion_fraction}
\end{equation}

Eqs.~\eqref{eq:evol_eqn_1}--\eqref{eq:evol_eqn_6} allow a calculation of the total ejecta mass $M_{\rm ejecta}$ obtained from a particular collapsar model. We evolve this set of coupled differential equations numerically until all stellar progenitor material has collapsed and has either been accreted onto the BH or been ejected into outflows.  Note that these equations explicitly conserve mass and angular momentum. Time stepping is equidistant in $\log t$ and chosen sufficiently high, such that i) the accuracy of the total fallback mass is dominated by the radial resolution of the provided stellar model (see Appendix \ref{app:convergence}) and that ii) conservation of total mass and angular momentum in Eqs.~\eqref{eq:evol_eqn_1}--\eqref{eq:evol_eqn_6} is achieved to better than $10^{-14}$ relative accuracy for all model runs.

Once a disk forms around the BH and its accretion rate $\dot{m}_{\rm acc}$ exceeds $10^{-4}\,M_\odot\,\text{s}^{-1}$, we assume that a relativistic jet emerges, powerful enough to drill through the remaining outer layers in the polar region. This threshold is motivated by typical GRB luminosities $L_\gamma\sim\!2\times 10^{50}\,\text{erg}\,\text{s}^{-1}$ \citep{goldstein_estimating_2016}, which, if accretion powered, require an accretion rate of at least $\dot{M}\sim L_\gamma / c^2 \sim 1.1\times 10^{-4}\,M_\odot\,\text{s}^{-1}$.  If this threshold is surpassed, we ignore any remaining material in the polar regions $\theta < \theta_{\rm jet}$ and $\theta > 180^\circ - \theta_{\rm jet}$ for the subsequent fallback process. This material has little effect on the total quantity of material accreted through the disk as it predominantly falls into the BH directly due to the low angular momentum in these regions. However, it has a slight indirect effect on nucleosynthesis by modifying the BH mass (see below). As a fiducial value, we take $\theta_{\rm jet}=30^\circ$.  We further justify the existence of such a successful jet {\it a posteriori} by the fact that our models reach the regime $\dot{M}>10^{-4}\,M_\odot\,\text{s}^{-1}$ favorable for powering typical observed long GRBs including the time necessary for the jet to drill through the stellar envelope (see Sec.~\ref{sec:fallback_results}).

The fallback process may in some cases give rise to massive, gravitationally unstable accretion disks. In this limit, the disk mass becomes comparable to the BH mass, and our assumption of a Kerr metric would not be justified anymore. We estimate this instability region by monitoring the ratio of self-gravity to external gravitational acceleration by the BH potential  \citep{paczynski_model_1978,gammie_nonlinear_2001},
\begin{equation}
	Q^{-1} \equiv \frac{2 \pi \Sigma r_{\rm disk}^2}{M_{\rm \bullet}h_{z,{\rm disk}}} \simeq \frac{2}{h_{z,{\rm disk}}} \frac{M_{\rm disk}}{M_{\rm \bullet}} > 1, \label{eq:gravitational_instability}
\end{equation}
where $\Sigma$ is the disk's surface density. If $Q<1$, we remove excess disk mass by enhancing accretion and wind production such as to restore $Q=1$. This is motivated by the fact that gravitationally unstable disks tend to self-regulate by increased angular momentum transport via gravitationally driven turbulence, thereby increasing the accretion rate and reducing the disk mass until $Q>1$ (e.g., \citealt{gammie_nonlinear_2001}).

The composition of the disk wind ejecta at a given time depends most sensitively on the instantaneous accretion rate \citep{siegel_collapsars_2019}. Following \citet{siegel_collapsars_2019}, \citet{Miller+20}, and \citet{Li&Siegel21}, we define the following accretion regimes:
\begin{equation}
	\frac{M_{\rm disk}}{t_{\rm visc}} = \left\{ \begin{array}{cc}
		> \dot{M}_{\nu,{\rm r-p}} & \text{weak $r$-process} \\
		\in [2\dot{M}_{\rm ign}, \dot{M}_{\nu,{\rm r-p}}] & \text{strong $r$-process}  \\
		\in [\dot{M}_{\rm ign}, 2\dot{M}_{\rm ign}] & \text{weak $r$-process} \\
		< \dot{M}_{\rm ign} & \text{no $r$-process,} \\
		&  ^{56}\text{Ni production}
	\end{array}\right. . \label{eq:accretion_regimes}
\end{equation}
Here, $\dot{M}_{\nu,{\rm r-p}}$ represents a threshold between production of lanthanides and first-to-second peak $r$-process elements only, accounting for the fact that increased neutrino irradiation at high accretion rates tends to raise the electron fraction above $\approx\!0.25$ required for lanthanide production (e.g., \citealt{lippuner_r-process_2015}). We assume this threshold scales with the accretion rate above which the inner disk becomes optically thick to neutrinos, which we estimate as (see Appendix \ref{sec:trapping})
\begin{equation}
	\dot{M}_{\nu, {\rm r-p}}\approx 0.1 M_\odot {\rm s}^{-1}  
    \left(\frac{\alpha}{0.02}\right) \left(\frac{M_{\rm \bullet}}{3M_{\odot}}\right)^{4/3}. \label{eq:Mdot_opaque}
\end{equation}
This expression has been normalized using numerical results by \citet{siegel_collapsars_2019} and \citet{Miller+20} for $M_{\rm \bullet}\approx 3 M_\odot$. Additionally including the effects of neutrino fast flavor conversions may increase $\dot{M}_{\nu,{\rm r-p}}$ significantly \citep{Li&Siegel21}, possibly up to $\approx\!1M_\odot$ s$^{-1}$ or higher for such light BHs.  We therefore treat the normalization as a free parameter and explore different scenarios in which the value is scaled up by a factor of ten.

Below the ignition rate $\dot{M}_{\rm ign}$, $r$-process production ceases abruptly and nucleosynthesis in the outflows with roughly equal numbers of neutrons and protons ($Y_e\simeq 0.5$) only proceeds up to iron-peak elements \citep{siegel_collapsars_2019}. A large fraction of the outflowing material in this epoch remains, however, as $^{4}$He instead of forming heavier isotopes. This is due to the slow rate of the triple-$\alpha$ reaction needed to create seed nuclei when $Y_{e} \approx 0.5$, relative to the much faster neutron-catalyzed reaction $^{4}$He($\alpha n,\gamma)^{9}$Be($\alpha,n)^{12}$C that operates when $Y_e \ll 0.5$ \citep{woosley_alpha-process_1992}. Here, we employ a simple model to estimate the yield of $^{56}$Ni in such $Y_e\simeq 0.5$ disk outflows, similar to \citet{siegel_collapsars_2019} (see Appendix \ref{sec:Ni56_production}).

A requisite for the synthesis of $^{56}$Ni in disk outflows is that nuclei from stellar fallback material are dissociated into individual nucleons once entering the inner part of the accretion disk. At late times during the accretion process, the disk densities and temperatures may not be high enough to ensure full dissociation. We estimate the transition time $t_{\rm diss}$ to this state by evaluating the conditions under which only 50\% of $\alpha$ particles are dissociated in the disk (see Appendix \ref{sec:Ni56_production}). For $t>t_{\rm diss}$ we ignore any potential further nucleosynthesis in disk outflows.

\subsection{Collapsar Model Results}
\label{sec:fallback_results}

We start in Sec.~\ref{sec:results_basic_model} by walking through the evolution of the collapse and mass-ejection process for a representative model corresponding to a star above the nominal PI mass gap.  Appendix~\ref{app:collapsars} presents the results of our model when applied to ``ordinary'' low-mass collapsars (with BH masses below the PI mass gap), demonstrating that for the fiducial range of parameters considered in this work, we obtain properties in agreement with observed GRBs and previously predicted $r$-process ejecta.  Using the same parameters (now ``calibrated'' to reproduce the properties of ordinary collapsars) we present in Sec.~\ref{sec:results_massgap_collapsars} a parameter exploration of ejecta masses and nuclear compositions for massive collapsars above the PI mass gap.

\subsubsection{Basic Model Evolution}
\label{sec:results_basic_model}

\begin{figure}
\centering
\includegraphics[width=0.99\linewidth]{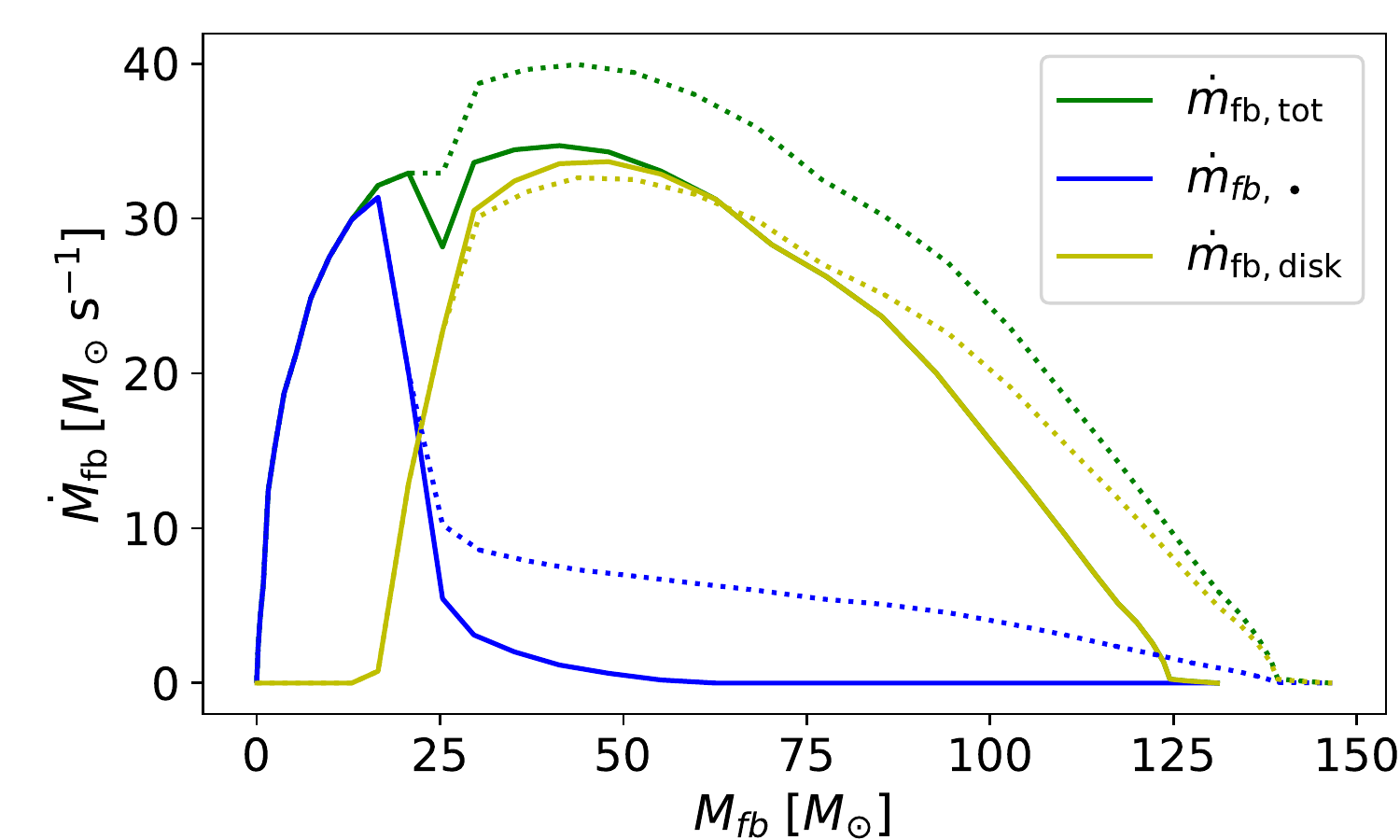}
\includegraphics[width=0.99\linewidth]{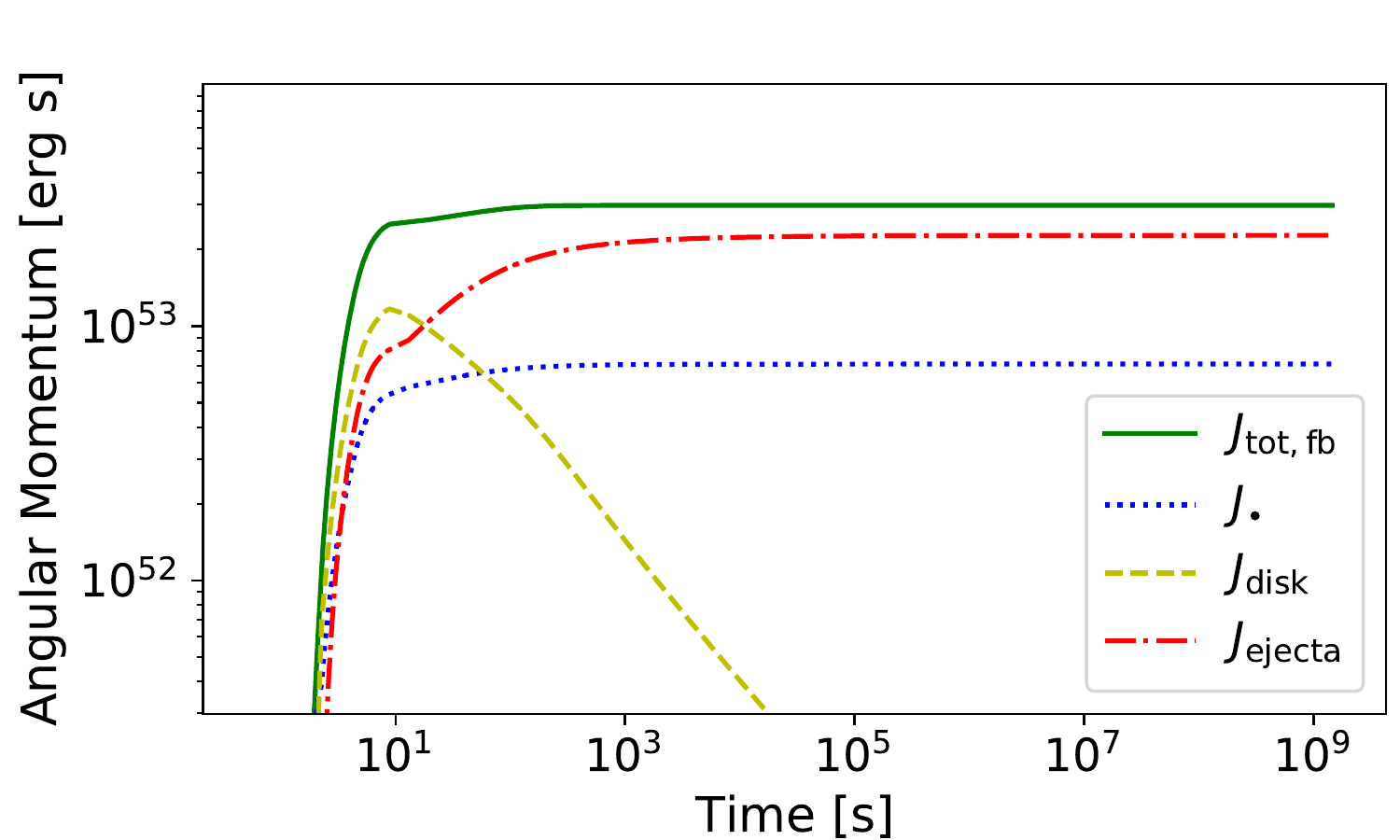}
\includegraphics[width=0.99\linewidth]{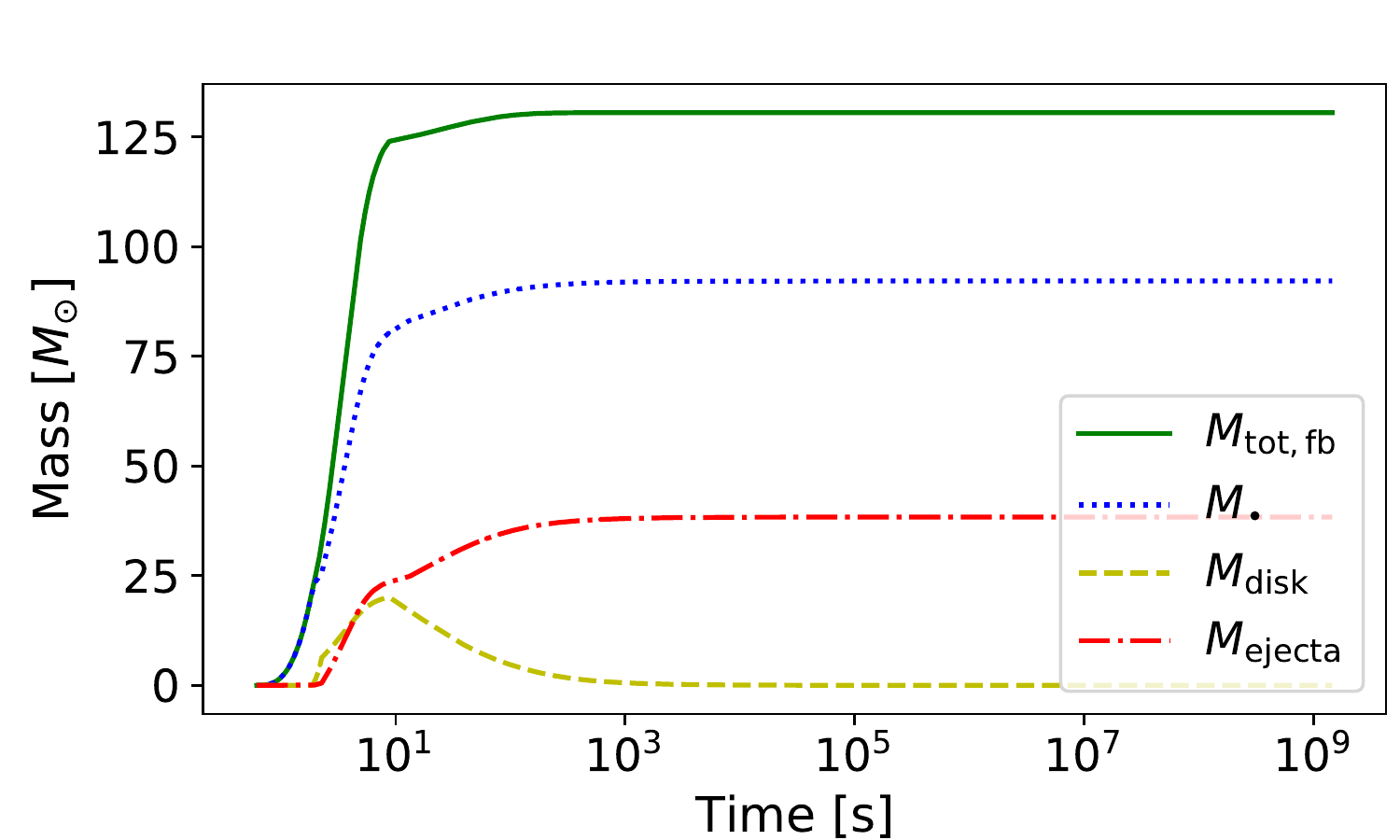}
\caption{Collapse evolution for a representative stellar model \texttt{250.25} with typical rotation parameters $p=4.5$, $f_{\rm K}=0.3$ and $r_{\rm b}=1.5\times 10^{9}$\,cm. Top: fallback rates $\dot{M}_{\rm fb}$ onto the BH (direct; blue), onto an accretion disk (yellow), and total (green), as a function of the total cumulative collapsed mass $M_{\rm fb}$. Dotted lines indicate the corresponding evolution when ignoring the effect of a jet. Center and bottom: evolution of angular momenta (center) and masses (bottom) as determined by Eqs.~\eqref{eq:evol_eqn_1}--\eqref{eq:evol_eqn_6}.}
\label{fig:Renzo_evolution}
\end{figure}

\begin{figure}[htbp]
\centering
\includegraphics[width=0.99\linewidth]{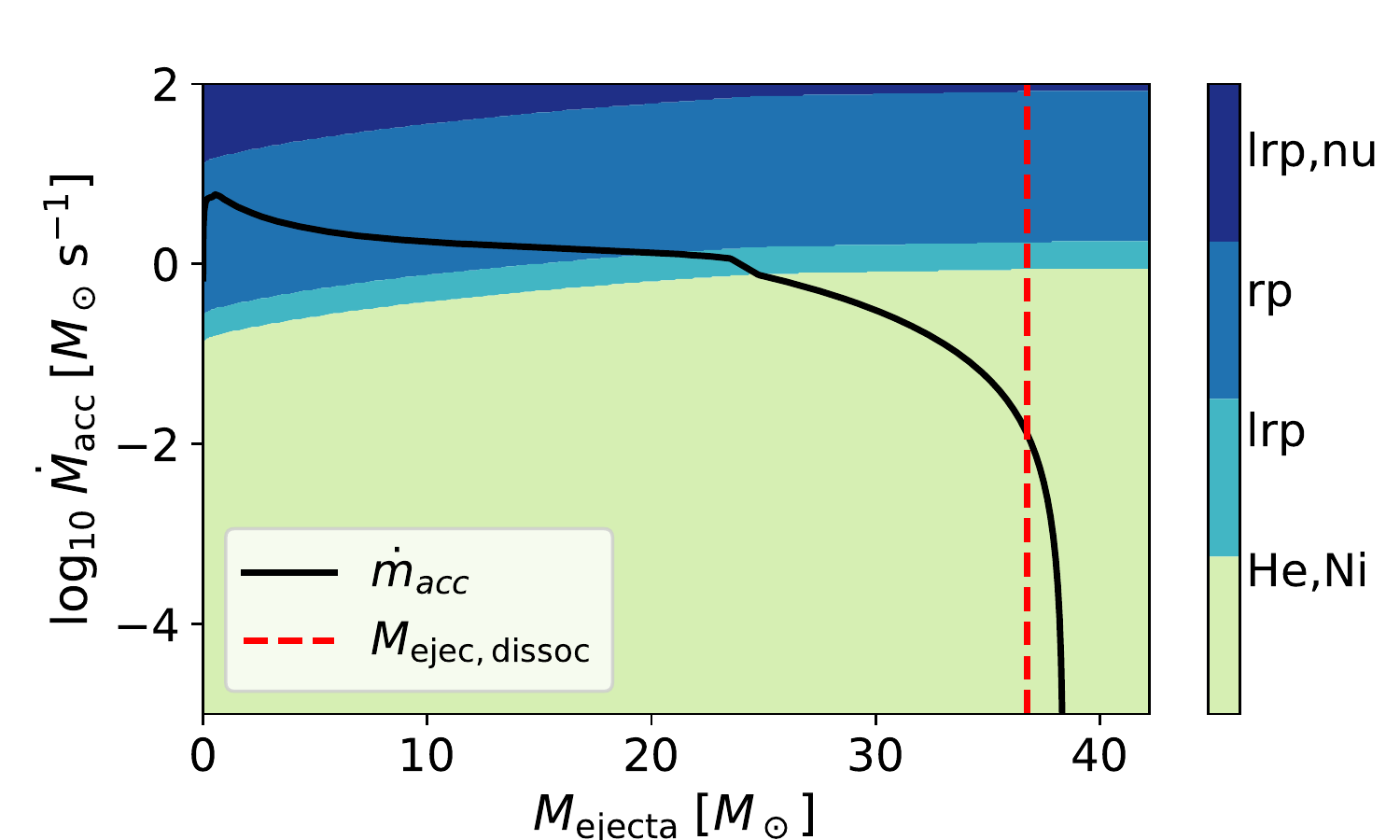}
\includegraphics[width=0.99\linewidth]{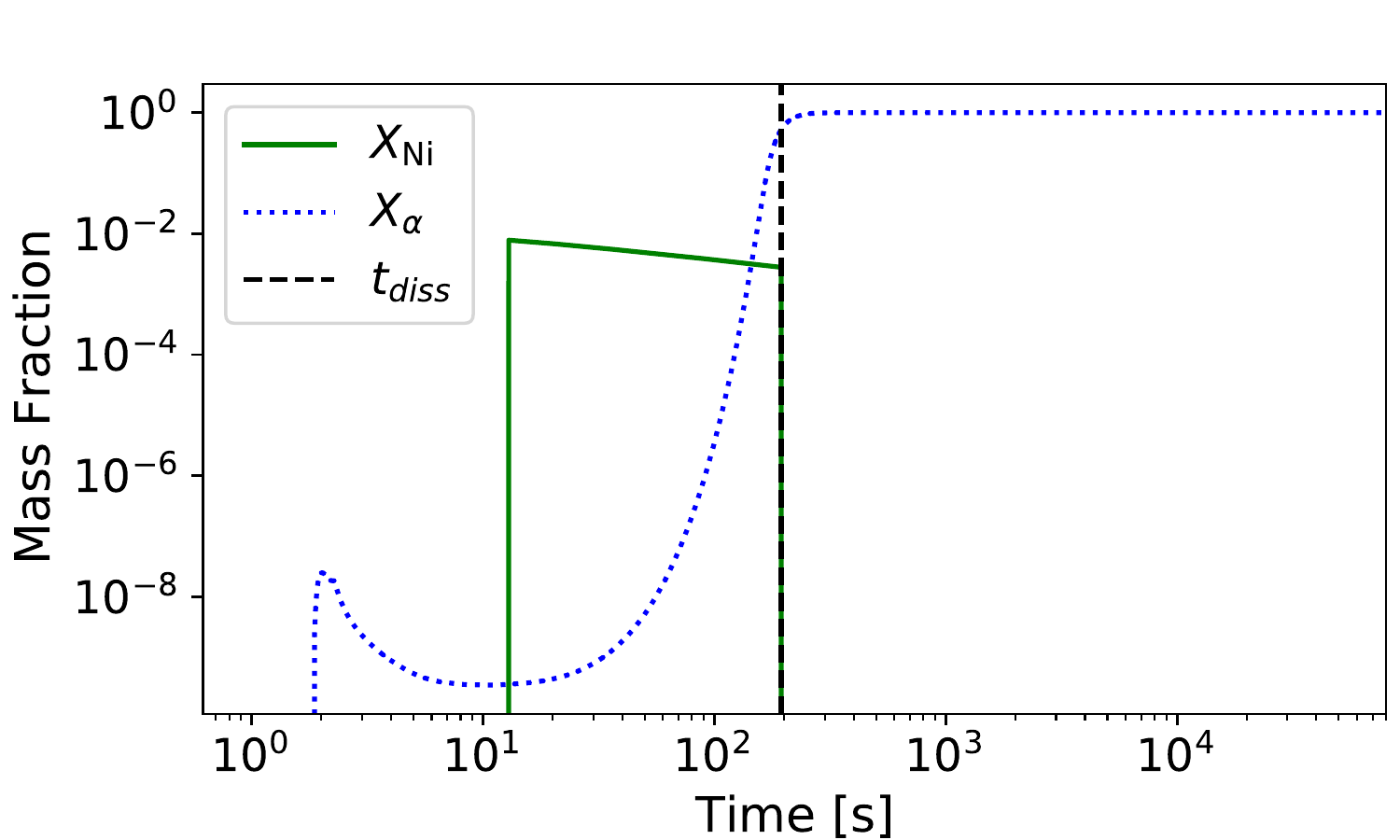}
\caption{Top: accretion rate at which ejecta is being produced as a function of cumulative ejecta mass for model \texttt{250.25} with $p=4.5$, $f_{\rm kep}=0.3$, and $r_{\rm b}=1.5\times 10^{9}$\,cm. The nucleosynthesis regimes according to Eq.~\eqref{eq:accretion_regimes} are color-coded. Bottom: corresponding mass fraction of $^{56}$Ni synthesized in disk outflows and of $^{4}$He in the accretion disk. The vertical dashed line refers to the time $t_{\rm diss}$ at which only 50\% of $\alpha$-particles are dissociated in the disk. For $t>t_{\rm diss}$ we ignore further $^{56}$Ni production in the outflows.}
\label{fig:Renzo_accretion_rate}
\end{figure}

Figure \ref{fig:Renzo_evolution} illustrates the collapse evolution of model \texttt{250.25} with representative rotation parameters of $p=4.5$, $f_{\rm K}=0.3$, and $r_{\rm b}=1.5\times 10^{9}$\,cm. Upon seed BH formation, the BH grows rapidly in mass and spin through low-angular momentum material at small radii directly falling into the seed BH before circularizing. Only after a few seconds and accretion of $\approx\!10\,M_\odot$ of the inner layers, first material starts to circularize outside the BH horizon to form an accretion disk (cf.~Fig.~\ref{fig:Renzo_evolution}, top and bottom panel), and initiate accretion onto the BH through a disk in addition to direct infall.

Direct fallback onto the BH subsides after the accretion of about $20\,M_\odot$ in this model (cf.~Fig.~\ref{fig:Renzo_evolution}, top panel) when a significant fraction of low angular momentum material residing in the polar region of the progenitor model has fallen into the BH. Further BH growth then proceeds almost entirely through disk accretion. This initial direct fallback episode partially clears up the polar regions for a relativistic jet to propagate through the outer stellar layers to eventually break out of the star and generate a long gamma-ray burst. Around the same time, a significant fallback rate onto the disk sets in (cf.~Fig.~\ref{fig:Renzo_evolution}, top panel) to establish a heavy $\sim\!15\,M_\odot$ accretion disk on a timescale of a few seconds (cf.~Fig.~\ref{fig:Renzo_evolution}, bottom panel). The disk accretion rate onto the BH, $\dot{m}_{\rm acc}$, quickly exceeds $\dot{M}_{\rm ign}$ and we assume that a relativistic jet forms. This removes the remaining low-angular momentum material in the polar regions and thus results in suppression of direct fallback onto the BH, which becomes negligible compared to disk fallback (cf.~Fig.~\ref{fig:Renzo_evolution}, top panel).

The top panel of Fig.~\ref{fig:Renzo_evolution} also shows that ignoring the effect of such a jet would lead to subdominant extended direct fallback of residual low-angular momentum material in polar regions onto the BH. While this does not have a direct impact on disk accretion, it has minor indirect consequences on nucleosynthesis in the disk winds due to its effect on the BH mass (cf.~Eq.~\eqref{eq:accretion_regimes}). For somewhat larger values of $r_{\rm b}$, the situation changes and direct fallback onto the BH may extend to late times even in the presence of a jet, due to the overall lower angular momentum budget of the progenitor star outside the polar cone with opening angle $\theta_{\rm jet}$. For more extreme scenarios, fallback onto the disk may become close to non-existent.

As soon as the disk forms, most angular momentum resides in the disk rather than the BH in this model (cf.~Fig.~\ref{fig:Renzo_evolution}, center panel). The majority of this is being blown off in the ejecta, while a subdominant amount is transferred to the BH as disk matter gradually accretes through the ISCO onto the BH. For significantly larger values of $r_b$ this trend reverses, and most angular momentum is transferred to the BH rather than the ejecta as less material accretes through a disk.

The top panel of Fig.~\ref{fig:Renzo_accretion_rate} shows the history of ejecta production in the model discussed above. Shown is the instantaneous accretion state $M_{\rm disk}/t_{\rm visc}$ of the disk as a function of the cumulative ejected wind mass, together with the nucleosynthesis regimes defined in Eq.~\eqref{eq:accretion_regimes}. This evolution shows a `sweep' through most nucleosynthesis regimes, typical of the models considered here. Nucleosynthesis regimes change during the evolution as a result of the BH mass growth and can be more dramatic in some cases than illustrated here. Outflows are first created in the regime of a main $r$-process with lanthanide production, during which the bulk of the wind ejecta is produced. The remaining $\approx10\,M_\odot$ of ejecta originate in a regime that mostly ejects $\alpha$-particles and $\sim\!0.1\,M_\odot$ of $^{56}$Ni. The bottom panel of Fig.~\ref{fig:Renzo_accretion_rate} illustrates $^{56}$Ni production in this regime. Shown are the mass fraction of $^{56}$Ni produced in disk outflows according to Eq.~\eqref{eq:Yseed} as well as the mass fraction of $\alpha$-particles in the accretion disk according to Eq.~\eqref{eq:Saha}. The vertical dashed line indicates the dissociation time $t_{\rm diss}$ after which $<50\%$ of $\alpha$-particles are dissociated into individual nucleons in the accretion disk (Sec.~\ref{sec:fallback}, Appendix \ref{sec:Ni56_production}). As a conservative estimate, for $t>t_{\rm diss}$, we ignore any further production of $^{56}$Ni according to Eq.~\eqref{eq:Yseed} as the required free nucleons become unavailable. However, this represents only a slight correction in most cases, as by far the dominant amount of $^{56}$Ni is typically produced before $t=t_{\rm diss}$.

\subsubsection{Parameter Study of Massive Collapsars}
\label{sec:results_massgap_collapsars}

Before systematically applying our model across the parameter space of massive collapsars, we first apply it to `ordinary' collapsars of stars well below the PI mass gap, the results of which we describe in Appendix \ref{app:collapsars}.  We use the progenitor models of \citet{heger_presupernova_2000} as representative of typical stellar progenitors of canonical long GRBs \citep{MacFadyen&Woosley99}.  Our results for the nucleosynthesis yields of the disk outflows as a function of the parameters $\{f_{\rm K}, r_{\rm b}\}$ which enter the progenitor angular momentum profile (Fig.~\ref{fig:Heger_ejecta}), broadly agree with those previously presented in \citet{siegel_collapsars_2019}, though some quantitative differences arise due to our more detailed treatment of different regimes of BH accretion (see Appendix \ref{app:collapsars} for a discussion).  Our low-mass collapsar models also exhibit BH accretion timescales and energetics of putative jet activity in agreement with long GRB observations.  We can therefore claim a rough ``calibration'' of our model across the adopted parameter space of progenitor angular momentum properties, allowing for more confidence when extrapolating to the regime of more massive collapsars described below.  

\begin{figure}
\centering
\includegraphics[width=0.98\linewidth]{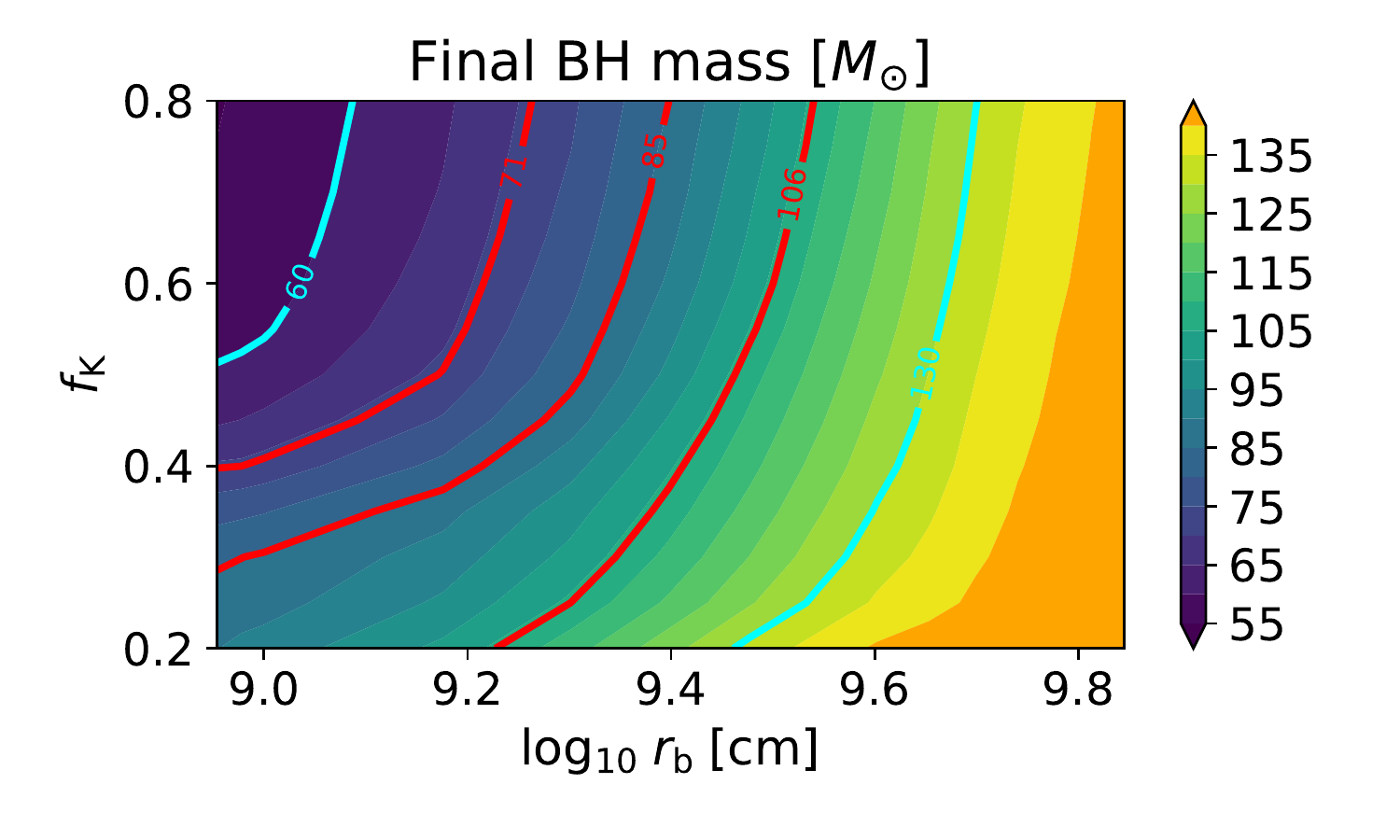}
\includegraphics[width=0.98\linewidth]{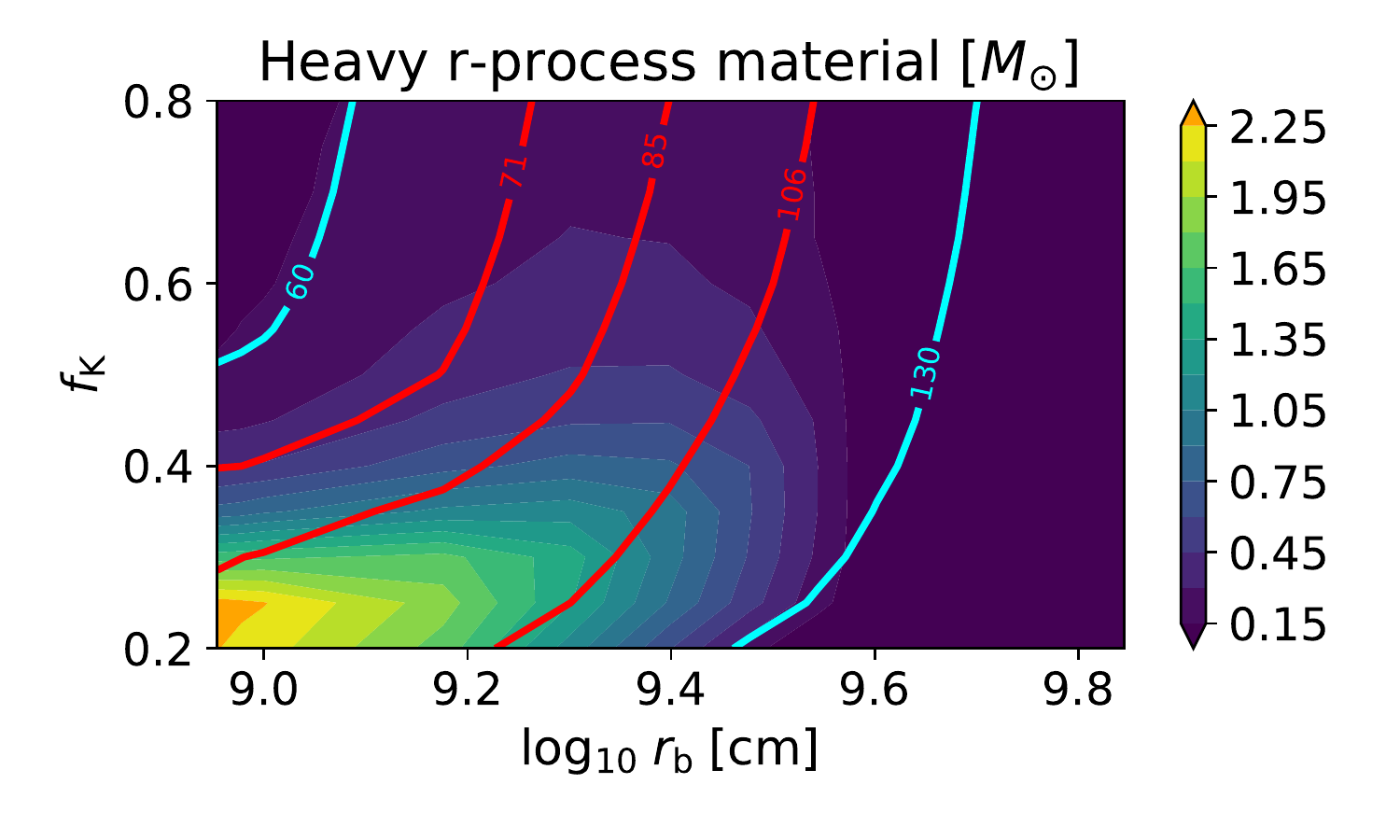}
\includegraphics[width=0.98\linewidth]{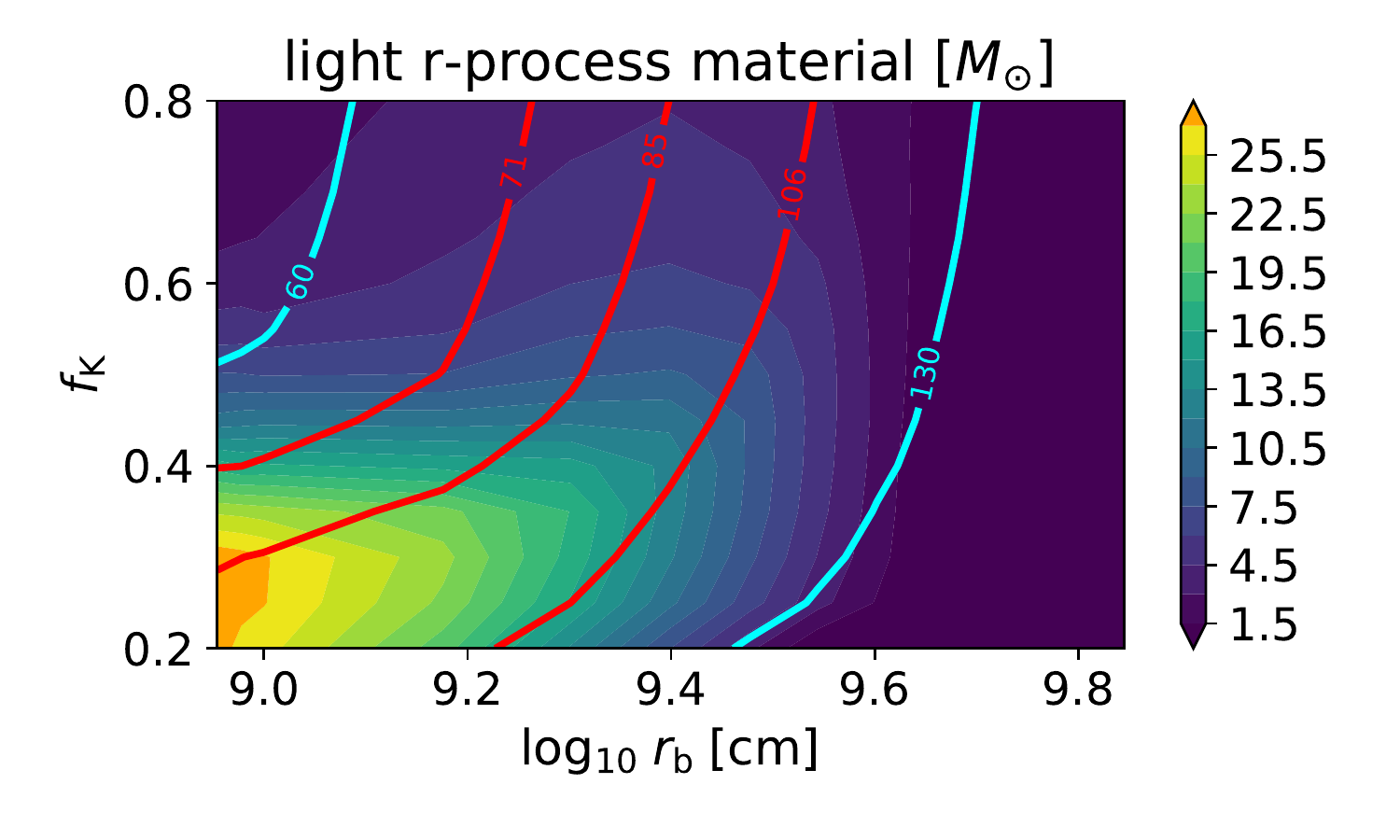}
\includegraphics[width=0.98\linewidth]{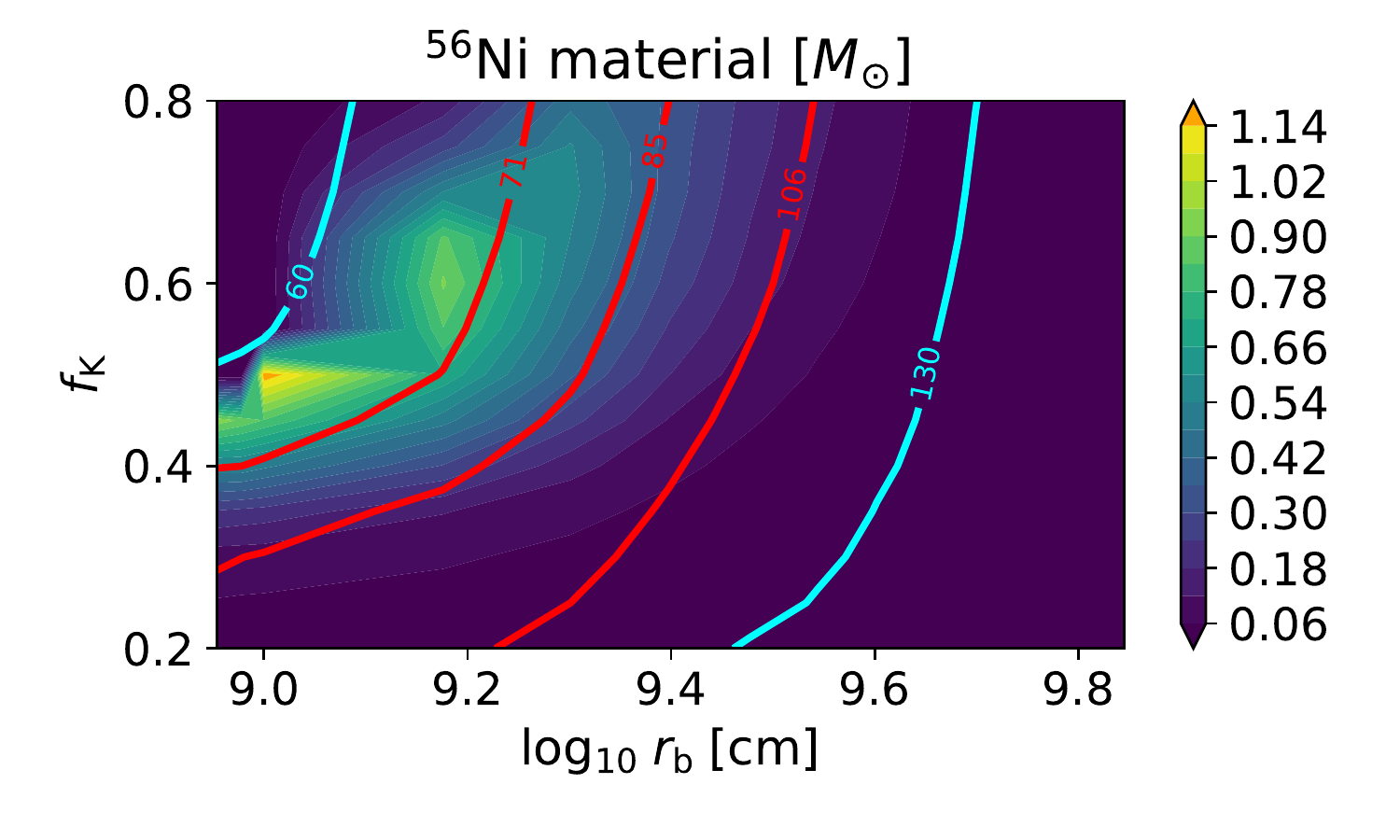}
\caption{BH masses and disk wind ejecta properties across the parameter space of progenitor rotational profiles (envelope Keplerian fraction $f_{\rm K}$ and break radius $r_{\rm b}$; see Fig.~\ref{fig:stellar_rotation_profile}, bottom panel) for model \texttt{250.25}.  Shown are the final BH mass (top), the total ejected mass in heavy ($A>136$) $r$-process elements including lanthanides (center top), light ($A<136$) $r$-process material (center bottom), and $^{56}$Ni (bottom). Red contours indicate the inferred primary mass of \gw{}, together with its 90\% confidence limits ($85^{+21}_{-14}\,M_\odot$; \citealt{Abbott+20_190521}). Cyan contours delineate final BH masses of 60\,$M_\odot$ and 130\,$M_\odot$, which approximately correspond to the lower and upper end of the PI mass gap.}
\label{fig:Renzo_ejecta}
\end{figure}


\begin{figure}
\centering
\includegraphics[width=0.98\linewidth]{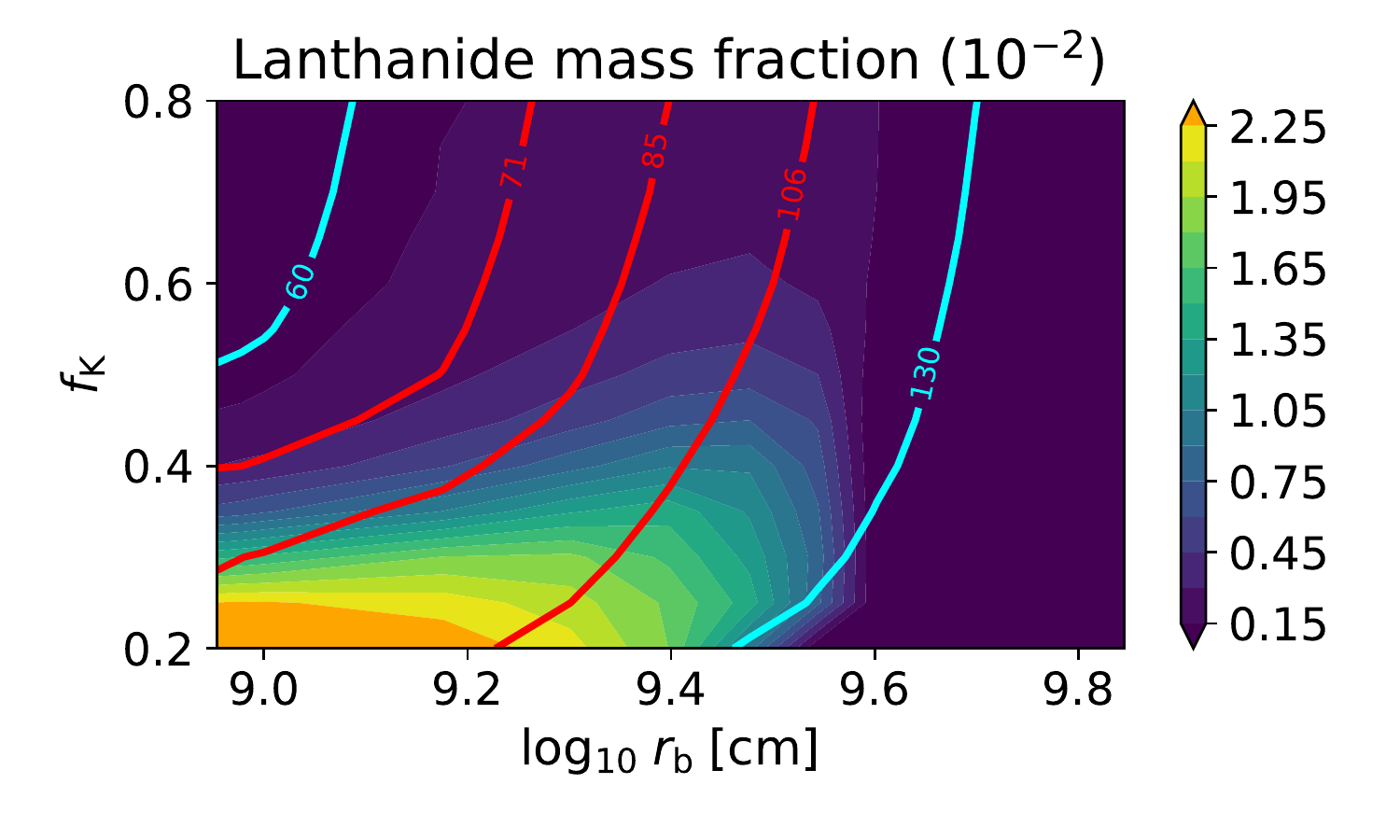}
\includegraphics[width=0.98\linewidth]{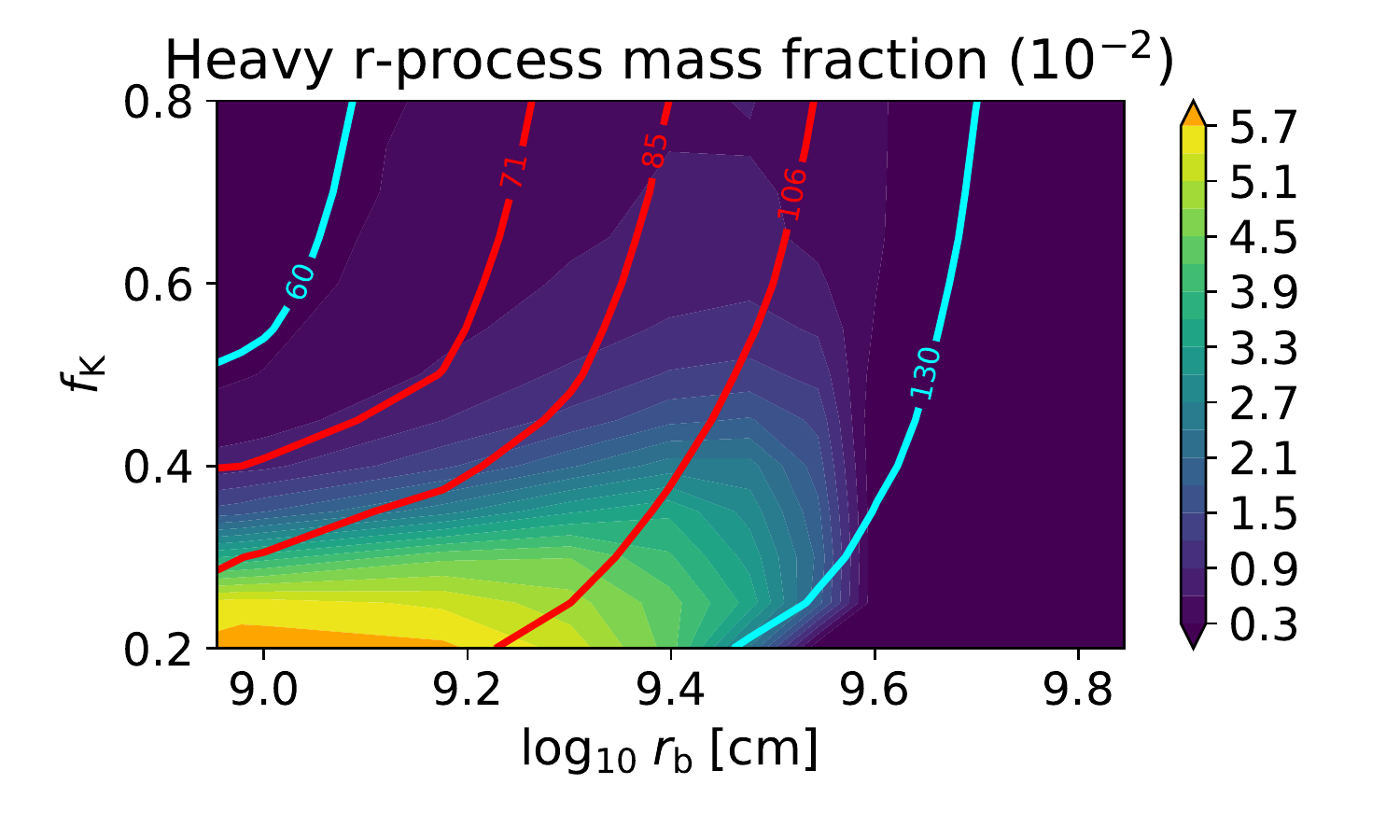}
\includegraphics[width=0.98\linewidth]{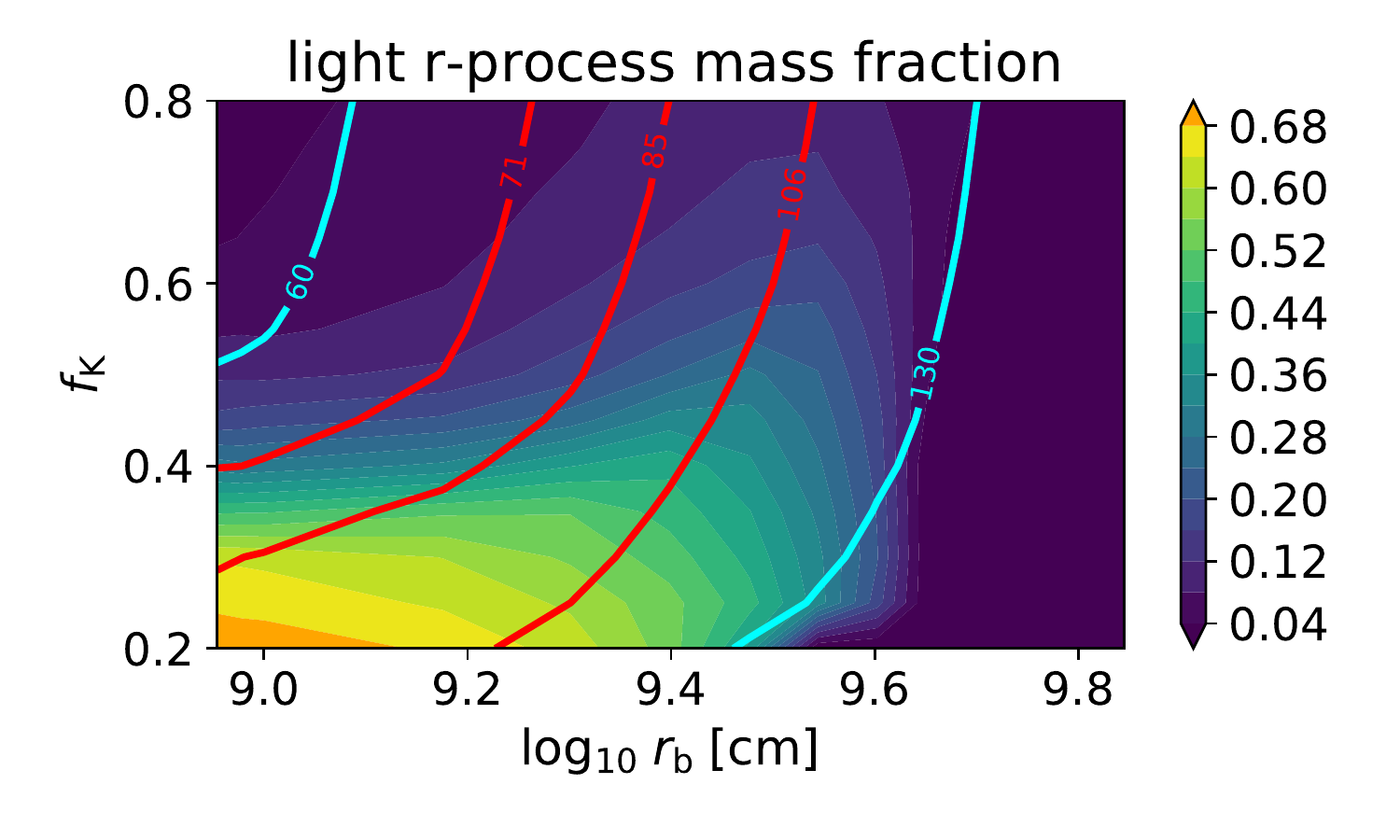}
\includegraphics[width=0.98\linewidth]{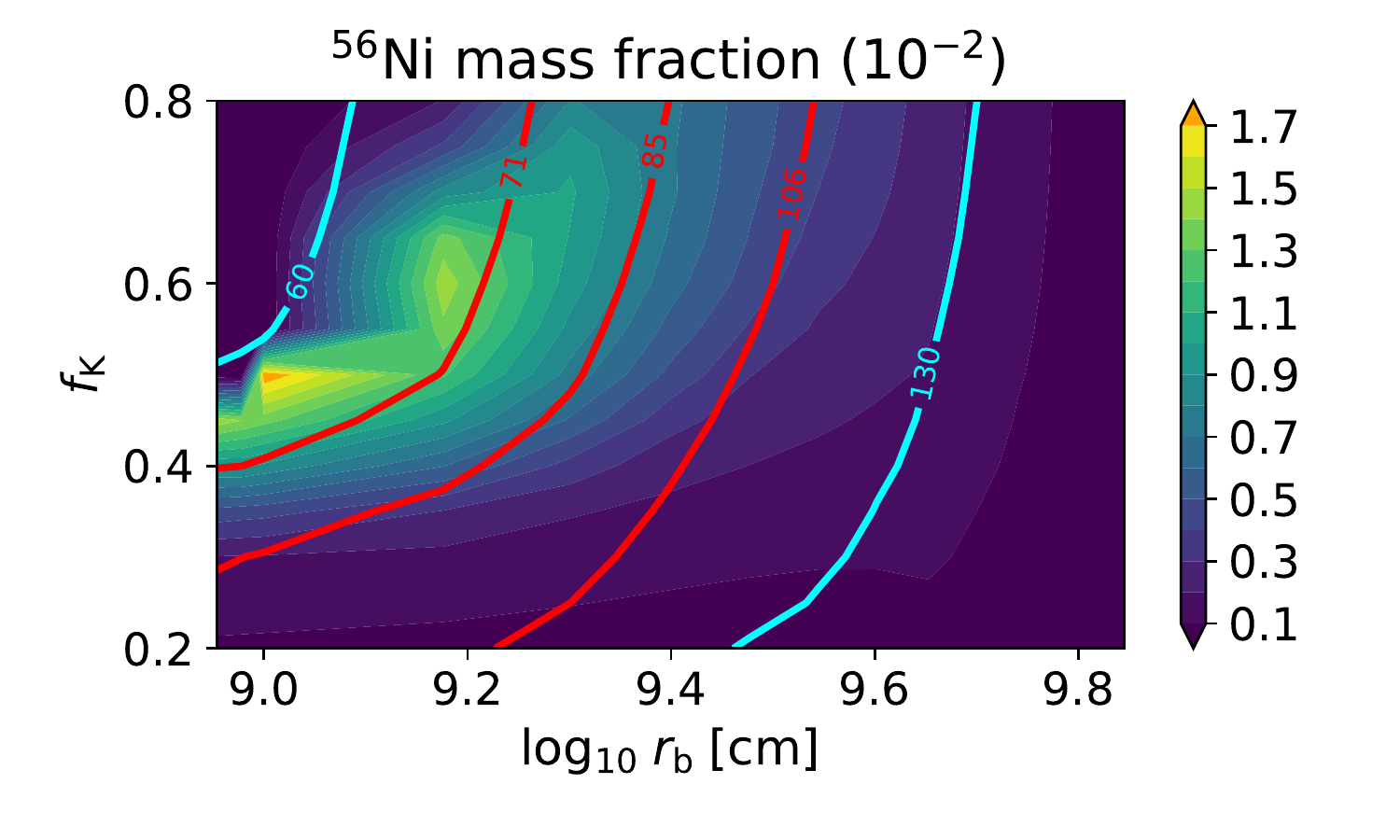}
\caption{Scan of the parameter space for model \texttt{250.25}. We show the mass fractions of lanthanides (top), light $r$-process elements ($A<136$; center), and of $^{56}$Ni (bottom) in the ejecta, assuming full mixing of ejecta components. Red contours indicate the inferred primary mass of \gw{}, together with its 90\% confidence limits. Cyan contours delineate final BH masses of 60\,$M_\odot$ and 130\,$M_\odot$, which approximately correspond to the lower and upper end of the PI mass gap.}
\label{fig:Renzo_mass_fractions}
\end{figure}

Figs.~\ref{fig:Renzo_ejecta} and \ref{fig:Renzo_GRB_accretion} summarize our results for the ejecta and GRB properties for model \texttt{250.25} as a representative example of a stellar model above the PI mass gap, in the parameter space $\{f_{\rm K}, r_{\rm b}\}.$  The top panel of Fig.~\ref{fig:Renzo_ejecta} shows that, even for a progenitor mass $M_{\star} =150\,M_\odot$ at the onset of collapse (that is, well above the PI mass gap), the final BH remnant can populate the entire mass gap between $\sim\!55\,M_\odot-130\,M_\odot$ (for typical parameter values), depending on the rotation profile at the onset of collapse. Labelled contours indicate the inferred primary mass of \gw{}, together with its 90\% confidence limits.  We focus on this region of the parameter space in what follows, insofar that superKNe generated from such events probe BHs formed in the PI mass gap.

As in case of the low-mass collapsars (Appendix~\ref{app:collapsars}), our results are not sensitive to the precise value of the power-law coefficient $p$, which we thus ignore in what follows.  We find ubiquitous $r$-process production throughout the parameter space, ranging between $\sim 0.1-2.3\,M_\odot$ of heavy ($A>136$) $r$-process material including lanthanides and $\sim 1-29\,M_\odot$ of light ($A<136$) $r$-process elements. Additionally, between $\sim 0.05-1\,M_\odot$ of $^{56}$Ni are synthesized in the ejecta.

Interestingly, the region of highest $r$-process production is well aligned with intermediate final BH masses in a range similar to the \gw{} confidence region (Sec.~\ref{sec:GW190521}). For large $r_{\rm b}$ the outer stellar layers possess too little angular momentum to form massive accretion disks that give rise to copious $r$-process ejecta, as most material directly falls into the BH. On the other hand, for small values of $r_{\rm b}$ and high values of $f_{\rm K}$ massive disks form; however, high angular momentum leads to large disk radii $r_{\rm disk}$ and associated viscous timescales, such that the accretion rate drops below the required thresholds for $r$-process production for most of the accretion process. This occurs despite the presence of spiral modes in this regime, which tend to increase the accretion rate (Sec.~\ref{sec:fallback}). Most $r$-process material (both light and heavy) is synthesized for small values of both $r_{\rm b}$ and $f_{\rm K}$, which represents the optimal compromise between high angular momentum and sufficient compactness of the accretion disk.  We discuss the possible contribution of massive collapsars to the long GRB population in Sec.~\ref{sec:GRB}.

For use in our subsequent light curve models (Sec.~\ref{sec:light_curve_models}), we decompose the ejecta content of the collapsar models further into mass fractions of several constituents of interest. Assuming full mixing of all ejecta content
 (see also Sec.~\ref{sec:analytic}), we calculate the mass fraction $X_{\rm La}$ of lanthanides (atomic mass number $136\lesssim A\lesssim 176$) based on the amount of main $r$-process material, assuming the solar $r$-process abundance pattern \citep{arnould_r-process_2007} motivated by the results of \citet{siegel_collapsars_2019}. A mass fraction $X_{\rm lrp}$ for light $r$-process elements is based on the combined mass fraction of light $r$-process material only plus the fraction of main $r$-process ejecta with $A<136$ when applying the solar $r$-process abundance pattern. Finally, we also compute the mass fraction $X_{\rm Ni}$ of $^{56}$Ni. Results are depicted in Fig.~\ref{fig:Renzo_mass_fractions}. For concreteness, we select several models along iso-mass contours for the final BH mass within the \gw{}\ confidence region and report the corresponding ejecta parameters in Tab.~\ref{tab:ejecta_parameters}.

\begin{table*}
	\centering
	\caption{Ejecta parameters for select mass gap collapsar models with $p=4.5$ along contours of constant final BH mass (cf.~Figs.~\ref{fig:Renzo_ejecta} and \ref{fig:Renzo_mass_fractions}).}
	\label{tab:ejecta_parameters}{}
	\begin{tabular}{ccccccccc} 
		\hline
		model & $M_{\bullet}$ & $M_{\rm ej}$ & $r_{\rm b}$ & $f_{\rm K}$ & $X_{\rm La}$ & $X_{\rm lrp}$ & $X_{\rm Ni}$ & $X_{\rm Ni}/X_{\rm rp}$\\
		\hline

		& ($M_\odot$) & ($M_\odot$) & ($10^9$ cm) &  & & & \\
		\hline
		\hline
		\texttt{250.25} & 106 &  27.24 & 1.94  & 0.25 & 0.020 & 0.59 & 0.0011 & 0.0018\\
                        & 106 &  29.67 & 2.35  & 0.35 & 0.014 & 0.47 & 0.0020 & 0.0041\\
                        & 106 &  30.51 & 2.67  & 0.45 & 0.008 & 0.36 & 0.0028 & 0.0075\\
                        & 106 &  31.53 & 3.04  & 0.60 & 0.004 & 0.21 & 0.0039 & 0.0179\\
                        & 85 &  45.57 & 1.20  & 0.35 & 0.012 & 0.47 & 0.0040 & 0.0079\\
                        & 85 &  46.25 & 1.69  & 0.45 & 0.007 & 0.32 & 0.0059 & 0.0175\\
                        & 85 &  47.20 & 1.96  & 0.55 & 0.005 & 0.22 & 0.0070 & 0.0303\\
                        & 85 &  47.48 & 2.06  & 0.60 & 0.004 & 0.19 & 0.0072 & 0.0359\\
                        & 71 &  58.06 & 1.19  & 0.50 & 0.004 & 0.19 & 0.0120 & 0.0601\\
                        & 71 &  58.22 & 1.37  & 0.55 & 0.003 & 0.16 & 0.0094 & 0.0562\\
                        & 71 &  58.00 & 1.50  & 0.60 & 0.003 & 0.13 & 0.0058 & 0.0417\\
                        & 71 &  59.78 & 1.53  & 0.65 & 0.002 & 0.11 & 0.0121 & 0.1035\\
        \texttt{200.25} & 106 &  11.41 & 2.14  & 0.25 & 0.011 & 0.42 & 0.0013 & 0.0029\\
                        & 106 &  13.87 & 2.59  & 0.35 & 0.008 & 0.34 & 0.0019 & 0.0054\\
                        & 85 &  28.77 & 1.32  & 0.35 & 0.016 & 0.53 & 0.0026 & 0.0046\\
                        & 85 &  29.74 & 1.59  & 0.45 & 0.010 & 0.39 & 0.0040 & 0.0096\\
                        & 71 &  40.93 & 1.07  & 0.50 & 0.007 & 0.29 & 0.0079 & 0.0254\\
                        & 71 &  40.64 & 1.21  & 0.55 & 0.006 & 0.25 & 0.0083 & 0.0308\\
		\hline
	\end{tabular}
\end{table*}

\section{Super-Kilonova Emission}
\label{sec:light_curve_models}

As the disk outflows expand away from the BH, the ejecta shell they form eventually gives rise to optical/infrared emission powered by radioactive decay (the ``superKN'').

\subsection{Analytic Estimates}
\label{sec:analytic}

We begin with analytic estimates of the superKN properties.  The total ejecta mass $M_{\rm ej}$ is comprised of up to three main components: (1) radioactive $r$-process nuclei, mass fraction $X_{\rm rp}$; (2) radioactive $^{56}$Ni, $X_{\rm Ni}$; (3) non-radioactive $^{4}$He, $X_{\rm He} = 1 - X_{\rm rp} - X_{\rm Ni}$ (also a placeholder for other non-radioactive elements).  Typical values for our fiducial models (Sec.~\ref{sec:fallback_results}) are $M_{\rm ej} \sim 10-60 M_{\odot}$, $X_{\rm rp} \sim 0.1-0.5$, $X_{\rm Ni} \sim 10^{-3}-10^{-2}$ ($M_{\rm Ni} \sim 10^{-2}-0.5M_{\odot}$).  As described in the previous section, the total $r$-process mass fraction can be further subdivided into that of light $r$-process nuclei $X_{\rm lrb}$ and of lanthanides $X_{\rm La}$.  For simplicity, throughout this section we assume the ejecta are mixed homogeneously into a single approximately spherical shell.  Physically, such mixing could result from hydrodynamic instabilities that develop between different components of the radial and temporally-dependent disk winds and or due to its interaction with the GRB jet (e.g., \citealt{Gottlieb+21}). 

\begin{figure}
\centering
\includegraphics[width=0.98\linewidth]{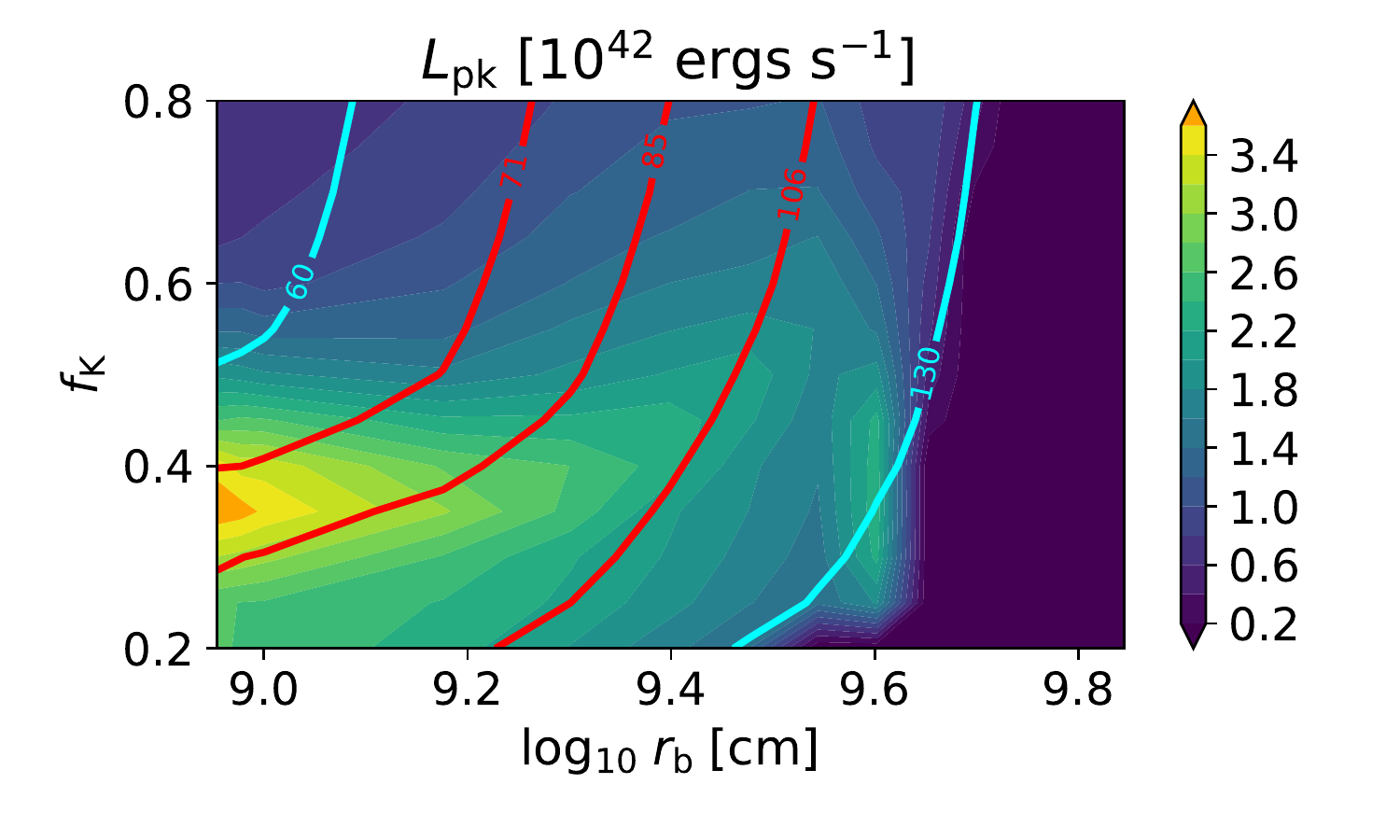}
\includegraphics[width=0.98\linewidth]{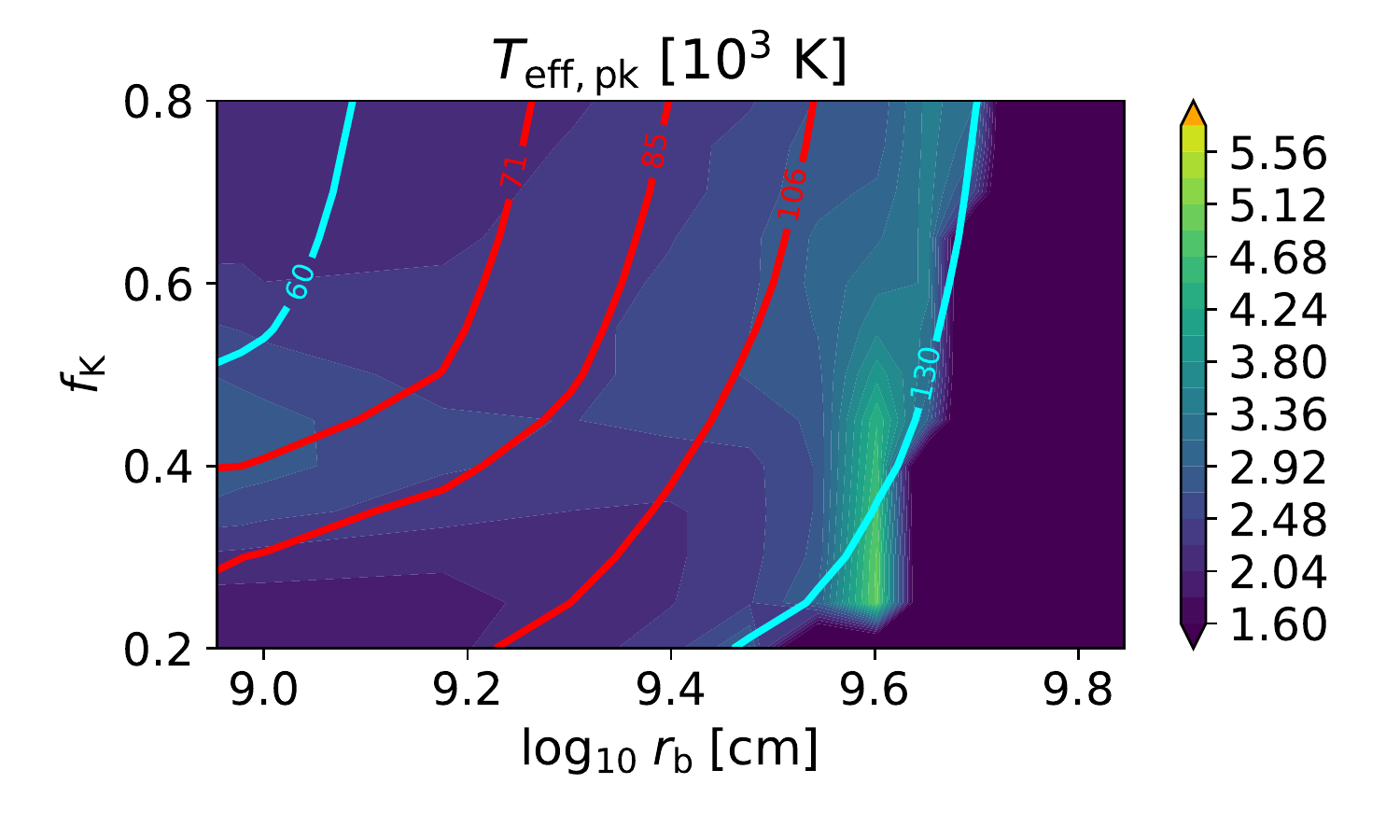}
\includegraphics[width=0.98\linewidth]{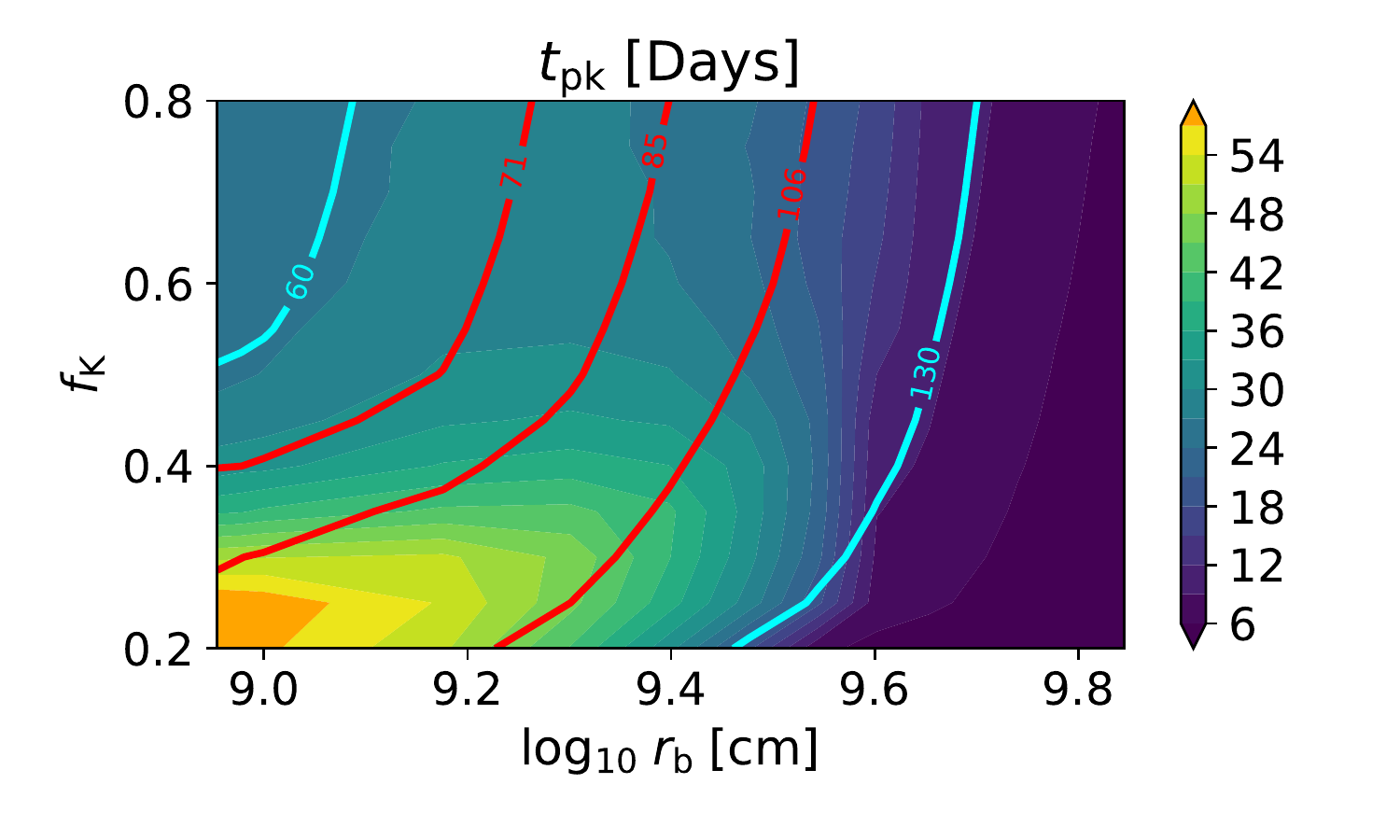}
\caption{Analytic light curve estimates across the parameter space for model \texttt{250.25}. Shown are the peak luminosity (top), peak effective temperature (center), and peak timescale  (bottom). Red contours indicate the inferred primary mass of \gw{}, together with its 90\% confidence limits. Cyan contours delineate final BH masses of 60\,$M_\odot$ and 130\,$M_\odot$, which approximately correspond to the lower and upper end of the PI mass gap.}
\label{fig:Renzo_KN}
\end{figure}

The light curve will peak roughly when the expansion timescale equals the photon diffusion timescale (e.g., \citealt{arnett_type_1982}),
\begin{eqnarray}
    t_{\rm pk} &\approx& \left(\frac{M_{\rm ej}\kappa}{4\pi v_{\rm ej} c}\right)^{1/2}  \\
    &\approx& 108\,{\rm d}\left(\frac{M_{\rm ej}}{50M_{\odot}}\right)^{1/2}\left(\frac{v_{\rm ej}}{0.1c}\right)^{-1/2}\left(\frac{\kappa}{1{\rm cm^{2}\,g^{-1}}}\right)^{1/2},
    \label{eq:tpk} \nonumber
\end{eqnarray}
where $v_{\rm ej}$ is the average ejecta velocity.  The effective gray opacity $\kappa$ varies in kilonovae from $\lesssim 1$ cm$^{2}$ g$^{-1}$ for ejecta dominated by light $r$-process species, to $\sim 10-30$ cm$^{2}$ g$^{-1}$ for ejecta containing a sizable quantity of lanthanide atoms and ions (e.g., \citealt{Kasen+13,Tanaka+20}).  However, $\kappa$ will be smaller than these estimates in the superKN case due to the large mass fraction of light elements, $X_{\rm He} \sim 0.5-0.9$, which contribute negligibly to the opacity.  For our analytical estimates below, we linearly interpolate $\kappa$ between 0.03 cm$^{2}$ g$^{-1}$(at $X_{\rm La}=10^{-4}$) and 3 cm$^{2}$ g$^{-1}$ (at $X_{\rm La}\ge 0.2$), which we find results in reasonable agreement with the detailed radiation transport calculations present in Sec.~\ref{sec:Supernova_radiation_transport_results}.

The peak luminosity and effective temperature can also be estimated using analytic formulae (e.g., \citealt{Metzger19}),
\begin{eqnarray}
    L_{\rm pk} &\approx& 4\times 10^{41}{\rm erg\,s^{-1}}\left(\frac{X_{\rm rp}}{0.2}\right)  \\ &&\times\left(\frac{M_{\rm ej}}{50M_{\odot}}\right)^{0.35}\left(\frac{v_{\rm ej}}{0.1c}\right)^{0.65}\left(\frac{\kappa}{{\rm cm^{2}\,g^{-1}}}\right)^{-0.65}, \nonumber
\label{eq:Lpk}
\end{eqnarray}
\begin{eqnarray}
T_{\rm eff,pk}
&\approx& 900\,{\rm K}\left(\frac{X_{\rm rp}}{0.2}\right)^{0.25}  \\
&&\times\left(\frac{M_{\rm ej}}{50M_{\odot}}\right)^{-0.16}\left(\frac{v_{\rm ej}}{0.1c}\right)^{0.41}\left(\frac{\kappa}{{\rm cm^{2}\,g^{-1}}}\right)^{-0.41}, \nonumber
\end{eqnarray}
where we have used the radioactive heating rate of $r$-process nuclei from \citet{Metzger+10} with an assumed thermalization efficiency of 50\%.  Near peak light $t \sim t_{\rm pk} \sim$ 100 d, the specific radioactive heating rate of $^{56}$Ni is $\sim 10-30$ times higher than that of $r$-process elements (e.g., \citealt{siegel_collapsars_2019}).  Given values $X_{\rm Ni}/X_{\rm rp} \sim 0.01-0.05$ for most of our disk outflow models, $L_{\rm pk}$ is moderately underestimated by Eq.~(\ref{eq:Lpk}), which neglects $^{56}$Ni heating.

Fig.~\ref{fig:Renzo_KN} shows the predicted peak timescale, luminosity, and effective temperature of the superKN emission in the parameter space $\{f_{\rm K}, r_{\rm b}\}$ for the fiducial model \texttt{250.25}.  For the same parameters which generate remnant BHs with masses in the PI gap, we predict peak luminosities $L_{\rm pk} \sim 10^{42}$ erg s$^{-1}$ and characteristic durations of months.  Though similar to other types of SNe in duration, superKNe are characterized by significantly cooler emission ($T_{\rm eff} \approx 1000$ K), as confirmed by radiative transfer calculations presented in the next section.

\subsection{SuperKN Light Curves and Spectra}\label{sec:Supernova_radiation_transport_results}

\begin{table*}
	\centering
	\caption{SuperKN Light Curve Models and Survey Detection Rates}
	\label{tab:LCmodels}
	\begin{tabular}{cccccccc} 
		\hline
		\multirow{2}{*}{Model} & $M_{\rm ej}$ & $v_{\rm ej}$ & {$M_{\rm Ni}$} &  {$M_{\rm lrp}$} & {$X_{\rm La}$} & $R_{\rm Rubin}^{(a)}$ & $R_{\rm Roman}^{(b)}$ \\[1 mm]
		& ($M_\odot$) & ($c$) & ($M_\odot$) & ($M_{\odot}$) & ($10^{-3}$) & (yr$^{-1}$) & (yr$^{-1}$)  \\
		\hline
		\hline
		a & 8.6 & 0.1 & 0.019 & 0.83 & 1.4 & 0.01 & 0.02 \\
		b & 31.0 & 0.1 & 0.012 & 8.28 & 17.0 & 0.03 & 0.4 \\
		c & 35.6 & 0.1 & 0.087 & 23.2 & 4.0  & 0.1 & 2 \\
		d & 50.0 & 0.1 & 0.53 & 9.59 & 0.53  & 0.1 & 4 \\
		e & 60.0 & 0.1 & 0.0 & 5.6 & 0.17 & 0.2 & 0.01 \\
		\hline
	\end{tabular}
	\\$^{(a)},^{(b)}$Detection rates per year by {\it Rubin Observatory} and {\it Roman}, respectively for an assumed $z = 0$ superKN rate of 10 Gpc$^{-3}$ yr$^{-1}$ (see Sec.~\ref{sec:survey} for details).
\end{table*}

\subsubsection{Model Selection and Parameters}

To elaborate on the estimates of \S \ref{sec:analytic}, we carried out detailed radiation transport simulations for five
ejecta models whose properties (\mej, \mni, \mlrp, and $M_{\rm La}$)
span the space defined by the subset of simulations that produced BHs
within the mass gap ($60 \lesssim M_{\bullet}/\msun \lesssim 130$),
i.e., models that fall between the two cyan contours of
Fig.~\ref{fig:Renzo_mass_fractions}. (See also \citealt{woosley:17, farmer:19,
  renzo:20csm, farmer:20, costa:21, mehta:21}).
The parameters of the mass gap models are largely confined to a plane in \mej-\mni-\mlrp-$M_{\rm La}$ space, making it straightforward to select a handful of characteristic parameters from the full set.
We used the KMeans routine of \texttt{sklearn} \citep{scikit-learn} to divide our models into four clusters, and adopted the positions of the cluster centers as four representative super-kilonova models.
However, a small fraction of the mass-gap models occupy a distinct region of the parameter space, having large \mej, but little to no nucleosynthetic products heavier than He.
Since these models were not captured by our clusters, we added a fifth model to explore the edge case of a high-mass, nickel-free outflow.
Our five models are listed in Tab.~\ref{tab:LCmodels}.

We performed for the models of Tab.~\ref{tab:LCmodels} one-dimensional radiation transport calculations carried out with Monte Carlo radiation transport code \texttt{Sedona}
\citep{kasen_time-dependent_2006,kasen.ea.inprep_sedona.paper}.
We adopted for each model a density profile such that the mass external to the velocity coordinate $v$ follows a power-law,
\begin{equation}
 M_{>v} \propto \left(\frac{v}{v_{\rm min}}\right)^{-\alpha}, \:\: v \geq v_{\rm min}.
\end{equation}
Above, the minimum ejecta velocity $v_{\rm min}$ is determined by the characteristic velocity $v_{\rm ej} = (2E_{\rm kin}/M_{\rm ej})^{1/2}$ (with $E_{\rm kin}$ the ejecta kinetic energy), and the choice of power-law index $\alpha$,
\begin{equation}
v_{\rm min} = \left(\frac{\alpha-2}{\alpha}\right)^{1/2}v_{\rm ej}.
\end{equation}
We take $\alpha=2.5$ and $v_{\rm ej} = 0.1c$ for all models, consistent with predictions of accretion disk outflow velocities \citep[e.g.][]{fernandez_outflows_2015,siegel_collapsars_2019}.

The opacity of the outflowing gas, and therefore the nature of the transients' electromagnetic emission, is sensitive to the abundance pattern in the ejecta.
Specifically, lanthanides and actinides, and to a lesser extent elements in the \emph{d}-block of the periodic table, contribute  a high opacity, while the opacities of \emph{s}- and \emph{p}- block elements is significantly lower \citep{Kasen+13,Tanaka+20}.

In this work, we predict the synthesis of helium, \iso{Ni}{56}, and light and heavy \rp{} material, but do not carry out detailed nucleosynthesis calculations, e.g. by post-processing fluid trajectories.
The composition of each model is then solely a function of its \mej, \mni, \mlrp, and \Xla.  We assume that heavy ($A>136$) \rp{} material is 41\% lanthanides and actinides by mass, equal to the solar value of $M_{\rm La}/M_{\rm A>136}$.
The remainder is split between \emph{d}-block and \emph{s}/\emph{p}-block elements (54\% and 5\% by mass, respectively).
For light \rp{} material, $\Xla=0$.
We estimated it comprises 95\% (5\%) \emph{d}-block (\emph{s-}/\emph{p}-block) elements by mass.

The composition adopted for our radiation transport models is limited by both our imperfect knowledge of the details of nucleosynthesis and incomplete atomic data of the sort necessary to calculate photon opacities in the ejecta.
Lanthanide and actinide mass ($M_{\rm La}$) is divided among lanthanide elements following the solar pattern, with one adjustment:
because the required atomic data is not available for atomic number Z=71, we redistribute the solar mass fraction of Z=71 to Z=70.

Atomic data is also unavailable for most of the \emph{d}-block elements produced by \rp{} (whether heavy or light).
We thus distribute \emph{d}-block mass evenly among elements with $Z=21-28$ (excluding $Z=23$ for lack of data), artificially increasing the mass numbers to $A \sim 90$ to avoid overestimating the ion number density.
All \rp{} \emph{s}- and \emph{p}-block material is modeled by the low-opacity filler Ca ($Z=20$).
$^{4}$He and \iso{Ni}{56} (as well as its daughter products \iso{Co}{56} and \iso{Fe}{56}) are straightforward to incorporate into the composition.

Our radiation transport simulations include radioactivity from both the \iso{Ni}{56} decay chain and from the \rp. 
We explicitly track energy loss by $\gamma$-rays from \iso{Ni}{56} and \iso{Co}{56}, and assume that positrons from \iso{Co}{56} decay thermalize immediately upon production.
We model \rp{} radioactivity using the results of \citet{lippuner_r-process_2015} for an outflow with $(Y_{\rm e}, s_{\rm B}, \tau_{\rm exp})=(0.13, 32 k_{\rm B}, 0.84 \; \text{ms})$, with $s_{\rm B}$ the initial entropy per baryon and $\tau_{\rm exp}$ the expansion timescale.
To account for thermalization, we adjust the absolute radioactive heating rate following the analytic prescription of \citet{barnes_radioactivity_2016}.

\subsubsection{Radiation Transport Results}

\begin{figure}
    \centering
    \includegraphics[width=\columnwidth]{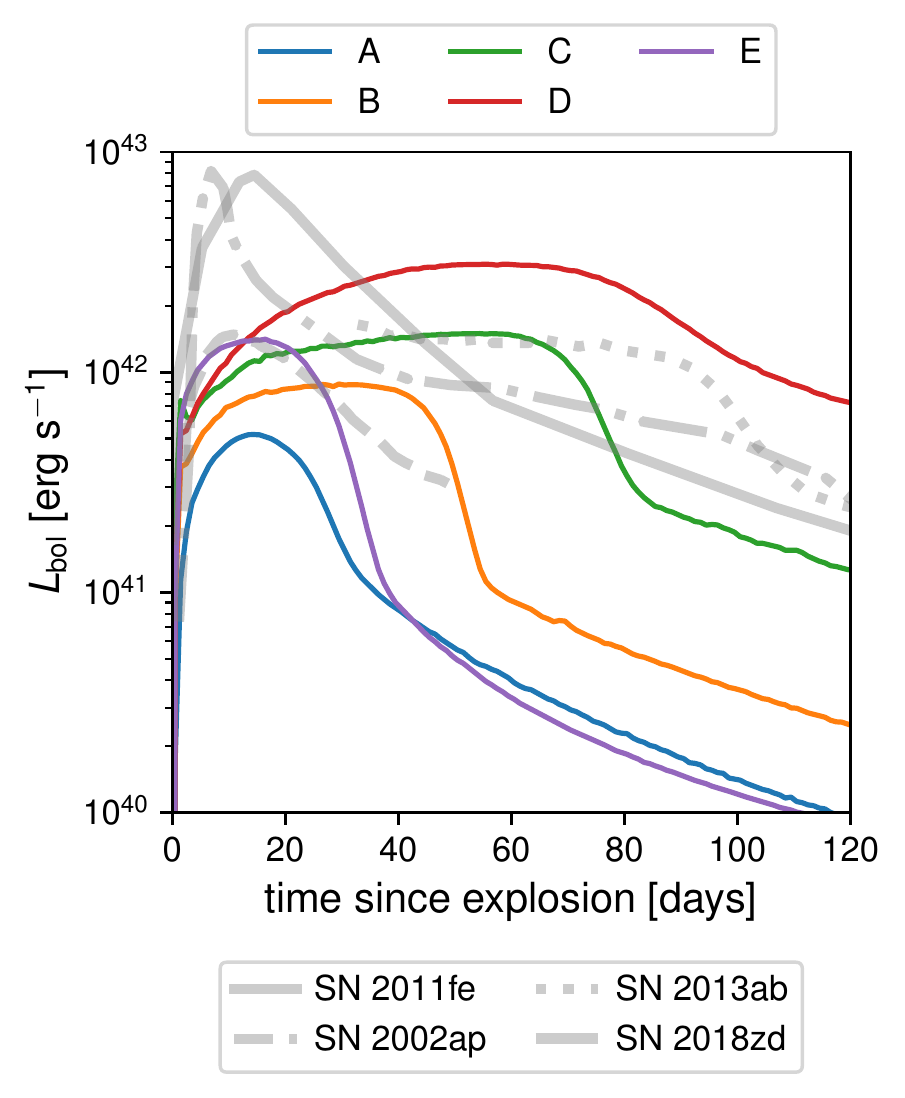}
    \caption{The bolometric light curves of the models in
      Tab.~\ref{tab:LCmodels}, compared to prototypical SNe 2011fe
      (Type Ia), 2002ap (Type Ic-bl), 2013ab (Type II-p), and 2018zd (electron-capture). The
      superKN light curves are dimmer than SNe Ia, but at some epochs
      can approximate the light curves of SNe Type Ic-bl and Type
      IIp.
    }
    \label{fig:sn_bollc}
\end{figure}

The bolometric light curves of models A through E are presented in Fig.~\ref{fig:sn_bollc}.
For comparison, we also show the light curves of typical SNe of various subtypes: Type Ia SN 2011fe \citep{Tsvetkov.ea_2013.CoSka_2011fe.obs}, Type Ic-bl SN 2002ap \citep{Tomita.ea_2006.ApJ_sn2002ap.lcs}, Type IIp SN 2013ab \citep{Bose.ea_2015MNRAS_2013ab.iip}, and the electron-capture SN 2018dz \citep{Hiramatsu.ea_2021.Nature_ECsn2018dz}
.

The superKN light curves exhibit considerable diversity, which is not surprising given the large ranges of ejecta and radioactive masses these systems may produce.
As would be expected from simple Arnett-style \citep{arnett_type_1982} arguments, higher masses are generally associated with longer light-curve durations.
This can be seen in the progression from model A to model D.

However, as model E demonstrates, the shape of the light curve also depends on the presence of \iso{Ni}{56} in the ejecta.
While the mass of \rp{} material burned in superKN outflows greatly exceeds that of \iso{Ni}{56}, the energy generated by the \iso{Ni}{56} decay chain, per unit mass, exceeds that of \rp{} decay by orders of magnitude (e.g., \citealt{Metzger+10, siegel_collapsars_2019}).
When \iso{Ni}{56} is present, it can be the main source of radiation energy for the transient.
As a result of the long half-life of the \iso{Ni}{56} daughter \iso{Co}{56} ($\tau_{1/2}^{\rm Co} \approx 77$ days), the energy generation rate for \iso{Ni}{56}-producing systems is declining slowly just around the time the light curves reach their maxima.
The effect is a more extended light curve (see \citealt{Khatami.Kasen_2019ApJ_lc.relations} and \citealt{Barnes.ea_2021.ApJ_nuc.kne} for more detailed discussions).

Model E, which produces no \iso{Ni}{56}, has a relatively short (${\sim}$month) duration, despite its high mass ($\mej = 60\msun)$, owing to the steep decline of the \rp{} radioactivity that is its only source of energy.
The qualitative difference between models that burn even small amounts of \iso{Ni}{56} and models that burn none points to the importance of a careful treatment of nucleosynthesis in disk outflows.

\begin{figure}
    \centering
    \includegraphics[width=\columnwidth]{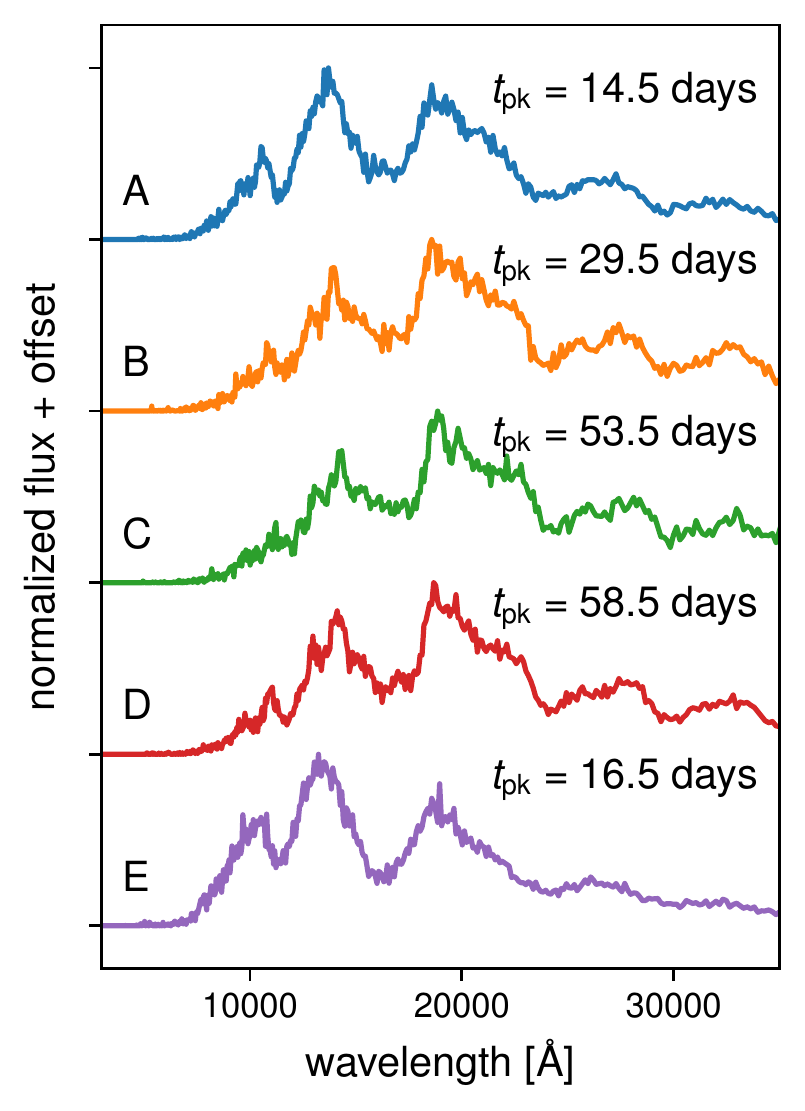}
    \caption{The flux per unit wavelength at bolometric peak for each of the five models defined in Tab.~\ref{tab:LCmodels}. All spectra have broad absorption features consistent with a high-velocity outflow, and a low-temperature, pseudo-black body spectrum, consistent with a high-opacity composition.
    These spectra distinguish superKNe from other classes of SNe,
    which are much bluer, and from other dust-enshrouded explosions, in which broad absorption features are absent.}
    \label{fig:sn_pkspec}
\end{figure}

As is apparent from Fig.~\ref{fig:sn_bollc}, the diversity of superKN light curves allows them to mimic other types of SNe.
While superKNe do not produce sufficient \iso{Ni}{56} to approach the luminosity of SNe Ia, they can, at various epochs, mimic the bolometric light curves of SNe Ic-bl, SNe IIp as well as electron-capture SNe.
However, the high opacity of the \rp-enriched ejecta pushes the superKN emission to redder wavelengths than what is observed for other classes of SNe.
This is illustrated in Fig.~\ref{fig:sn_pkspec}, which shows the normalized spectra for models A through E at bolometric peak.

Unlike other types of SNe, most of the superKN flux emerges at near- and even mid-infrared wavelengths.
This is likely due to a combination of lower radioactive heating per unit ejecta mass, as well as the high opacity from \rp{} elements (particularly lanthanides and actinides) and the high \mej, which work in concert to increase the optical depth across the ejecta and push the photosphere out to the exterior where temperatures are cooler.

A second distinguishing feature of superKNe is their broad absorption features.
These are a product of our assumed ejecta velocities ($v_{\rm ej} = 0.1c$), which are higher than what is inferred for all supernova other than the hyper-energetic SNe Ic-bl.
And while SNe Ic-bl produce spectra with similarly wide absorption features, in the case of Ic-bl these features are found at much bluer (4000 \AA $\lesssim \lambda \lesssim 8000$ \AA) wavelengths.
Thus, despite their bolometric similarities, superKNe are spectroscopically unique among SNe.

The peak photospheric temperatures of superKNe $\sim 1000$ K are also similar to those required for solid condensation, suggesting the possibility of dust formation in the ejecta (e.g., \citealt{Takami+14,Gall+17}).  Insofar as the optical/NIR opacity of ${\sim}\mu$m sized dust is roughly comparable to that of lanthanide-enriched ejecta, dust formation would not qualitatively impact the appearance of the transient.  However, this does imply potential degeneracy between the photometric signatures of superKNe and other dust-enshrouded explosions unrelated to $r$-process production, including stellar mergers (e.g., \citealt{Kasliwal+17}).  This degeneracy with dusty transients can generally be broken by the predicted broad spectral features of superKNe $(v_{\rm ej} \sim 0.1c$).

\section{Discovery Prospects}
\label{sec:discovery}

In this section we explore the discovery prospects of superKNe with future optical/infrared transient surveys and via late-time infrared follow-up observations of energetic long GRBs.  We then discuss how superKN emission could be enhanced by circumstellar interaction for collapsars embedded in AGN disks.

\subsection{Volumetric Rates}
\label{sec:rates}

We begin by estimating the volumetric rate of superKNe. One approach
is to scale from the observed rates of ordinary collapsars.  The local
(redshift $z \simeq 0$) volumetric rate of classical long GRBs is
$\approx 0.6-2$ Gpc$^{-3}$yr$^{-1}$ \citep{wanderman_luminosity_2010},
which for an assumed gamma-ray beaming fraction $f_{\rm b} = 0.006$
\citep{goldstein_estimating_2016}, corresponds to a total collapsar
rate of $\approx 100-300$ Gpc$^{-3}$ yr$^{-1}$.  Under the assumption
that ordinary collapsars originate from stars of initial mass
$M_{\rm ZAMS} \gtrsim 40 M_{\odot}$, then the more massive stars
$M_{\rm ZAMS} \gtrsim 250M_{\odot}$ which generate helium core masses
above the PI mass gap ($M_\mathrm{BH}\gtrsim 130M_{\odot}$) will be
less common by at least a factor
$\sim (40/250)^{\alpha-1} \sim 0.1-0.3$ for an initial-mass function
(IMF) $dN_{\star}/dM_{\star} \propto M_{\star}^{-\alpha}$, where we
consider values for the power-law index between $\alpha = 2.35$ for a
Salpeter IMF and a shallower value $\alpha \approx 1.8$
(\citealt{Schneider+18b}). This optimistically assumes that (i) stars
that massive exist \citep[e.g.,][]{dekoter:97,crowther:16}, and that (ii)
these can form helium cores such that
$M_\mathrm{He}\simeq M_\mathrm{ZAMS}$, for instance because of
rotational mixing \citep[e.g.,][]{maeder:00, marchant:16, demink:16} or
continuous accretion of gas \citep[e.g.,][]{Jermyn+21,
  dittmann:21}. Various processes act to remove mass from a massive
star during its evolution, and generally the more massive the star,
the larger its mass loss rate. Some of these mechanisms
(e.g., continuum-driven stellar winds and eruptive mass
loss phenomena, \cite{} see also \citealt{renzo:20merger}) might occur even at low
metallicity.

With the above estimate and caveats, we obtain an optimistic maximum
local rate of superKNe from massive collapsars of $\sim 10-100$
Gpc$^{-3}$ yr$^{-1}$.  On the other hand, the long GRB rate increases
with redshift in rough proportion to the cosmic star-formation rate
(SFR $\propto (1+z)^{3.4}$ for $z \lesssim 1$; e.g.,
\citealt{Yuksel+08}) and hence the maximum rate of superKNe is larger
at redshift $z \gtrsim 1$ by a factor $\sim 10$ than at $z \simeq 0$,
corresponding to a maximum superKN rate of $\sim 100-1000$ Gpc$^{-3}$
yr$^{-1}$ at $z \gtrsim 1$.

The superKN rate question can be approached from another perspective:
What is the minimum birth-rate of BHs in the PI mass gap to explain
\gw{}-like merger events (Sec.~\ref{sec:GW190521}) via the massive
collapsar channel?  The rate of \gw{}-like mergers at $z \simeq 0$
was estimated by LIGO/Virgo to be $\sim 0.5-1$ Gpc$^{-3}$ yr$^{-1}$
\citep{Abbott+20_190521}.  This rate is smaller than the maximum superKN rate estimated above, consistent with only a small
fraction of BHs formed through this channel ending up in tight
binaries that merge due to gravitational waves at $z \approx 0$.

\subsection{Discovery with Optical/Infrared Surveys}
\label{sec:survey}

We now evaluate the prospects for discovering superKNe with impending wide-field optical/infrared surveys.

First, we explore the expected observable rates within the Legacy Survey of Space and Time (LSST) conducted with the {\it Vera Rubin Observatory}. LSST is currently set to commence in early 2024 and will explore the southern sky in optical wavelengths to a $5\sigma$ stacked nightly visit depth of $\sim24$ mag. We inject the set of SEDONA light curves of models described in Tab.~\ref{tab:LCmodels} into the publicly available LSST operations simulator, OpSim \citep{delgado2016lsst}. We use the most recent baseline scheduler ({\tt baseline v1.7}) to calculate LSST pointings, limiting magnitudes, and expected sky noise across a full simulated 10 year survey in $ugrizY$ bands. We additionally apply dust reddening following the dust maps of \cite{schlegel1998maps}. For each model, we inject a superKN randomly 300 times within the full LSST simulation (including both the wide-fast-deep survey and deep-drilling fields) at redshift bins of 0.01.

We find that superKNe discovered with LSST are confined to the local
universe, with $z<0.1$.  Assuming that the superKN rate traces
star-formation with a local rate of 10 Gpc$^{-3}$yr$^{-1}$, we expect
LSST to discover $\sim 0 - 0.2$ superKNe annually, resulting in up to
$2$ events over its 10-year nominal duration.  We note that the larger
the ejecta masses (i.e., Models b and e) the most likely the detection
with LSST.

Given the expected red colors of the superKN emission (Fig.~\ref{fig:sn_pkspec}), we additionally explore the possibility of discovering superKN with the \textit{Nancy Grace Roman Space Telescope}, expected to launch in the mid 2020s. Although not fully defined, \textit{Roman} expects to conduct a $\sim5$ year, 10 deg$^2$ SN survey, primarily targeted at Type Ia SNe for cosmological distance measurements.  We assume a survey cadence of 30 days and single-visit, stacked $5\sigma$ depth of 27th magnitude, corresponding to roughly an hour of integration time (in F158 band). We inject the same set of models using the \textit{Roman} F062, F158 and F184 filters, corresponding to central wavelengths of 0.62, 1.58 and 1.83 $\mu$m, respectively. We assume observations are taken in each filter at the same epoch, and consider superKNe with three or more $3\sigma$ detections to be detectable. Assuming the \textit{Roman} wide-field survey footprint is chosen to minimize galactic dust, we do not account for any galactic reddening. 

We find that \textit{Roman} is most sensitive to models with the largest Lanthanide fractions. Assuming that the superKN rate traces star-formation with a local rate of 10 Gpc$^{-3}$yr$^{-1}$, we expect a 5 year \textit{Roman} survey as described would find roughly 1--20 superKNe, most favoring the Lanthanide-rich Model B. These superKNe will be observable out to a redshift of $z\sim0.9$. We note that longer cadences significantly decrease the number of superKN detections possible with \textit{Roman}, at least within the 3-detection discovery criterion we have adopted.

\subsection{Energetic Long GRB Accompanied by SuperKNe}
\label{sec:GRB}

SuperKNe could also be detected following a subset of long GRBs.  Figure \ref{fig:Renzo_GRB_accretion} summarizes the GRB properties for our fiducial massive collapsar model \texttt{250.25}.  We find accretion timescales comparable to those of ordinary collapsars from lower mass progenitor stars (Appendix \ref{app:collapsars}). These mass gap collapsars are therefore candidates for contributing to the observed population of long GRBs, except that they may be a factor of $\sim\!10$ times more luminous and energetic than typical GRBs if the gamma-ray luminosity tracks the BH accreted mass.  Furthermore, if the fraction of massive stars above the PI mass gap which form or evolve into collapsar progenitors is greater at lower metallicity, this could imprint itself on the redshift evolution of the long GRB luminosity function (for which there is claimed evidence; \citealt{Petrosian+15,Sun+15,Pescalli+16}).

In the local universe, long GRBs with supernovae are commonly
accompanied with the luminous hyper-energetic Type Ic SNe with broad lines (Ic-BL; e.g.,
\citealt{Woosley&Bloom06, japelj:18, modjaz:20}).  The superKN transients we predict from the birth of more massive BHs are of comparable or moderately lower peak luminosities than ordinary collapsar SNe (e.g., \citealt{Cano16}) but significantly redder (Figs.~\ref{fig:sn_bollc}, \ref{fig:sn_pkspec}).  Luminous optical SNe have been ruled out to accompany a few nominally long duration GRBs (\citealt{Fynbo+06,Gehrels+06}).  One of these events, GRB 060614, was found to exhibit a red excess which \citet{jin2015light} interpreted as a kilonova.  However, the luminosity and timescale of the excess could also be consistent with superKN emission from a massive collapsar of the type described here.  We encourage future deep infrared follow-up observations of energetic long GRB with {\it Roman} or {\it JWST} on timescales of weeks to months after the burst to search for infrared superKN emission.

\subsection{SuperKNe Embedded in AGN Disks}
\label{sec:AGN}

The optical emission from superKNe could be significantly enhanced by circumstellar interaction if they are embedded in a gas-rich environment.

\citet{Graham+20} reported a candidate optical wavelength counterpart
to \gw{} in the form of a flare from an active galactic nucleus
(AGN).  The flare reached a peak luminosity $L_{\rm pk} \sim 10^{45}$
erg s$^{-1}$ in excess of the nominal level of AGN emission and lasted
a timescale $t_{\rm pk} \sim 50$ days, over which it radiated a total
energy of $E_{\rm rad} \sim 10^{51}$ erg.  \citet{shibata_alternative_2021} propose
a scenario for \gw{} as a massive stellar core collapse generating
a single BH and a massive accretion disk $\gtrsim 10-50M_{\odot}$
rather than a binary BH merger.  Although our results in
Secs.~\ref{sec:GW} and~\ref{sec:GW190521} challenge this interpretation, our present work shows that a prediction of this scenario is a superKN counterpart with $M_{\rm ej} \sim 3-20M_{\odot}$ and $v_{\rm ej} \sim 0.1$ c.  Though the predicted peak timescale, $t_{\rm pk} \sim 50$ days  (Eq.~\ref{eq:tpk}), of the superKNe emission roughly agrees with that observed by \citet{Graham+20}, the luminosity powered by radioactivity $\sim 10^{42}$ erg s$^{-1}$ (Fig.~\ref{fig:sn_bollc}) is too small to explain the observations by several orders of magnitude.

This problem could be alleviated if the collapsing star is embedded in a
dense gaseous AGN disk (e.g., \citealt{Jermyn+21, dittmann:21}). If
the density of the AGN disk at the star location is sufficiently
high, $\rho\gtrsim10^{-15}\,\mathrm{g\ cm^{-3}}$, runaway accretion of
mass might help building up very massive and fast rotating helium cores. The mass accretion might be interrupted as the AGN turns off
(on a few Myr timescale), and depending on the balance between mass
loss processes and the previous accretion phase one might obtain a
superKN progenitor. At its collapse, the shock-mediated collision
between the superKN ejecta and the surrounding disk material could
power a more luminous optical signal than from radioactive decay
alone, akin to interaction-powered super-luminous SNe (e.g.,
\citealt{Smith+07}).

Given the large kinetic energy of the superKN
ejecta,
$E_{\rm kin} \sim M_{\rm ej}v_{\rm ej}^{2}/2 \sim 1-5\times 10^{53}$
erg, the \citet{Graham+20} transient could be powered by tapping into
only $\sim 1\%$ of $E_{\rm kin}$ by shock deceleration.  Insofar as
such luminous shocks are radiative and momentum-conserving, the
swept-up gaseous mass in the AGN disk required to dissipate
$E_{\rm rad} \sim 10^{51}$ erg is only
$M_{\rm sw} \sim (E_{\rm rad}/E_{\rm kin})M_{\rm ej} \sim 0.1-1M_{\odot}$.
Treating the swept-up material as being approximately spherical and
expanding at $\sim v_{\rm ej}$, the optical diffusion time through
$M_{\rm sw}$ is (Eq.~\ref{eq:tpk}),
\begin{equation}
t_{\rm pk,min} \approx 5\,{\rm d}\left(\frac{M_{\rm sw}}{0.3M_{\odot}}\right)^\frac{1}{2}\mskip-5mu\left(\frac{v_{\rm ej}}{0.1c}\right)^{-\frac{1}{2}}\mskip-5mu\left(\frac{\kappa}{0.3\,{\rm cm^{2}\,g^{-1}}}\right)^\frac{1}{2},
\end{equation}
where $\kappa$ is now normalized to a value more appropriate to AGN disk material.  Insofar that $t_{\rm pk,min}$ is significantly shorter than the observed $\sim 50$ d rise time of the \citet{Graham+20} counterpart, this implies the rise of the putative counterpart would instead need to be limited by photon diffusion through the unshocked external AGN disk material (e.g.,~\citealt{Graham+20,Perna+21}).  

A bigger challenge for this scenario is the typically much closer source distance for \gw{} that would be predicted if this resulted from a self-gravitating collapsar disk instead of a binary BH merger (redshift $z \lesssim 0.05$; Sec.~\ref{sec:GW}), compared to that of the AGN identified by \citet{Graham+20} at redshift $z = 0.438$.

\section{Other Observable Implications}
\label{sec:implications}

\subsection{Luminous Slow Radio Transients}
\label{sec:radio}

In addition to their prompt optical/IR signal, superKNe produce synchrotron radio emission as the ejecta decelerates by driving a shock into the circumburst medium (e.g., \citealt{nakar_detectable_2011,Metzger&Bower14}).  This emission can be particularly luminous because the kinetic energy of the superKN ejecta $E_{\rm kin} \approx 1-5\times 10^{53}$ erg can be one to two orders of magnitude higher than those of ordinary collapsar SNe.

The radio transient rises on the timescale required for the ejecta to sweep up a mass comparable to their own, 
\begin{equation}
t_{\rm radio} \approx 200\,{\rm yr}\left(\frac{E_{\rm kin}}{5\times 10^{53}\,\rm erg}\right)^{\frac{1}{3}}\mskip-5mu \left(\frac{v_{\rm ej}}{0.1c}\right)^{-\frac{5}{3}}\mskip-5mu\left(\frac{n}{\rm 1\,cm^{-3}}\right)^{-\frac{1}{3}},
\end{equation}
where $n$ is the particle density of the external medium.  The peak luminosity at a frequency $\nu = $ 1 GHz can be estimated as (e.g., \citealt{nakar_detectable_2011})
\begin{eqnarray}
\nu L_{\nu} &\approx& 5\times 10^{39}\,{\rm erg\,s^{-1}}\left(\frac{E_{\rm kin}}{5\times 10^{53}\,\rm erg}\right)\left(\frac{v_{\rm ej}}{0.1c}\right)^{2.3}\nonumber \\
&& \times \left(\frac{n}{\rm 1\,cm^{-3}}\right)^{0.83}\left(\frac{\epsilon_e}{0.1}\right)^{1.3}\left(\frac{\epsilon_{B}}{0.01}\right)
\end{eqnarray}
where the fraction of the shock energy placed into relativistic electrons $\epsilon_{e}$ and magnetic fields $\epsilon_{B}$ are normalized to characteristic values, respectively, and we have assumed a power-law index $p = 2.3$ for the energy distribution of the
shock accelerated electrons, $dN/dE \propto E^{-p}$. 

For characteristic circumstellar densities $n \sim 0.1-10$ cm$^{-3}$ the peak radio luminosity is comparable to that of rare energetic transients, such as those from binary neutron star mergers that generate stable magnetar remnants (e.g., \citealt{Metzger&Bower14,Schroeder+20}).  However, the predicted timescale of the radio evolution of decades to centuries is much longer in the superKN case due to the large ejecta mass.  This slow evolution makes it challenging to uniquely associate the radio source with a known GRB or gravitational wave event, or to even identify it as a transient in radio time-domain surveys (e.g., \citealt{Metzger+15}).  We note that luminous radio point sources are in fact common in the types of dwarf galaxies which host collapsars (e.g., \citealt{Eftekhari+20}).  \citet{Ofek17} place an upper limit on the local volumetric density of persistent radio sources in dwarf galaxies of luminosity $\gtrsim 3\times 10^{38}$ erg s$^{-1}$ of $\mathcal{N} \lesssim 5\times 10^{4}$ Gpc$^{-3}$.  Assuming the superKN radio emission remains above this luminosity threshold for a time $t_{\rm det} \sim 10t_{\rm radio} \sim 10^{3}$ yr, this constrains the local rate of superKNe to obey $\lesssim 10-100$ Gpc$^{-3}$ yr$^{-1}$, consistent with the estimates given in Sec.~\ref{sec:rates}.

\subsection{Gravitational Wave Emission}
\label{sec:GW}

\begin{figure}
\centering
\includegraphics[width=0.98\linewidth]{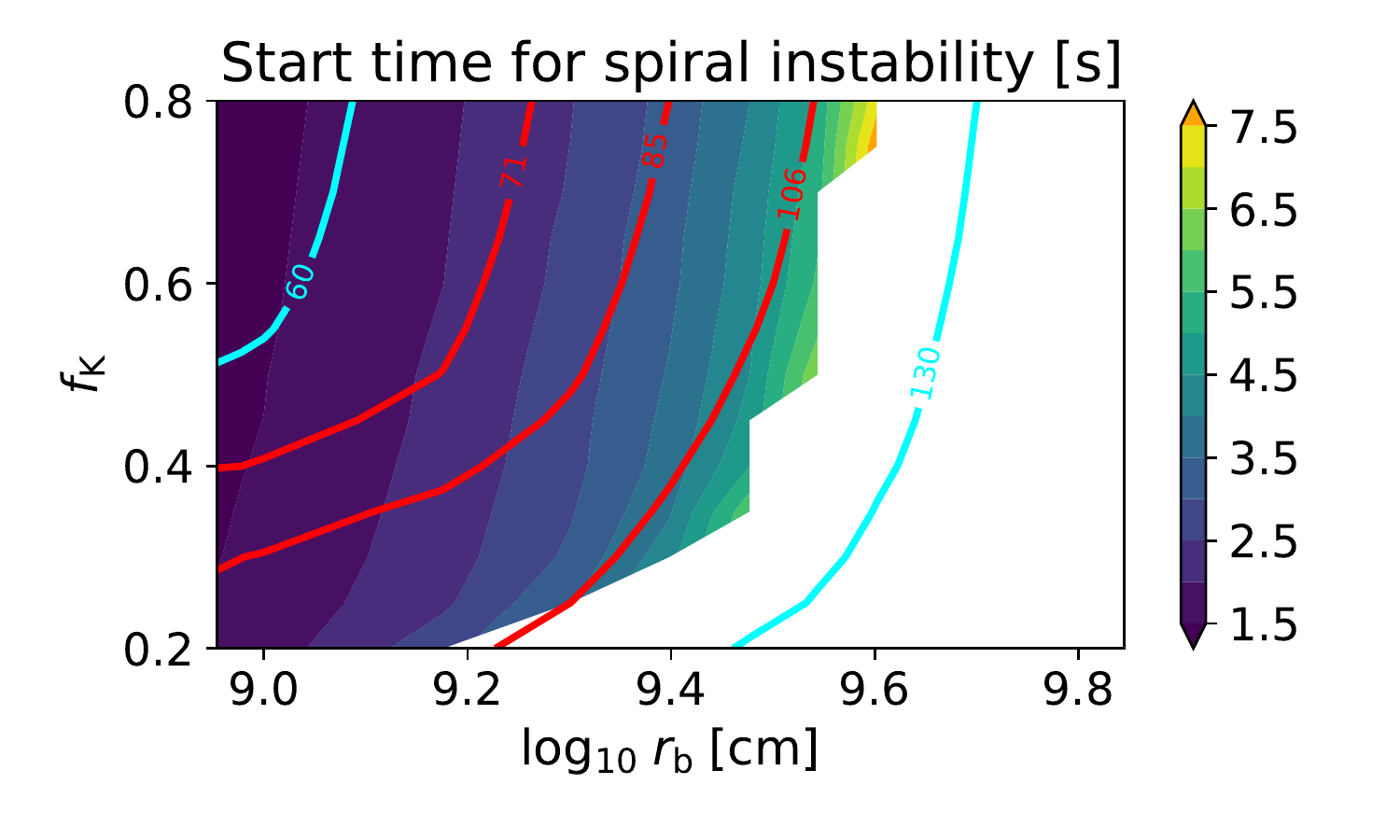}
\includegraphics[width=0.98\linewidth]{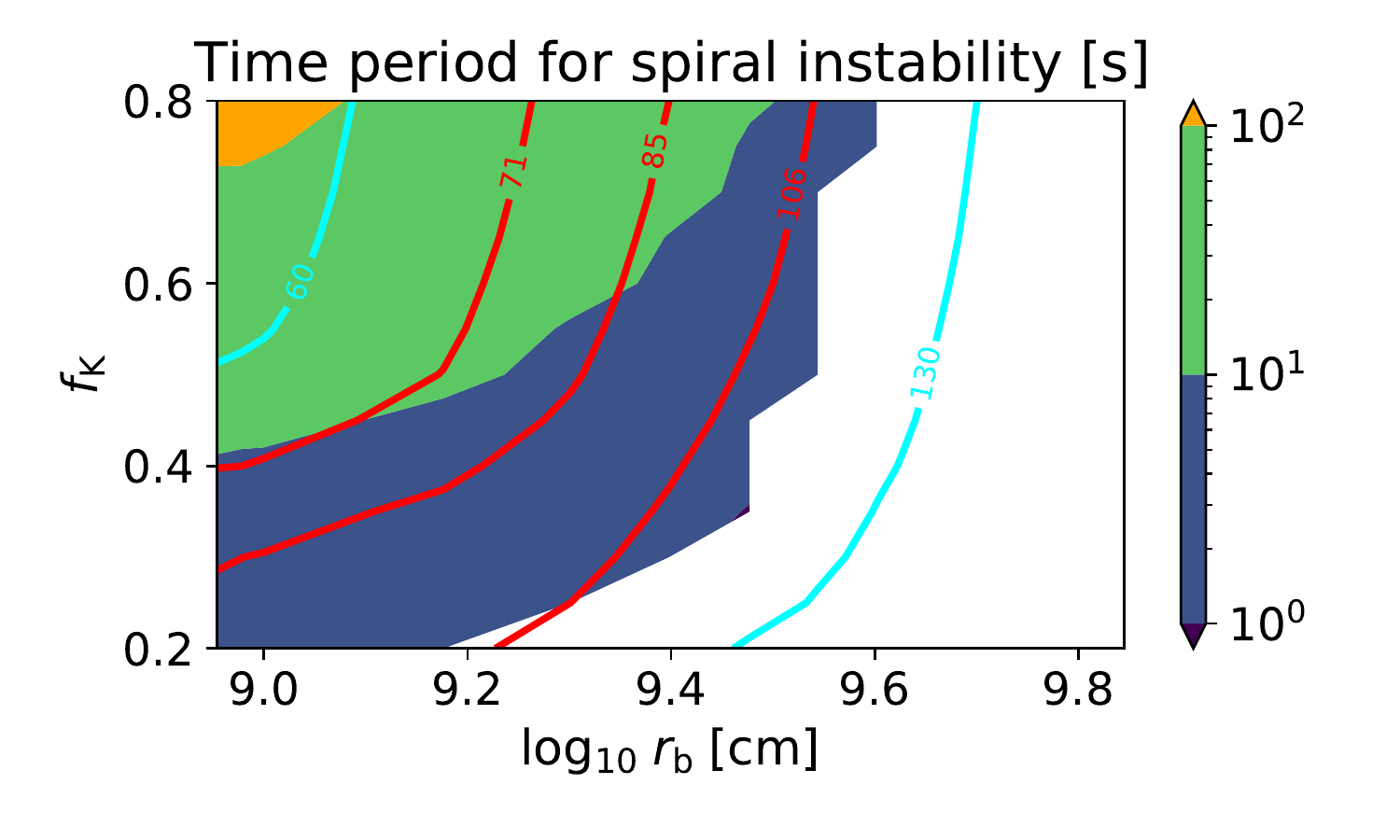}
\caption{Time after core collapse at which gravitational instabilities in the collapsar disk first set in (top panel) and the duration over which gravitational instabilities are continuously excited during the fallback process (bottom panel), shown in the space of $\{r_{\rm b},f_{\rm K}\}$ for the fiducial progenitor model \texttt{250.25}.  White space indicates models that do not experience gravitational instabilities during fallback accretion. Red contours indicate the inferred primary mass of \gw{} [$M_\odot$], together with its 90\% confidence limits. Cyan contours delineate final BH masses of 60\,$M_\odot$ and 130\,$M_\odot$, which approximately correspond to the lower and upper end of the PI mass gap.}
\label{fig:gravitational_instabilities}
\end{figure}

The accretion disks formed in superKN collapsars may become susceptible to gravitational instabilities if their disk mass approaches an order-unity fraction of the BH mass during the fallback evolution process (Sec.~\ref{sec:fallback}). As shown in Fig.~\ref{fig:gravitational_instabilities}, only progenitor cores with high angular momentum (small $r_{\rm b}$ and/or high $f_{\rm K}$) lead to fallback accretion that result in gravitational instabilities. Low-angular momentum cores instead form heavier BHs with relatively smaller disk masses.

The onset time of the instability of typically a few seconds (Fig.~\ref{fig:gravitational_instabilities}), representative of all superKN progenitor models investigated here, is determined by the progenitor structure, its rotation profile, and the free-fall timescale. Once triggered, subsequent fallback material collapsing onto the disk continues to excite these instabilities in the collapsar disk for a timescale of seconds to hundreds of seconds (Fig.~\ref{fig:gravitational_instabilities}), until viscous draining of the disk becomes fast compared to the free-fall timescale of the remaining outer layers of the progenitor star (roughly $\sim\!10$\,s for our fiducial model in Fig.~\ref{fig:Renzo_evolution}).

The onset of the instability manifests itself as the exponential growth of a non-axisymmetric one-arm ($m=1$) density mode in the disk with growth time on the order of the orbital period of the disk, typically followed by exponential growth of an $m=2$ mode (e.g., \citealt{kiuchi_nonaxisymmetric_2011,shibata_alternative_2021,wessel_gravitational_2021}). These non-axisymmetric density perturbations give rise to gravitational-wave emission with dominant frequency at the orbital and twice the orbital frequency, respectively (e.g., \citealt{wessel_gravitational_2021}). 

\begin{figure}
\centering
\includegraphics[width=0.98\linewidth]{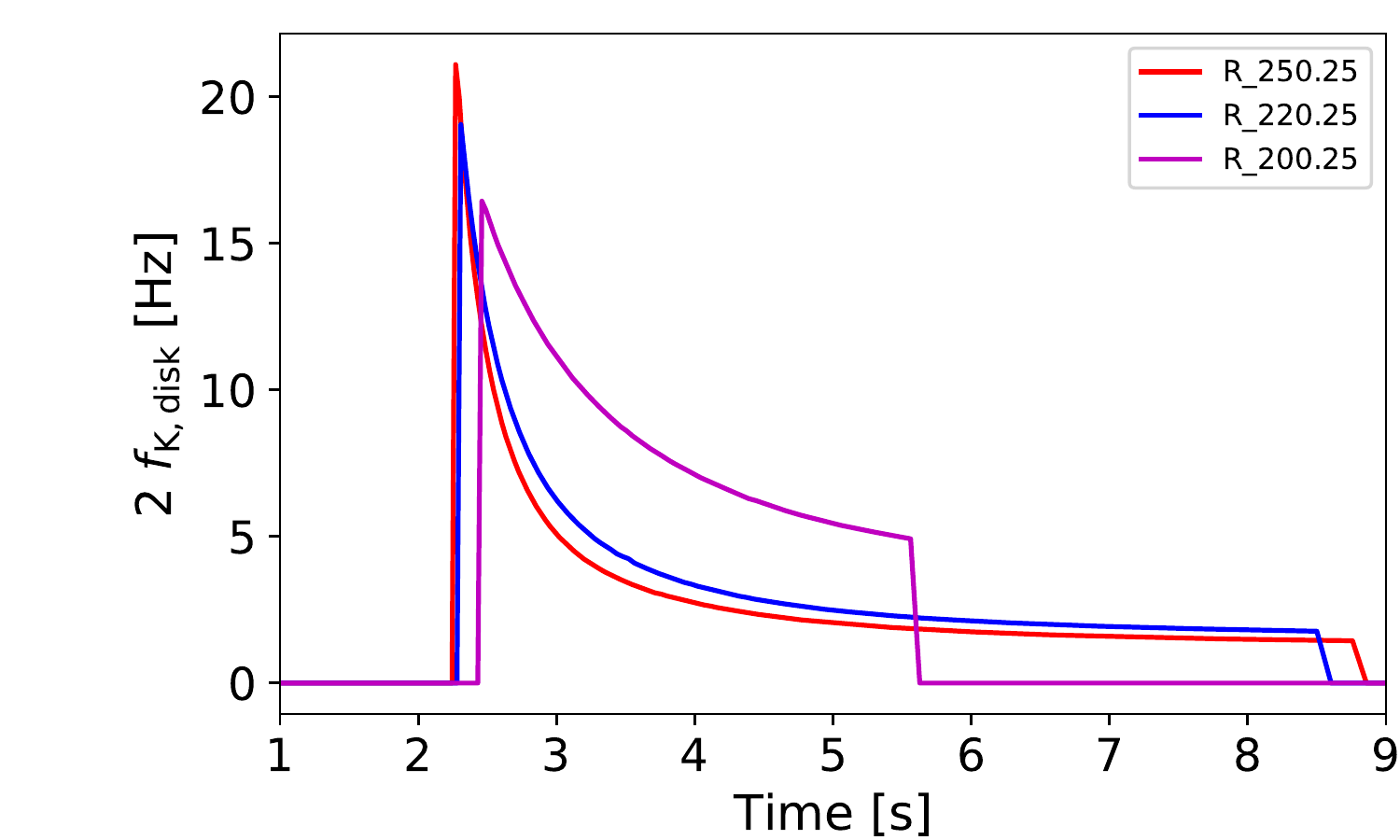}
\includegraphics[width=0.98\linewidth]{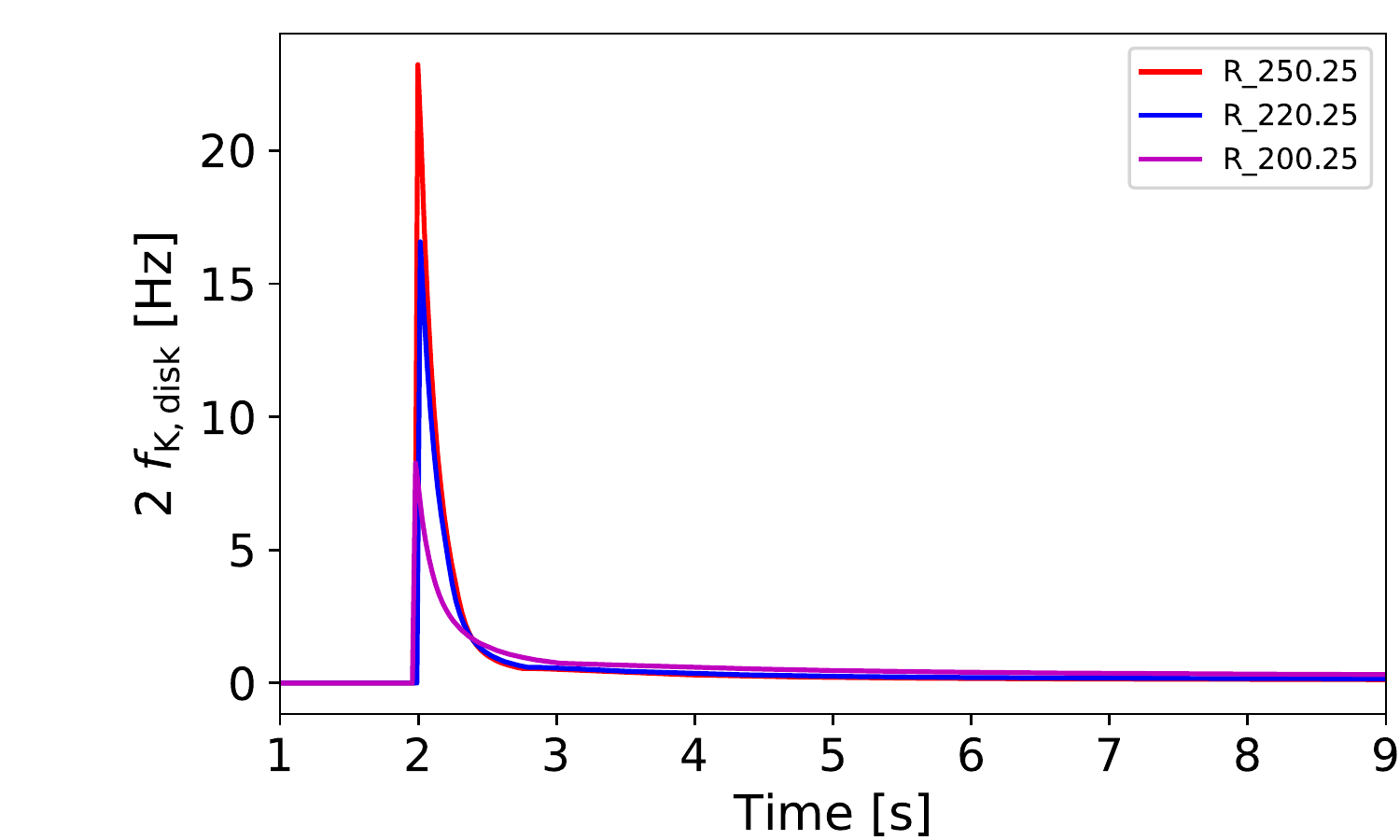}
\caption{Disk frequency evolution (Eq.~\ref{eq:Omega_disk}) for three progenitor models (\texttt{250.25}, \texttt{220.25}, \texttt{200.25}) with rotation parameters $p=4.5$, $r_{\rm b}= 1.5\times10^{9}$\,cm, and overall small ($f_{\rm K} = 0.3$; top) or large ($f_{\rm K}=0.6$; bottom) Keplerian angular momentum parameter. Plotted is twice the orbital angular frequency, which corresponds to the gravitational-wave frequency of the $m=2$ density mode of the gravitationally unstable disk. The frequency evolution is largely controlled by $f_{\rm K}$, with all models reflecting the `sad-trombone' nature of the gravitational-wave signal.}
\label{fig:gravitational_wave_frequency_evolution}
\end{figure}

As long as further fallback keeps the disk in the instability regime defined by Eq.~\eqref{eq:gravitational_instability}, we assume that the dominant gravitational-wave frequencies of these modes are determined by the evolving angular frequency $\Omega_{\rm K, disk}$ of the disk (Eq.~\ref{eq:Omega_disk}) with radius $r_{\rm disk}(t)$ (Sec.~\ref{sec:fallback}). Since $r_{\rm disk}(t)$ monotonically increases with time as the black hole grows and material with larger specific angular momentum enters the disk, the gravitational-wave frequency decreases, sweeping down with a rate and amplitude that depends on the density and angular momentum structure of the progenitor star envelope. The gravitational-wave signal thus exhibits a ``sad-trombone'' pattern in the time-frequency spectrogram, as opposed to a ``chirp'' signal generally associated with gravitational waves from compact binary mergers. Examples of the frequency evolution of the disk for different mass models and for high and low specific angular momentum of the progenitor envelope are shown in Fig.~\ref{fig:gravitational_wave_frequency_evolution}. Over a large range of the parameter space and progenitor models explored here, superKN collapsars are strong emitters of quasi-monochromatic gravitational waves of duration $\sim\!1-100$\,s with a decreasing frequency trend (between $\sim\!0.1-40$\,Hz for the $l=m=2$ and $\sim\!\text{few}\times 10^{-2} -25$\,Hz for the $l=2$, $m=1$ mode) characteristic of their progenitor stellar structure (see Figs.~\ref{fig:gravitational_instabilities} and \ref{fig:gravitational_wave_frequency} for a representative example). If detected, such gravitational-wave signals could reveal information about the rotation profiles of and angular momentum transport in evolved massive stars. The ``sad-trombone'' feature simultaneously followed by typically two dominant modes separated in frequency space by the instantaneous characteristic disk rotation frequency may prove useful in searching and detecting such sources with gravitational-wave detectors.

\begin{figure}
\centering
\includegraphics[width=0.98\linewidth]{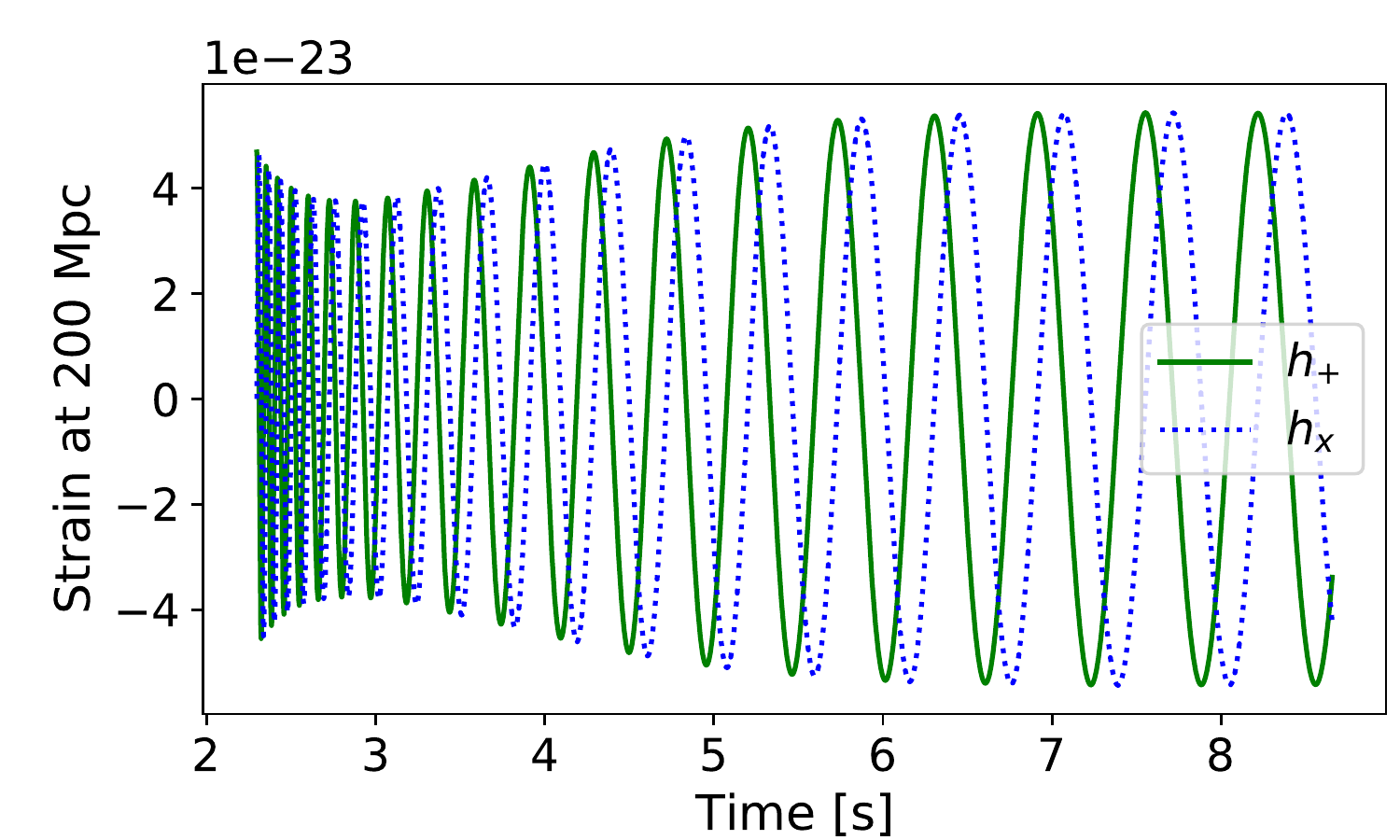}
\caption{Plus and cross polarization strain amplitudes of the $l=m=2$ mode of gravitational waves resulting from the gravitationally unstable collapsar disk of the fiducial model shown in Fig.~\ref{fig:Renzo_evolution} with $p=4.5$, $f_{\rm K}=0.3$ and $r_{\rm b}=1.5\times 10^{9}$\,cm, assuming a face-on orientation of the accretion disk ($\iota=0$). The emission starts a few seconds after the onset of collapse and persists for several seconds until viscous draining of the disk dominates fallback accretion and the disk becomes stable again at around 9\,s after the onset of collapse.}
\label{fig:gravitational_wave_strain}
\end{figure}

\begin{figure}
\centering
\includegraphics[width=0.98\linewidth]{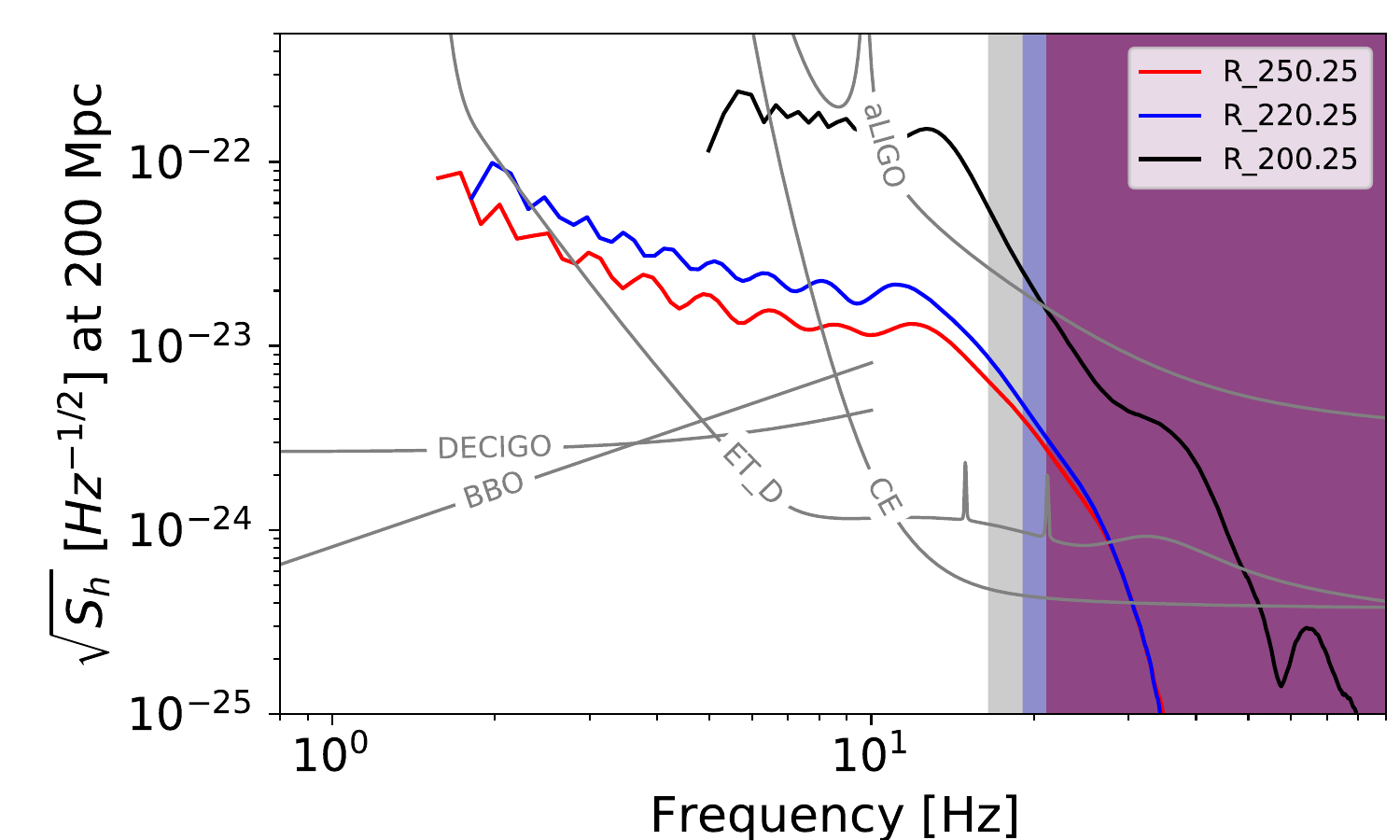}
\caption{Amplitude spectral density (ASD) of gravitational-wave
  emission from the collapsar disk, shown for three progenitor models
  (\texttt{250.25}, \texttt{220.25}, \texttt{200.25}) and stellar
  rotation parameters $p=4.5$, $f_{\rm K}=0.3$,
  $r_{\rm b}=1.5\times 10^{9}$\,cm at an assumed source distance of 200\,Mpc. The shaded region for each curve shows the unphysical frequency regime above the maximum disk frequency as plotted in Fig.~\ref{fig:gravitational_wave_frequency_evolution}, which is ignored in the SNR calculations. Shown for comparison are the measured or predicted noise curves for aLIGO, CE, ET, DECIGO, and BBO with sensitivity curve data from
  \url{https://dcc.ligo.org/LIGO-T1500293/public} and \citet{DECIGO_ASD}.}
\label{fig:gravitational_wave_frequency_sensitivity}
\end{figure}

\begin{figure}
\centering
\includegraphics[width=0.98\linewidth]{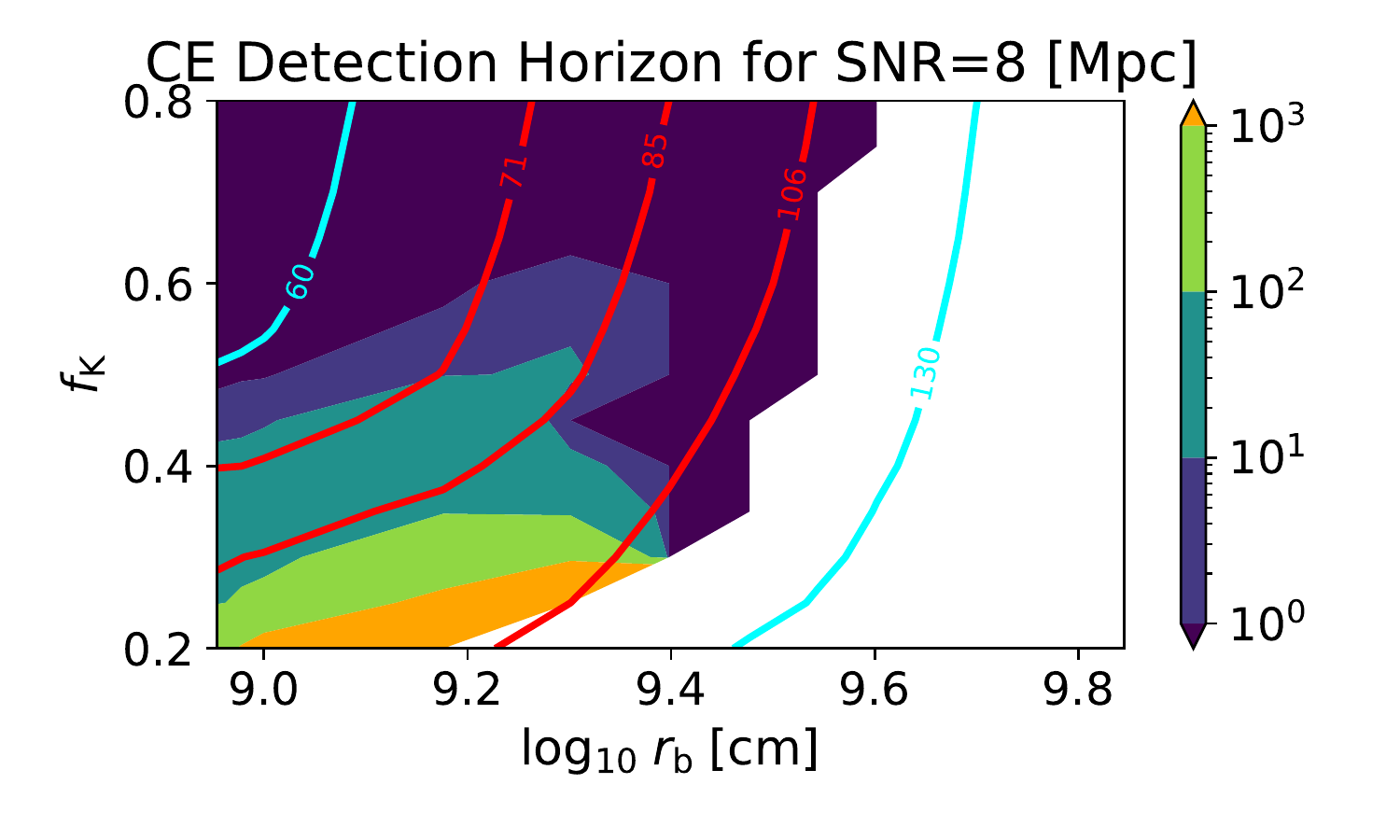}
\includegraphics[width=0.98\linewidth]{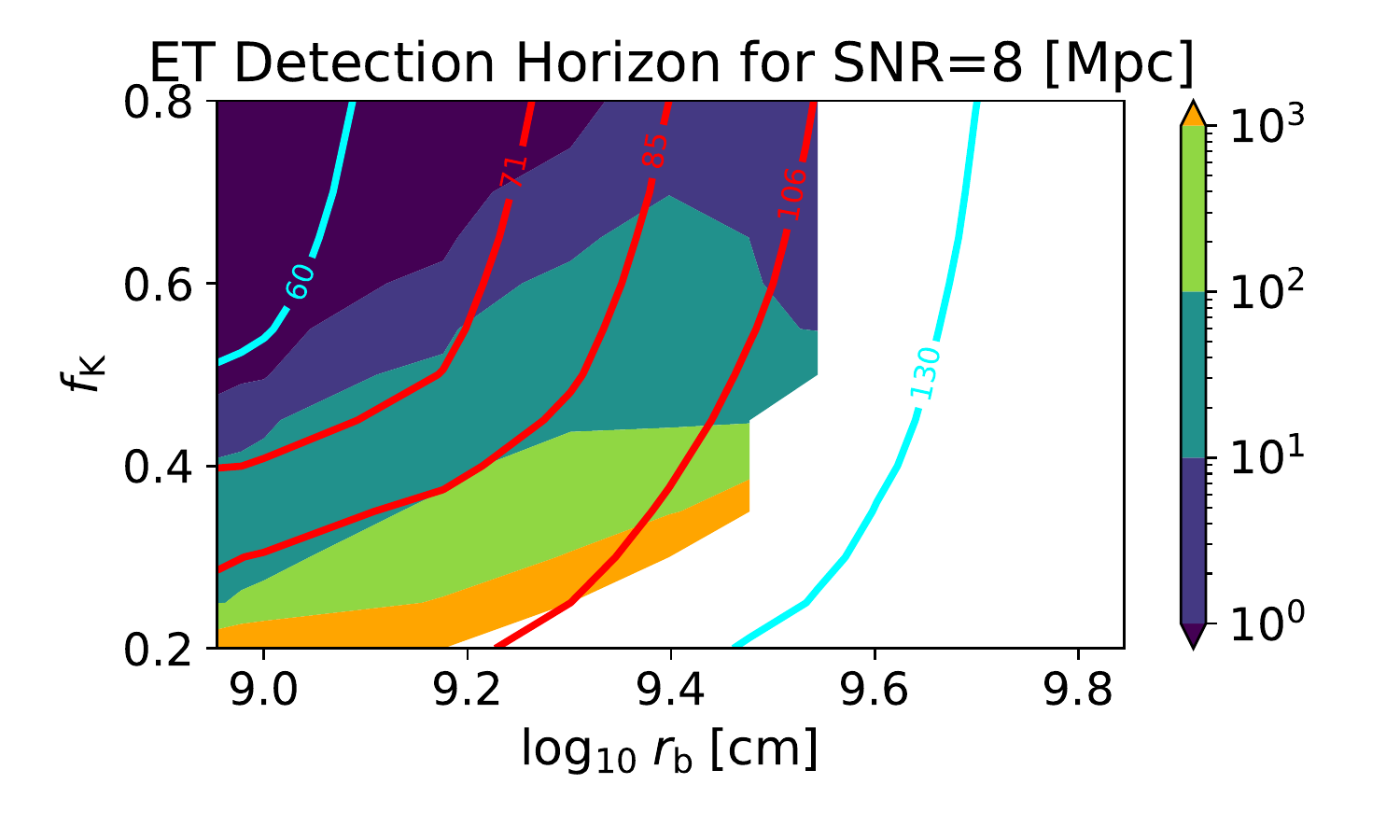}
\includegraphics[width=0.98\linewidth]{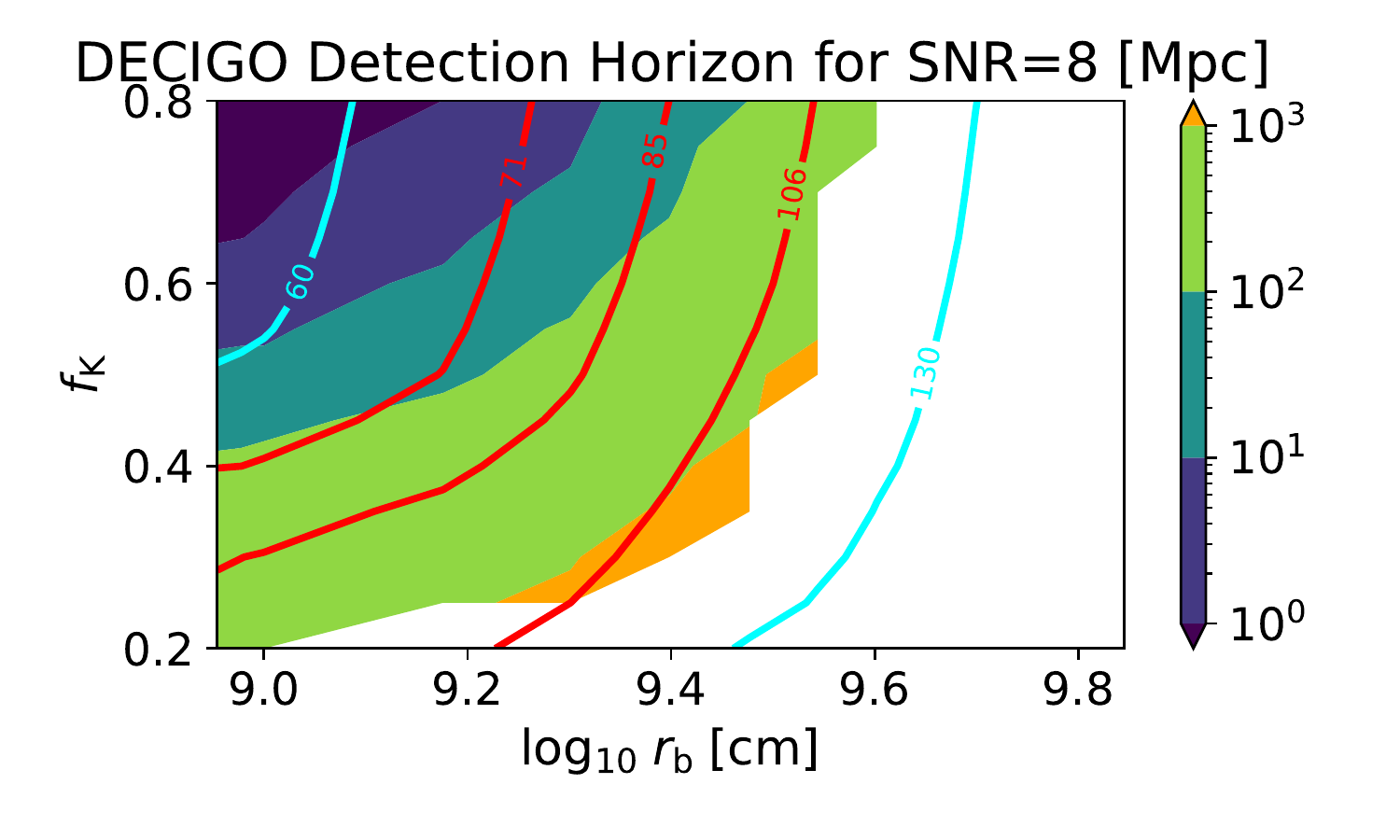}
\caption{Detection horizons of gravitational waves from our fiducial model shown in Fig.~\ref{fig:Renzo_evolution} with $p=4.5$, $f_{\rm K}=0.3$ and $r_{\rm b}=1.5\times 10^{9}$\,cm for Cosmic Explorer (top), the Einstein Telescope (center), and DECIGO (bottom), assuming optimal matched filtering and a signal-to-noise ratio of 8. For progenitors with medium to low rotation, these detectors may be able to detect gravitational waves from superKN collapsars at distances of typically a few hundred Mpc up to a few Gpc. These estimates are based on the corresponding physical frequency regime as indicated in Fig.~\ref{fig:gravitational_wave_frequency_evolution}. The sharp decrease in detection horizon for CE at $\log r_{\rm b}\gtrsim 9.4$ is due to low-frequency emission below 10\,Hz (below CE's sensitivity band). Contours delineate final BH masses as in previous figures.}
\label{fig:gravitational_wave_detection_horizon}
\end{figure}

\begin{figure}
\centering
\includegraphics[width=0.98\linewidth]{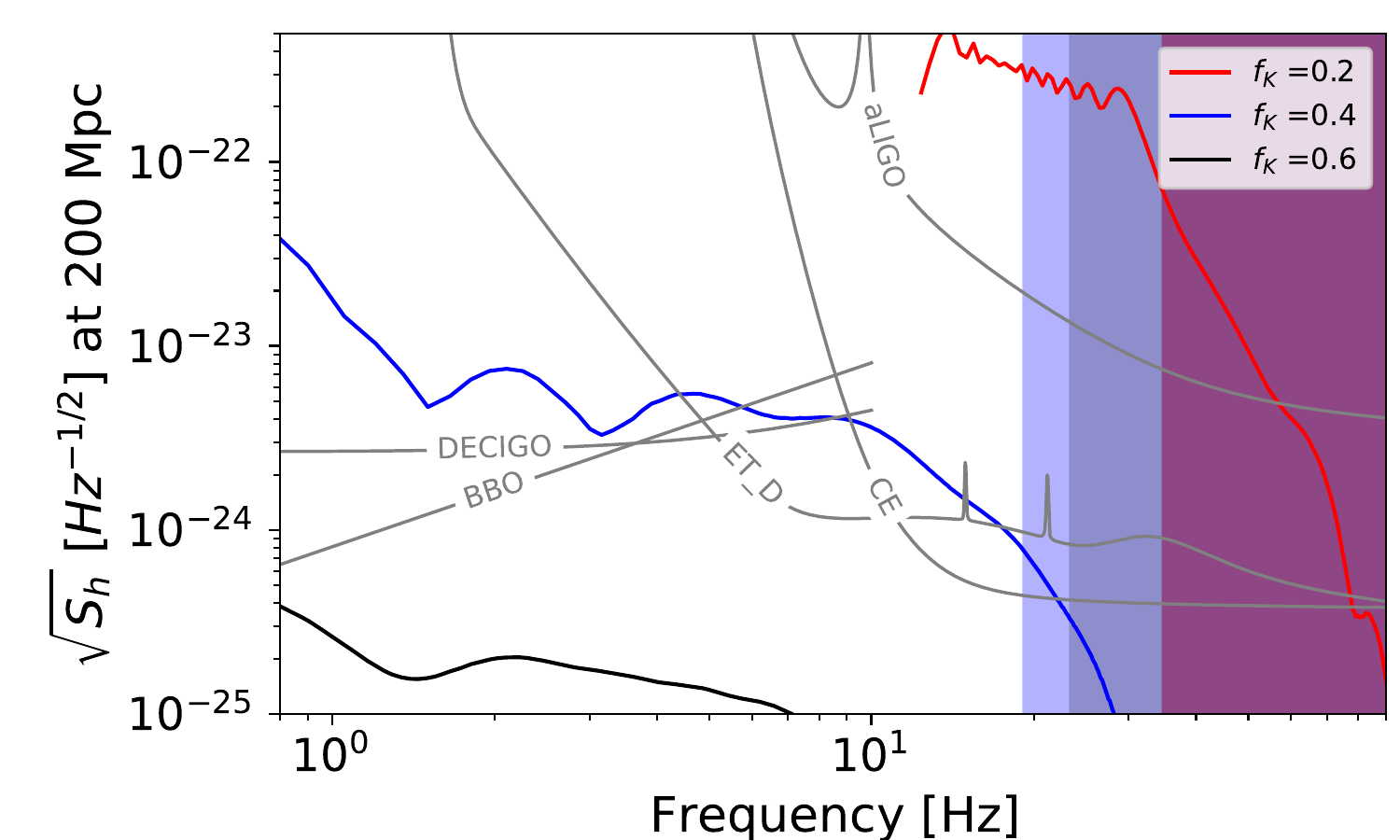}
\caption{Same as
  Fig.~\ref{fig:gravitational_wave_frequency_sensitivity} but for the
  progenitor model \texttt{250.25} with $p=4.5$,
  $r_{\rm b}= 1.5\times10^{9}$\,cm and different values of the Keplerian
  fraction, $f_{\rm K} = 0.2, 0.4$ and $0.6$.  Relatively low to medium angular momentum models
  (here $f_{\rm K}\approx 0.2$) generate a stronger signal in all detectors (with detection horizon $\sim$ 400 Mpc with aLIGO at an SNR of 8)
  compared to cases with higher angular momentum (large values of $f_{\rm K}$). The $f_{\rm K} = 0.6$ model would only be detectable by BBO with a distance up to 100 Mpc at an SNR of 8 (cf.~Fig.~\ref{fig:gravitational_wave_detection_horizon_2}).}
\label{fig:gravitational_wave_frequency_sensitivity_diff_rotation}
\end{figure}

We calculate the gravitational wave strain of emitted gravitational waves as described in Appendix \ref{app:GW_emission}. Figures \ref{fig:gravitational_wave_strain}--\ref{fig:gravitational_wave_frequency_sensitivity_diff_rotation} present results for gravitational-wave emission, evaluated for a typical distance of 200\,Mpc, at which superKN events are expected to occur once every $\sim\!3$ years for our fiducial local superKN rate of 10 Gpc$^{-3}$ yr$^{-1}$ (Sec.~\ref{sec:rates}). Figure~\ref{fig:gravitational_wave_strain} shows the time evolution of the plus and cross polarization strain calculated for the fiducial progenitor model (Fig.~\ref{fig:Renzo_evolution}) assuming a face-on orientation of the collapsar disk ($\iota=0$). The maximum characteristic strain $h_c$ (typically $h_c\sim 10^{-24} - 10^{-22}$) and the frequency range of the gravitational wave emission vary considerably across the $\{f_{\rm K},r_{\rm b}\}$ parameter space (Figs.~\ref{fig:gravitational_wave_frequency} and \ref{fig:gravitational_wave_strain_contour}, Appendix \ref{app:GW_emission}). 

SuperKN collapsars are multi-band gravitational-wave sources. Figures \ref{fig:gravitational_wave_frequency_sensitivity} and \ref{fig:gravitational_wave_frequency_sensitivity_diff_rotation} compare the  gravitational-wave signal in frequency space to the sensitivity of advanced LIGO (aLIGO), {\it Cosmic Explorer} (CE), {\it Einstein Telescope} (ET), {\it DECi-hertz Interferometer Gravitational wave Observatory} (DECIGO), and {\it Big Bang Observer} (BBO). Gravitational-wave emission typically starts at a few tens of Hz in the frequency band of aLIGO, CE, and ET, and subsequently drifts into the deciherz regime of DECIGO and BBO as the disk expands. The relative strain amplitude in these two different bands encodes information about the total mass and mass profile of the progenitors (Fig.~\ref{fig:gravitational_wave_frequency_sensitivity}). Lighter progenitors typically give rise to louder gravitational-wave signals over a narrower frequency band for the same rotation profile.

The overall magnitude of the amplitude spectral density is largely determined by the progenitor angular momentum as illustrated in Fig.~\ref{fig:gravitational_wave_frequency_sensitivity_diff_rotation}. In the limit of high angular momentum (large value of the parameter $f_{\rm K}$) for fixed $r_{\rm b}$, the instability and gravitational-wave emission are triggered earlier than for smaller values of $f_{\rm K}$ (cf.~Fig.~\ref{fig:gravitational_instabilities}). This is because matter deposition in the disk at early times is enhanced (rather than direct fallback onto the black hole). Under these conditions, the gravitational-wave signal is relatively weak due to the small disk and BH mass. Owing to enhanced viscosity and enhanced accretion during the instability epoch, disks that become unstable early on tend to stay relatively light; the gravitational-wave signal thus remains relatively weak throughout the fallback process. As a result, these signals tend to peak late and thus in the decihertz regime, which may only render them detectable there for Mpc distances. A non-detection in the high-frequency band may thus be indicative of the angular momentum budget of the progenitor star. In the other limit of low angular momentum (small value of the parameter $f_{\rm K}$ and large $r_{\rm b}$), the accretion disk may never become susceptible to the instability and gravitational-wave emission may be negligible (cf.~Fig.~\ref{fig:gravitational_instabilities}). Hence, there exists an intermediate regime of progenitor angular momentum (intermediate values of $f_{\rm K}$) in which the gravitational wave strain becomes maximal. For the given parameters of our fiducial progenitor model, this optimum is reached for $f_{\rm K}\approx 0.2$, which is also reflected by the detection horizons (Figs.~\ref{fig:gravitational_wave_detection_horizon}, \ref{fig:gravitational_wave_detection_horizon_2}).

We calculate a detection horizon for these events assuming an optimal matched filter and an SNR of 8 (see Appendix \ref{app:GW_emission} for details). We find a detection horizon of $\sim$5\,Mpc (aLIGO), $\sim$300\,Mpc (ET), $\sim$250\,Mpc (CE), and $\sim$425\,Mpc (DECIGO) for our fiducial model with mass $250.25M_{\odot}$, $f_{\rm K}=0.3$, and $r_{\rm b}=1.5\times 10^9$\,cm. A parameter space study of the detection horizons is presented in Fig.~\ref{fig:gravitational_wave_detection_horizon}, showing that third-generation detectors (ET, CE) as well as DECIGO are able to detect gravitational waves from superKN collapsars at distances of typically a few hundred Mpc up to a few Gpc. Detection horizons for aLIGO are typically limited to $\lesssim 100$\,Mpc (Fig.~\ref{fig:gravitational_wave_detection_horizon_2}, Appendix~\ref{app:GW_emission}). BBO will be particularly sensitive to the lowest-frequency sources with low angular momentum in the progenitor `core' (medium to large values of $r_{\rm b}$) and typically reach several hundred Mpc to several Gpc. (Fig.~\ref{fig:gravitational_wave_detection_horizon_2}, Appendix~\ref{app:GW_emission}).

\subsection{GW190521}
\label{sec:GW190521}

Our work has several potential implications for the gravitational wave
event \gw{} \citep{Abbott+20_190521}.  Firstly, as already
discussed, in the standard interpretation of \gw{} as a binary BH
coalescence, mass loss associated with the birth of one or both of the
constituent BHs can place them in the nominal PI mass gap ``from
above'', even if they would have been above the PI gap if all of the
star's mass were accreted at the time of core collapse
(Fig.~\ref{fig:Renzo_ejecta}, top panel).  To generate a BH with a
mass consistent with the more massive member of \gw{} of
$\sim 88 M_{\odot}$ \citep{Abbott+20_190521} from a star with a helium
core nominally above the gap, would require the ejection of
$\gtrsim 50M_{\odot}$ of ejecta (most of it $r$-process enriched;
Sec.~\ref{sec:fallback}).  In a direct sense, superKNe probe one
channel for forming BHs in the PI mass gap.

Our scenario requires a fast rotating pre-collapse star and predicts
that the magnitude of the BH spin would be nearly maximal
($a_{\rm BH,fin} \sim 1$; Fig.~\ref{fig:Renzo_GRB_accretion}, bottom
panel).  Although a low orbit-aligned spin
$\chi_{\rm eff} \lesssim 0.35$ (90\% confidence) was measured for
\gw{}, there is some evidence for a large spin component in the
binary plane \citep{Abbott+20_190521}.  However, assuming that the
progenitor stars can retain large rotation rates
\citep[see however][]{spruit:02,fuller:19} and the
progenitor of \gw{}~formed from an isolated stellar binary through
common envelope evolution \citep[e.g.,][]{belczynski:16Nat}, stable
mass transfer \citep[e.g.,][]{vandenheuvel:17, vanson:21}, or via
chemically homogeneous evolution driven by tidal interactions
\citep[e.g.,][]{maeder:00, demink:16, marchant:16}, one would
expect the stellar angular momentum vector---and hence that of the
BHs formed from the collapse---to be aligned with the orbital angular
momentum \citep{mandel:16b}.


In the case of rapidly rotating progenitors, we speculate that misaligned spins could arise from a kick imparted to
the BH by mass loss in the disk winds.  Our calculations in
Sec.~\ref{sec:GW} indicate that the formed disks can become
self-gravitating and hence will be subject to bar-mode like
instabilities, generating non-axisymmetric spiral density waves.  The
latter could impart a non-axisymmetric component to the wind mass
loss, which would endow the BH with an effective kick.  To significantly
misalign the spins without breaking the binary, the natal kick must be
comparable to the pre-collapse orbital velocity of the system,
$v_{\rm kick} \sim 300$ km s$^{-1}$ (e.g.,
\citealt{kalogera:96, callister:20}). Given the characteristic wind ejecta speed
$v_{\rm ej} \sim 0.1$ c, from momentum conservation an asymmetry in
the disk mass-loss rate or velocity at the level of
$\gtrsim v_{\rm kick}/v_{\rm ej} \sim 10^{-2}$ would be sufficient to
impart significant spin-orbital misalignment.  Although
self-gravitating instabilities result in non-axisymmetric disk density
fluctuations at the level $\delta \rho/\rho \gtrsim 0.1$, quantifying the
extent to which these impart non-axisymmetric mass-loss will require
additional GRMHD simulations of the disk outflows in the regime of
massive, self-gravitating disks.

In an alternative approach, \citet{shibata_alternative_2021} interpret \gw{} as gravitational waves from a non-axisymmetric instability similar to those discussed in Sec.~\ref{sec:GW} in a massive BH--disk system, thought to originate from the collapse of a massive star. In contrast to the BH--torus systems of fixed mass numerically evolved by \citet{shibata_alternative_2021}, in an astrophysical setting mass is continuously fed to the disk at a rate $\dot{m}_{\rm fb, disk}$ due to fallback of the progenitor envelope (Sec.~\ref{sec:fallback}). If the accretion disk approaches the gravitationally unstable regime its mass evolution $M_{\rm disk}(t)$ is dominated by the addition of fallback material through $\dot{m}_{\rm fb, disk}$. The associated timescale over which the fallback rate changes is $\tau_{\dot{m}_{\rm fb, disk}}\propto t^\alpha$, where $\alpha\approx 1$ (\citealt{siegel_collapsars_2019} and Sec.~\ref{sec:fallback}). This is reflected by the fact that disks for the progenitor models considered here become gravitationally unstable on a timescale of a few seconds after BH formation, and this instability phase then lasts for a few seconds to tens of seconds (Fig.~\ref{fig:gravitational_instabilities}, Sec.~\ref{sec:GW}). Generating a gravitational-wave signal of only a few cycles and duration $t_{\rm GW}\sim 0.1$\,s as required by \gw{} \citep{Abbott+20_190521} with comparatively negligible amplitude thereafter thus requires a fallback rate given by the total mass of the BH--disk system divided by the duration of the signal of $\dot{m}_{\rm fb, disk} > (M_\bullet + M_{\rm disk})/t_{\rm GW}\sim 650-1000\,M_\odot\,\text{s}^{-1}$ for the configurations considered by \citet{shibata_alternative_2021}. While fallback rates of the order of up to $\sim\!30-40\,M_\odot\,\text{s}^{-1}$ may be reached realistically (cf., e.g., Fig.~\ref{fig:Renzo_evolution}), fallback rates that are larger by one or two orders of magnitude seem implausible even with the most compact progenitor models possible (Sec.~\ref{sec:stellar_models}). These fallback rates also set limits on the compactness and thus on the gravitational-wave frequency of possible BH--disk systems---the frequency of gravitational-wave emission may be in tension with \gw{} as well. While a frequency around $\sim\!60$\,Hz of \gw{} may not be impossible per se, our unstable BH--disk systems are typically not compact enough to reach such high frequencies even for an $l=m=2$ mode and instead strongly prefer maximum frequencies below 20-30\,Hz (Fig.~\ref{fig:gravitational_wave_frequency_evolution}).  Another consequence of the weaker gravitational wave signals that we predict from massive collapsars are much closer detection horizons, which for Advanced LIGO at O3 sensitivity amount to $\lesssim 100$ Mpc (Fig.~\ref{fig:gravitational_wave_detection_horizon_2}, Appendix \ref{app:GW_emission}); it is unlikely that a superKN transient from \gw{} would have gone undetected by existing wide-field optical surveys at such close distances.

\subsection{Implications for Galactic $r$-Process Enrichment}

Although the massive collapsars studied here are probably less common by a factor $\gtrsim 10-30$ than the bulk of ordinary collapsars (Sec.~\ref{sec:rates}), they can in principle generate $\sim 10$ times more $r$-process ejecta mass for similar progenitor angular momentum structure.  The contribution of superKNe to total $r$-process production in the Universe could therefore be non-negligible relative to that of ordinary collapsars.

SuperKNe are probably too rare to explain the occurrence of individual $r$-process pollution events in small dwarf galaxies (e.g., \citealt{Ji+16}), but this does not exclude them from contributing to larger stellar systems.  The total mass of $r$-process elements in the Milky Way is only $\sim 10^{4}M_{\odot}$, so given an $r$-process yield of $\gtrsim 10M_{\odot}$ per superKNe, the number of contributing events must be $\lesssim 1000$ and probably $\lesssim 100$ (accounting for the dominant contribution likely coming from other channels such as lower mass collapsars and neutron star mergers).

Depending on the efficiency of gas mixing and retention in the environments of superKNe, subsequent generations of star formation could produce a modest number of extremely $r$-process-enriched stars.  The dilution mass of the interstellar medium into which the superKN ejecta is mixed, can be estimated as (e.g., \citealt{Macias&RamirezRuiz19})
\begin{equation}
    M_{\rm dil} \approx 2\times 10^{7}M_{\odot}\left(\frac{E_{\rm kin}}{5\times 10^{53}
\rm erg}\right)^{0.97}\left(\frac{n}{\rm cm^{-3}}\right)^{-0.062},\end{equation}
where we have assumed $10$ km s$^{-1}$ for the sound speed of the interstellar medium.  The value of $M_{\rm dil}$ in superKNe is larger than in ordinary SNe ($E_{\rm kin} \sim 10^{51}$ erg) or ordinary collapsars ($E_{\rm kin} \sim 10^{52}$ erg) by a factor of $\gtrsim 10-100$.

The total amount of $r$-process material generated by superKNe is
larger than ordinary collapsars by a similar factor $\sim 10$, while
the production $\sim 0.1-0.5M_{\odot}$ of $^{56}$Ni and hence
$^{56}$Fe (Fig.~\ref{fig:Renzo_mass_fractions}) are similar to
ordinary collapsars.  If a superKN were to occur in otherwise pristine
material at very low metallicity (perhaps an questionable idealization
given that SNe tend to be spatially and temporally clustered), the
next generation of stars which form from this material could possess a
metallicity as low as [Fe/H] $\sim -5$ and a Europium abundance as
high as [Eu/Fe] $\sim 5$, much higher than the current record holder (\citealt{Reichert+21}).  This abundance combination would also contrast
strongly with the chemical signatures of PI SNe (e.g.,
\citealt{woosley:02,Aoki+14}), for which a large quantity of iron
group elements but no $r$-process elements are produced.

\section{Conclusions}
\label{sec:conclusions}

We have explored the collapse of rotating very massive
$\gtrsim 130M_{\odot}$ helium stars and predicted their nucleosynthetic, electromagnetic, and gravitational waves signatures.  Our conclusions can be summarized as follows.

\begin{itemize}

\item Building on \citet*{siegel_collapsars_2019}, we present a semi-analytic model for the BH accretion disk in collapsars and its associated outflows which predict the quantity and composition of the disk wind ejecta, as well as the final BH mass and spin, given an assumed angular momentum structure of the progenitor star.  The accretion regimes are calibrated based on the results of numerical GRMHD simulations and analytic scaling relations (Appendix \ref{sec:accretion_rates}).  Although the radial angular momentum structure of the progenitor star at collapse is uncertain theoretically, our approach allows us to cover a wide portion of the physically allowed parameter space.  Applied to ``ordinary'' low-mass collapsars, the model predicts accretion luminosities and durations broadly consistent with long GRB observations.

\item Our main application is to massive collapsars, originating from
  progenitor stars with final helium core masses
  $\sim 125-150M_{\odot}$, which avoid pair instability SNe and
  nominally (in the case of zero mass ejection) would create BHs above
  the PI mass-gap.  Analogous to lower-mass collapsars, as the fall-back accretion rate declines in time, the composition of the disk outflows systematically evolve from heavier to lighter elements (Fig.~\ref{fig:Renzo_accretion_rate}).  Across a wide parameter space of progenitor rotational properties, we find total wind ejecta masses $\sim 10-50M_{\odot}$, of which $\sim 10-60\%$ is composed of $r$-process nuclei, including a sizable quantity of lanthanide elements associated with heavy $r$-process production.  The remaining ejecta is primarily unprocessed material (assumed to be $^{4}$He in our models) and a modest quantity $\sim 0.1-1M_{\odot}$ of $^{56}$Ni, formed from the brief hot, proton-rich phases of the disk evolution.

\item The radioactive decay of $r$-process nuclei and $^{56}$Ni in the ejecta of massive collapsars powers a months-long transient with a peak luminosity $\sim 10^{42}$ erg s$^{-1}$ (Fig.~\ref{fig:sn_bollc}), which we refer to as a ``superKN''.  The spectral energy distribution of superKNe near maximum light peaks at several microns due to the large opacity of the lanthanide elements (Fig.~\ref{fig:sn_pkspec}), similar to lanthanide-rich kilonovae from neutron star mergers.  Although the bolometric light curves of superKNe are broadly similar to common types of core-collapse SNe, their combination of extremely red colors and high-velocity spectral features ($v_{\rm ej} \sim 0.1$ c) should render superKNe distinguishable from other transient classes.  

Our radiative transfer calculations have assumed a homogeneous ejecta structure; if the ejecta instead exhibits significant radial stratification, particularly a low lanthanide abundance in the highest velocity outermost layers, then the early light curve could be substantially brighter and bluer than our baseline predictions.

\item Even for a progenitor stars well above the PI mass gap at
  collapse, the final BH remnant can populate the entire mass gap
  between $\sim 55-130M_{\odot}$ due to disk wind mass-loss (e.g.,
  Fig.~\ref{fig:Renzo_ejecta}; Tab.~\ref{tab:ejecta_parameters}).
  SuperKNe therefore probe one mechanism for populating the PI mass gap
  ``from above''.  The BHs formed through this channel are predicted to be rapidly spinning due to the large accretion of angular momentum, with final Kerr parameter $a_{\rm BH,fin} \sim 1$.  If the BH is formed in a binary, we speculate that its spin angular momentum axis could become misaligned with that of the binary angular momentum due to non-asymmetric mass-outflows associated with the gravitationally-unstable phases of the accretion (Sec.~\ref{sec:GW}).  Future numerical simulation work is necessary to explore this possibility quantitatively.

\item One avenue to discover SuperKNe is via wide-field optical/infrared surveys.  A 5-year survey with the {\it Roman Space Telescope} similar to that planned for Type Ia SNe, could potentially detect $\sim 1-20$ superKNe out to redshift $z \simeq 1$, for an assumed $z = 0$ superKN rate of $\sim 10$ Gpc$^{-3}$ yr$^{-1}$.  SuperKNe could also be discovered by LSST, but the detection rate is lower because the predicted emission peaks at redder wavelengths than covered by the LSST bands.  Measurements or limits on the occurrence rate of superKNe would constrain the birth rate of PI mass gap BHs via this channel (for comparison, the local rate of \gw{}-like mergers is $\sim 1$ Gpc$^{-3}$ yr$^{-1}$; \citealt{Abbott+20_190521}).  SuperKNe may also be detectable following (particularly energetic) GRBs with {\it JWST} after the GRB afterglow has faded.

\item The large kinetic energies of the superKN ejecta $\gtrsim 10^{53}$ erg results in a bright, long-lived synchrotron radio transient as the ejecta decelerates via shock interaction with the circumstellar medium (Sec.~\ref{sec:radio}).  However, the slow evolution of the radio emission for typical circumstellar densities will render these radio sources challenging to identify as radio transients (they may appear as luminous persistent sources in star-forming dwarf galaxies, for example; \citealt{Eftekhari+20}).  

If superKNe occur inside gaseous AGN disks, shock interaction with the dense disk material could substantially enhance the optical luminosity of the event relative to that powered by radioactivity alone.  This offers a speculative explanation for the claimed optical counterpart of \gw{}~\citep{Graham+20}, provided it represents a gravitational wave burst from a core collapse event \citep{shibata_alternative_2021} instead of a black hole merger (however, see Sec.~\ref{sec:GW190521}).

\item The massive accretion disks from massive collapsars can become gravitationally unstable, generating gravitational wave emission as a result of non-axisymmetric density fluctuations.  The predicted duration of the gravitational waves is several seconds or longer (calling into question the core-collapse origin for \gw{} proposed by \citealt{shibata_alternative_2021}, Sec.~\ref{sec:GW190521}), while the frequency range overlaps the sensitivity window of ground-based (e.g., LIGO/CE/ET) and space-based intermediate-frequency gravitational-wave detectors (e.g., DECIGO, BBO).  Unlike the gravitational wave signal of compact binary mergers, which increase in frequency and amplitude with time (``chirp''), the gravitational wave signals of collapsar disks {\it decreases} in frequency as the disk radius grows (``sad-trombone'').  Our simple estimates suggest that gravitational waves from massive collapsar disks are detectable by CE/ET/DECIGO to distances of up to several hundred Mpc (Figs.~\ref{fig:gravitational_wave_frequency_sensitivity}, \ref{fig:gravitational_wave_detection_horizon}, \ref{fig:gravitational_wave_frequency_sensitivity_diff_rotation}), interior to which the event rate could be as high as once every few years.

\item SuperKNe are unlikely to contribute dominantly to the total
  production of $r$-process elements in the Universe, compared to
  neutron star mergers or ordinary low-mass collapsars, because the
  progenitors are disfavored by the initial mass function of stars.
  However, their extremely $r$-process-rich but iron-poor ejecta could
  in principle seed the creation of a small fraction of stars with abundance ratios more
  extreme than currently known metal-poor $r$-process-enhanced stars
  (e.g., [Eu/Fe] $\sim 5$).

\end{itemize}

\section*{Acknowledgements}

DMS and AA acknowledge discussions with R.~Essick. This research was enabled in part by support provided by SciNet (www.scinethpc.ca) and Compute Canada (www.computecanada.ca). DMS acknowledges the support of the Natural Sciences and Engineering Research Council of Canada (NSERC), funding reference number RGPIN-2019-04684. Research at Perimeter Institute is supported in part by the Government of Canada through the Department of Innovation, Science and Economic Development Canada and by the Province of Ontario through the Ministry of Colleges and Universities.  AA acknowledges support through a MITACS Globalink Graduate Fellowship. JB, BDM, and MR acknowledges support from the National Science Foundation (grant AST-2002577). VAV acknowledges support by the Simons Foundation through a Simons Junior Fellowship (\#718240) during the early phases of this project.

\appendix
\section{Dependence of Critical Accretion Rates on BH Mass}
\label{sec:accretion_rates}

Here we estimate various critical accretion rates which enter our calculations in Sec.~\ref{sec:fallback}.  We focus on deriving the scaling laws as a function of the effective $\alpha$ viscosity parameter and the BH mass $M_{\bullet}$, normalizing our final results to those obtained by one-dimensional (e.g., \citealt{chen_neutrino-cooled_2007}) and three-dimensional MHD simulations (e.g., \citealt{de_igniting_2020}).

\subsection{Ignition Accretion Rate}
\label{sec:ignition}

We begin with the critical accretion rate $\dot{M}_{\rm ign}$, below which neutrinos can no longer efficiently cool the inner regions of the disk and lead to its self-neutronization.

At the transition between efficient neutrino cooling and adiabatic evolution, the midplane temperature $T$ is determined by balancing the specific rate of neutrino cooling $\dot{q}_{\nu} \propto T^{6}$ in the optically-thin limit (due to the capture of relativistic electrons and positrons on free nuclei; e.g., \citealt{qian_nucleosynthesis_1996,di_matteo_neutrino_2002}), with the rate of viscous heating,
\begin{eqnarray}
\dot{q}_{\rm visc} \approx \frac{9}{4} m_{\rm p} \nu \Omega^{2} \approx \frac{9}{4}\alpha m_{\rm p}r^{2}\Omega^{3}\left(\frac{H}{r}\right)^{2},
\end{eqnarray}
where $\nu = \alpha c_{\rm s}H \simeq \alpha R^{2}\Omega(H/r)^{2}$ is the kinematic viscosity and $m_p$ is the proton mass.  We have assumed a Keplerian disk with angular velocity $\Omega \simeq \Omega_{\rm K} = (GM_{\bullet}/r^{3})^{1/2}$, midplane sound speed $c_{\rm s} = H\Omega$ and aspect ratio $H/r$.  Since $H/r \sim \mathcal{O}(1)$ once advective cooling competes with radiative cooling (e.g., \citealt{di_matteo_neutrino_2002}), we have $H/r \sim \mathrm{constant}$ at the transition to an advective disk from an efficiently neutrino-cooled disk.  This gives,
\begin{eqnarray}
T^{6} \propto \alpha r^{2}\Omega^{3}. \label{eq:1}
\end{eqnarray}
We also assume that, near the transition point to an advective disk, radiation pressure dominates over gas pressure in the disk midplane, i.e. the midplane pressure obeys $P \propto T^{4}$.  From vertical hydrostatic equilibrium, we then have
\begin{eqnarray}
\text{const.} &\sim& \left(\frac{H}{r}\right)^{2} \approx \frac{c_{\rm s}^{2}}{r^{2}\Omega^{2}} \approx \frac{P/\rho}{r^{2}\Omega^{2}} \propto \frac{\alpha T^{4}r}{\Omega \dot{M}} \nonumber\\
&\Rightarrow& T^{4} \propto \frac{\dot{M}\Omega}{\alpha r} 
\label{eq:2}
\end{eqnarray}
where $\rho$ is the midplane density and in the final line we have used the fact that the local disk mass $M_{\rm d} \propto \rho r^{2}H$ accretes on the local viscous time $t_{\rm visc} \propto \nu/r \sim \alpha^{-1}\Omega^{2}(H/r)^{-2}$, i.e.
\begin{eqnarray}
\dot{M} &\propto& \frac{M_{\rm d}}{t_{\rm visc}} \propto \frac{\rho r^{2}H}{\alpha^{-1}\Omega^{-1}(H/r)^{-2}} \propto  \alpha \rho \Omega r^{3} \left(\frac{H}{r}\right)^{3}\nonumber \\
&\Rightarrow& \rho \propto \frac{\dot{M}}{\alpha \Omega r^{3}}.
\label{eq:3}
\end{eqnarray}
Combining Eqs.~(\ref{eq:1}), (\ref{eq:2}), we obtain
\begin{eqnarray}
\dot{M} \propto \alpha^{5/3}r^{7/3}\Omega.
\end{eqnarray}
Scaling $r$ to the radius of the innermost stable circular orbit, $r_{\rm ISCO} \propto M_{\bullet}$, we have $r \propto M_{\bullet}$ and $\Omega \propto M_{\bullet}^{-1}$, thus giving
\begin{eqnarray}
\dot{M}_{\rm ign} \propto \alpha^{5/3}M_{\bullet}^{4/3}.
\end{eqnarray}
A derivation of this scaling in the general-relativistic context is given in \citet{de_igniting_2020}. The scaling $\dot{M}_{\rm ign} \propto \alpha^{5/3}$ has been verified in 1D models \citep{chen_neutrino-cooled_2007} for BH masses $M_{\bullet} \approx 3M_{\odot}$.  However, the scaling should hold for the higher values of $M_{\bullet}$ of interest in this paper because the assumptions which enter the above derivation (i.e., optically thin cooling, radiation pressure dominating over gas pressure at the advective transition) only strengthen with increasing BH mass.  In particular, the ratio of radiation to gas pressure,
\begin{eqnarray}
\left.\frac{T^{3}}{\rho}\right|_{\dot{M}_{\rm ign}} \propto r^{5/3}\Omega^{3/2} \propto M_{\bullet}^{1/6},
\label{eq:entropy}
\end{eqnarray} increases with $M_{\bullet}$.  Other forms of neutrino cooling, particularly arising from the annihilation of electron-positron pairs into neutrino-antineutrino pairs, will dominate over the pair-capture rates entering the radiative cooling rate at sufficiently high $T^{3}/\rho$ (e.g., \citealt{qian_nucleosynthesis_1996}), changing the above scalings.  However, the weak $M_{\bullet}$-dependence in Eq.~(\ref{eq:entropy}) suggests that this is not likely to occur for even the highest BH masses $M_{\bullet} \sim 100M_{\odot}$ of interest in this paper.

\subsection{Neutrino Opaque and Trapping Thresholds}
\label{sec:opaque}

We now consider the accretion rate $\dot{M}_{\nu, \rm r-p}$, above which the inner disk is optically thick to neutrinos.  The vertical optical depth through the disk obeys
\begin{equation}
    \tau_{\nu} \propto \Sigma \kappa \propto \rho H T^{2},
    \label{eq:taunu}
\end{equation}
where $\sigma \propto T^{2}$ is the energy-dependent absorption cross section with which electron neutrinos or antineutrinos are absorbed by free neutrons or protons, respectively.  Using Eqs.~(\ref{eq:2}) and (\ref{eq:3}) for $T$ and $\rho$, and again taking $H/r \sim \text{const.}$, we find:
\begin{equation}
    \tau_{\nu} \propto \frac{\dot{M}^{3/2}}{\alpha^{3/2}\Omega^{1/2}r^{5/2}}.
\end{equation}
Evaluating the marginally thick condition $\tau_{\nu} = 1$ at $r \approx r_{\rm ISCO}$, we obtain
\begin{eqnarray}
\dot{M}_{\nu, \rm r-p } \propto \alpha M_{\bullet}^{4/3}.
\end{eqnarray}

\subsection{Neutrino Trapping Threshold}
\label{sec:trapping}

Finally, consider the ``trapping'' accretion rate $\dot{M}_{\rm tr}$, above which the thermal energy released by the accretion flow is advected into the BH faster than it can be radiated through neutrinos.  Equating the neutrino diffusion timescale out of the disk midplane $t_{\rm diff} \sim \tau_{\nu}(H/c)$ with the inwards flow time $\sim r/v$, we find that neutrinos are trapped for $\tau_{\nu} > (c/v)(r/H)$. Using $v \sim \nu/r$ for the inflow velocity of a steady disk and Eq.~(\ref{eq:taunu}), and again taking $H/r \sim \text{const.}$, the trapping condition can be written
\begin{eqnarray}
\Sigma \nu T^{2} > r.
\label{eq:trap}
\end{eqnarray}
Finally, recalling that $\nu \Sigma \propto \dot{M}$ for steady accretion, then using Eq.~(\ref{eq:2}) and evaluating Eq.~(\ref{eq:trap}) at $r = r_{\rm ISCO}$, we find
\begin{eqnarray}
\dot{M}_{\rm tr} \propto \alpha^{1/3}M_{\bullet}^{4/3}.
\end{eqnarray}



\section{Production of $^{56}$Ni in disk winds and dissociation threshold}
\label{sec:Ni56_production}

Assuming the dominant seed particle formation process in the disk wind outflow is $^{4}$He(2$\alpha,\gamma$)$^{12}$C, rather than the neutron-catalyzed reaction $^{4}$He($\alpha n,\gamma)^{9}$Be($\alpha,n)^{12}$C in neutron-rich environments $Y_e \ll 0.5$ \citep{woosley_alpha-process_1992}, the destruction of $\alpha$-particles in disk winds proceeds as \citep{roberts_integrated_2010}
\begin{equation}
	\frac{dY_{\alpha}}{d\tau} \approx -14 \rho^{2}Y_{\alpha}^{3}\lambda_{3\alpha}, \label{eq:dY_alpha}
\end{equation}
where $Y_{\alpha}$ is the abundance of $\alpha$-particles, $\lambda_{3\alpha}(T)$ is the temperature-dependent triple-alpha rate coefficient, and the factor of 14 is due to assuming $\alpha$-captures cease at $^{56}$Ni. Furthermore, $d \tau = -(\tau_{\rm d}/3T)dT$, where $\tau_{\rm d}$ is the expansion time of the outflow around the point of $\alpha$-particle formation. We take $\tau_{\rm d}\approx 5-30\,\text{ms}$ typical of expansion time scales of such accretion disk winds \citep{siegel_three-dimensional_2018,siegel_collapsars_2019}. Integration of Eq.~\eqref{eq:dY_alpha} yields the resulting abundance of seed particles \citep{roberts_integrated_2010}
\begin{equation}
Y_{\rm seed} = \frac{1}{56}\left\{1 - [1 + 35(\tau_{\rm d}/{\rm ms})s_{\rm f}^{-2}]^{-1/2}\right\}.
\label{eq:Yseed}
\end{equation}
Here, $s_{\rm f}$ is the final entropy in $k_{\rm B}$ per baryon, which we estimate by $s_{\rm f} \approx s_{\rm m} = s_{\rm m,rad} + s_{\rm m,N}$, where
\begin{eqnarray}
s_{\rm m,rad} &=& \frac{11\pi^{2}k^{3}m_\text{p}}{45 c^{3}\hbar^{3}}\frac{T_{\rm m}^{3}}{\rho_{\rm m}} \\
 &\approx& 21 \left(\frac{r_{\rm disk}}{5 r_{\rm g}}\right)^{-\frac{3}{8}}\left(\frac{\dot{M}}{10^{-4}M_{\odot}\,s^{-1}}\right)^{-\frac{1}{4}}\mskip-5mu\left(\frac{\alpha}{0.01}\right)^{\frac{1}{4}}\nonumber\\
 &&\times\left(\frac{M_{\bullet}}{3M_{\odot}}\right)^{\frac{7}{8}}\left(3h_{z,\text{disk}}\right)^{\frac{9}{4}},
\end{eqnarray}
is the entropy of radiation in the disk midplane and
\begin{equation}
s_{\rm m, N} \approx 13.3 + {\rm ln}\left(\frac{(T_{\rm m} / 1 \text{MeV})^{3/2}}{\rho_{\rm m}/10^7 \text{g}\,\text{cm}^{-3}}\right)
\end{equation}
is the entropy in the non-relativistic nucleons. The disk midplane density $\rho_{\rm m}$ and temperature $T_{\rm m}$ are calculated using an $\alpha$-disk model as in Appendix \ref{sec:accretion_rates} (see also \citealt{siegel_collapsars_2019}). We translate Eq.~\eqref{eq:Yseed} into a mass fraction $X_{^{56}\mathrm{Ni}}$ and $X_\mathrm{He}\approx 1 - X_{\rm seed}$ by assuming that seed particles are mostly within the iron peak (charge numbers $24\le Z \le 28$), using the nucleosynthesis results of \citet{siegel_collapsars_2019}.

At late times during the fallback process (low accretion rates), the accretion disk must be hot and dense enough to dissociate $\alpha$-particles as well as heavier nuclei of the infalling stellar material into individual nucleons, a requisite for the synthesis of $^{56}$Ni in disk outflows. We estimate the time of this transition to a state in which dissociation into individual nucleons becomes suppressed by considering a fluid of neutrons, protons, and $\alpha$-particles with a proton fraction $Y_e=0.5$ and determining the state in which half of the $\alpha$-particles are dissociated. Assuming that nuclear statistical equilibrium holds (a good assumption as our disk midplane temperatures typically stay above 0.5\,MeV), this can be estimated from the Saha equation \citep{shapiro_black_1983}
\begin{equation}
	X_{\rm nuc}^4 = 1.57 \times 10^{4} X_{\alpha} \rho_{{\rm m},10}^{-3} T_{{\rm m},10}^{9/2} \exp\left(-\frac{32.81}{T_{{\rm m},10}}\right), \label{eq:Saha}
\end{equation}
where $X_{\rm nuc}=2 X_{\rm p}=2 X_{\rm n}$ is the mass fraction of nucleons, and $\rho_{{\rm m},10}$ and $T_{{\rm m},10}$ denote the midplane density and temperature in $10^{10}\,\text{g}\,\text{cm}^{-3}$ and $10^{10}$\,K, respectively. Expressing $X_{\rm nuc} = 1 - X_{\alpha}$ and setting $X_\alpha=0.5$, Eq.~\eqref{eq:Saha} can be solved numerically for the transition time $t_{\rm diss}$ upon inserting $\rho_{\rm m}$ and $T_{\rm m}$ as obtained from the numerical evolution of Eqs.~\eqref{eq:evol_eqn_1}--\eqref{eq:evol_eqn_6}. The time-dependence of the functions $\rho_{\rm m}$ and $T_{\rm m}$ is determined by the evolution of $M_\bullet$, $M_{\rm disk}$, and $r_{\rm disk}$. At $t > t_{\rm diss}$ we set $Y_{\rm seed}=0$ in Eq.~\eqref{eq:Yseed}, discarding any potential $^{56}$Ni production after this time.

\section{Resolution studies}
\label{app:convergence}

\begin{figure}
\centering
\includegraphics[width=0.98\linewidth]{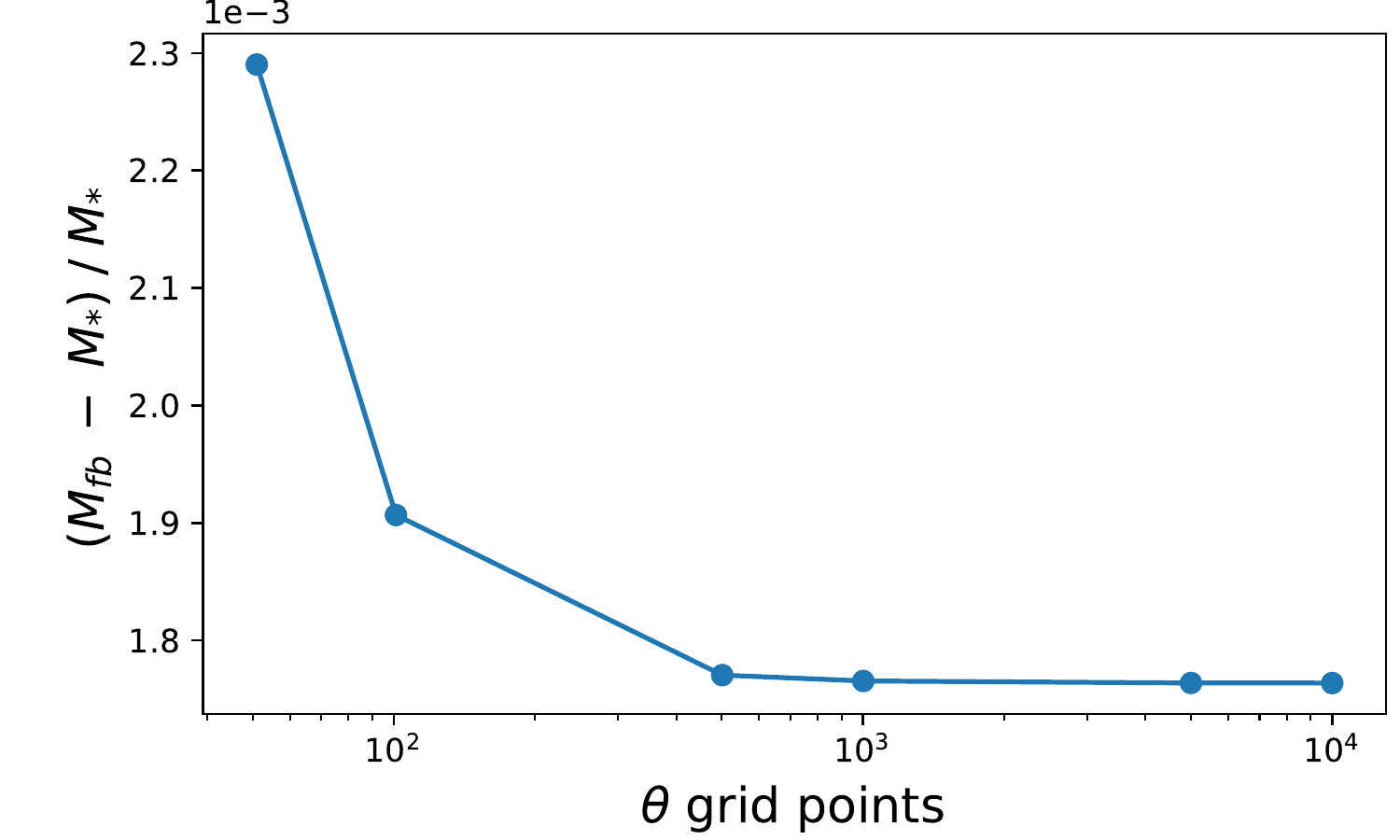}
\includegraphics[width=0.98\linewidth]{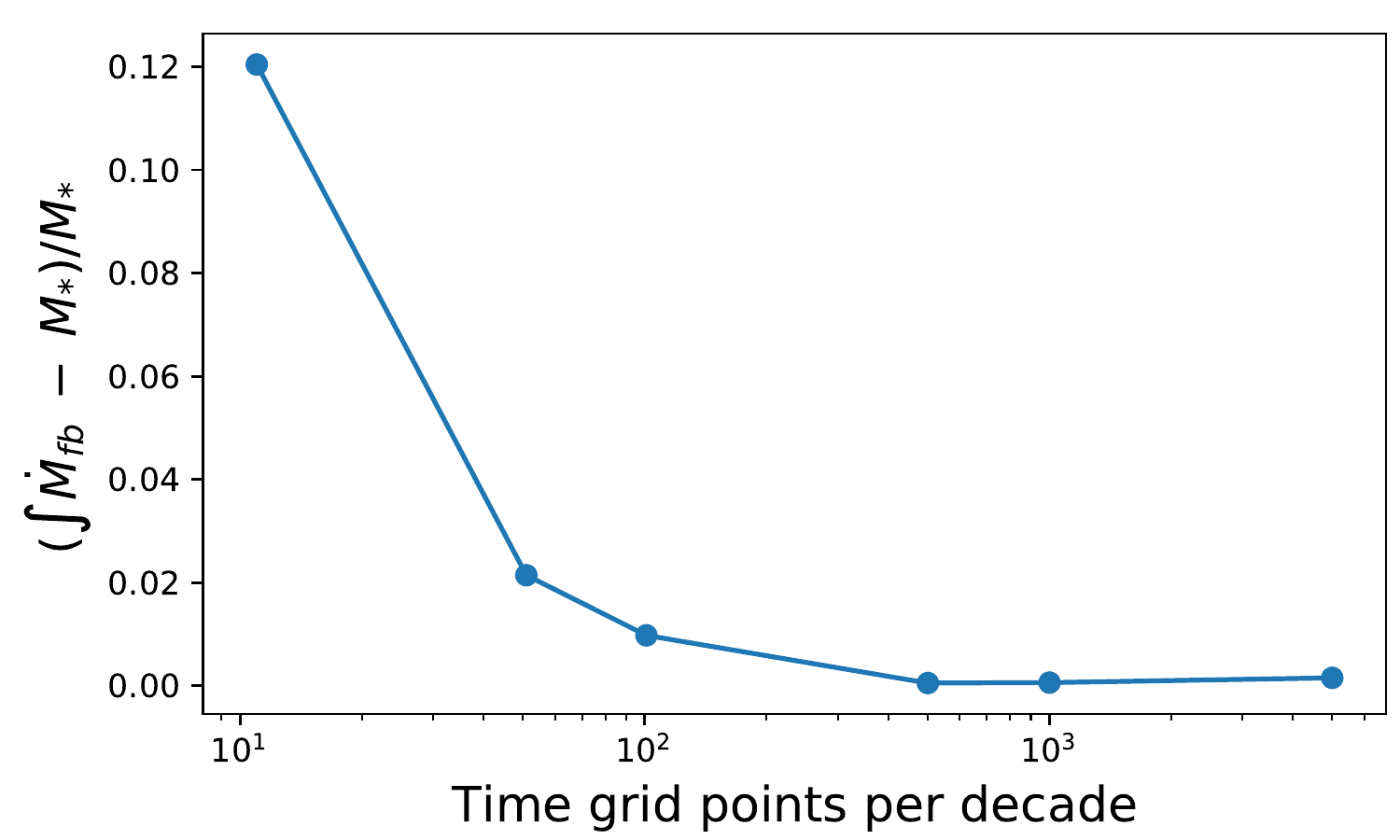}
\caption{Convergence test for model \texttt{250.25}. Top: relative error in computing the total mass of the star $M_\star$ using a grid discretized in both polar angle and radius, showing that at the fiducial angular resolution of $n_\theta=1001$ the relative error $\lesssim 0.0018$ is dominated by the radial resolution of the progenitor model. Bottom: total stellar mass as computed by numerical integration of Eqs.~\eqref{eq:evol_eqn_1}--\eqref{eq:evol_eqn_6} with fixed angular resolution $(n_\theta=1001)$, varying the time step. For sufficiently high temporal resolution, the relative error of $\lesssim 0.002$ is again dominated by the radial discretization of the progenitor model, consistent with the error budget shown in the top panel.}
\label{fig:Renzo_convergence}
\end{figure}

We have performed numerical convergence tests to check convergence of nucleosynthesis results from our collapsar model (Sec.~\ref{sec:fallback_results}) and to determine optimal resolution for our numerical collapsar evolution calculations (Sec.~\ref{sec:fallback}). Figure \ref{fig:Renzo_convergence} shows results of two convergence tests to determine optimal discretization for the polar coordinate and for time integration. The top panel illustrates that at our fiducial resolution in the polar angle of $n_\theta=1001$ the relative error in computing the total mass of the star by numerical integration is dominated by the radial resolution of the progenitor model (relative error of $\lesssim 0.0018$). The bottom panel indicates that for sufficiently high angular resolution ($n_\theta\sim 1001$) and sufficiently large number of time steps of several hundred to $10^{3}$, the relative error in computing the stellar mass by numerical integration of Eqs.~\eqref{eq:evol_eqn_1}--\eqref{eq:evol_eqn_6} is again dominated by the radial discretization of the progenitor model. The relative error of $\lesssim 0.002$ in this case is consistent with the error budget obtained for the corresponding convergence test in the polar angle. This shows that our results are converged with roughly $10^3$ grid points both in $\theta$ an in time, which we employ for all model runs.

\section{Results for ordinary collapsars and GRB properties}
\label{app:collapsars}

\begin{figure}
\centering
\includegraphics[width=0.98\linewidth]{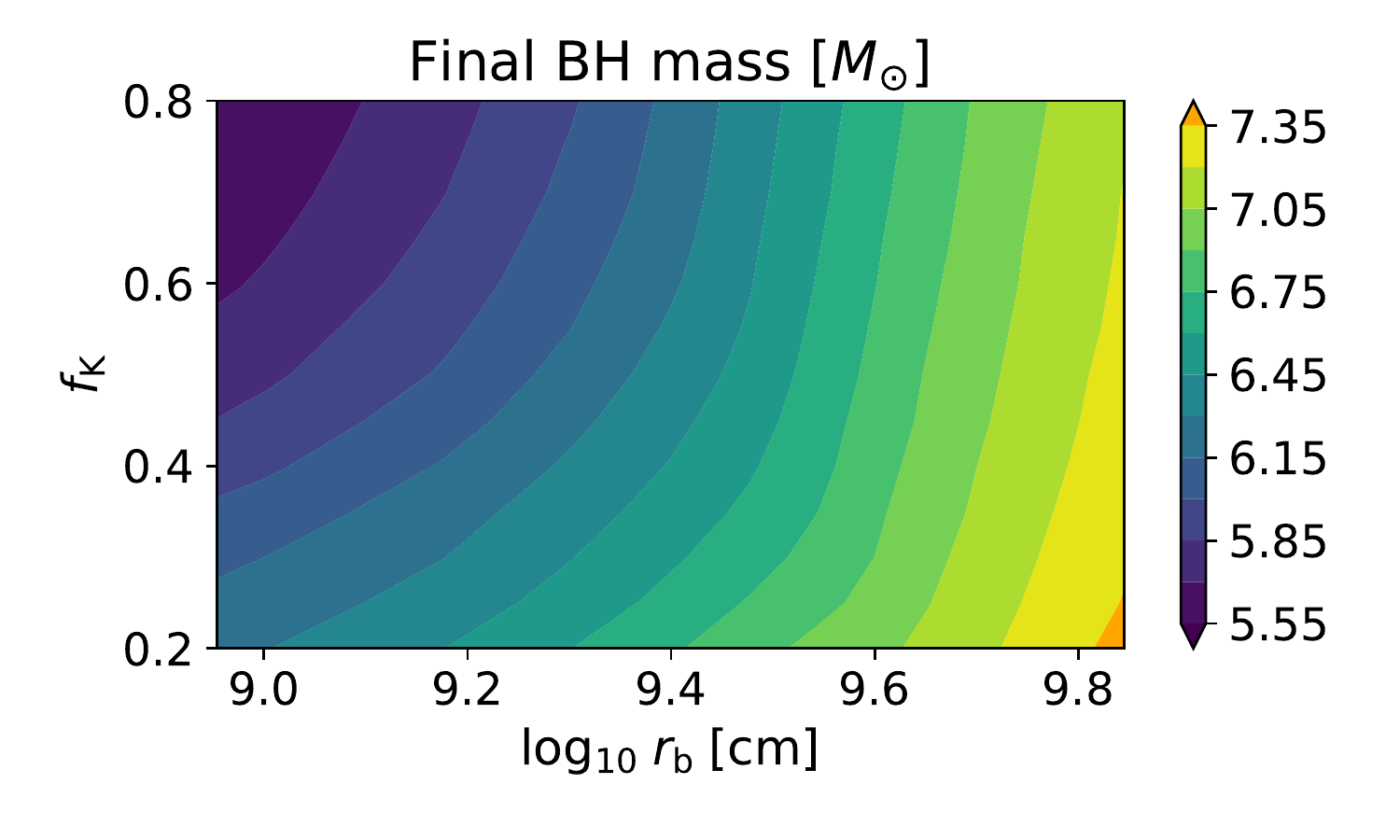}
\includegraphics[width=0.98\linewidth]{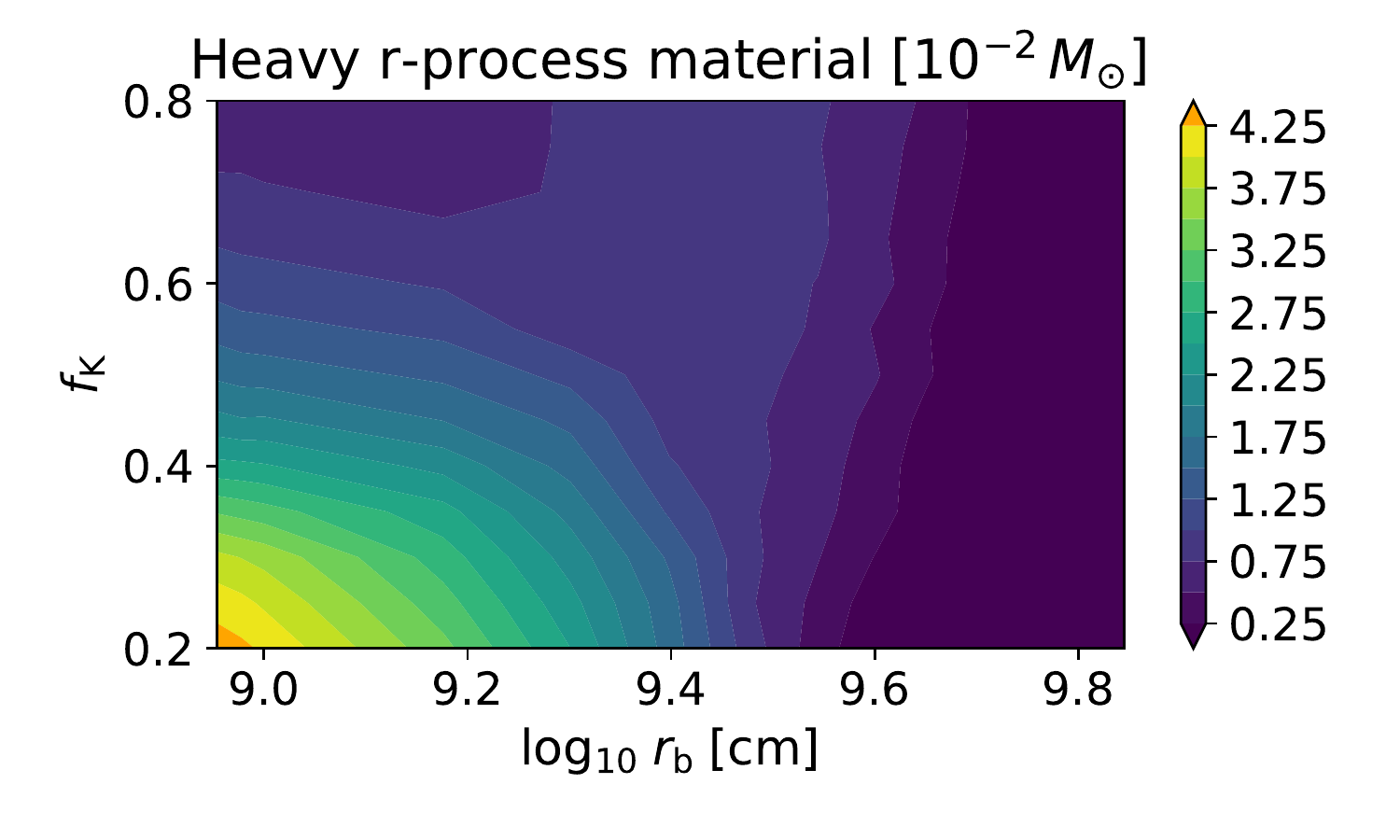}
\includegraphics[width=0.98\linewidth]{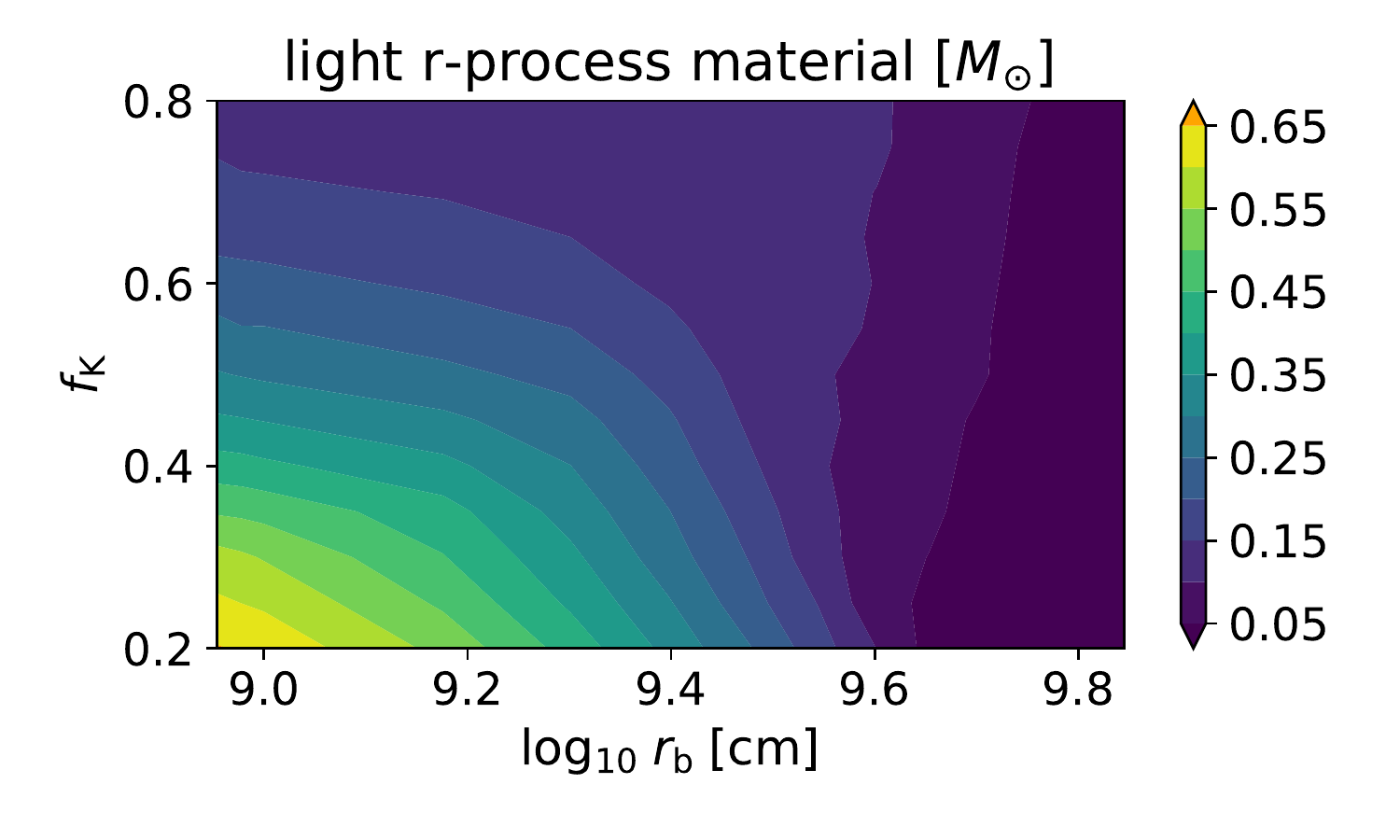}
\includegraphics[width=0.98\linewidth]{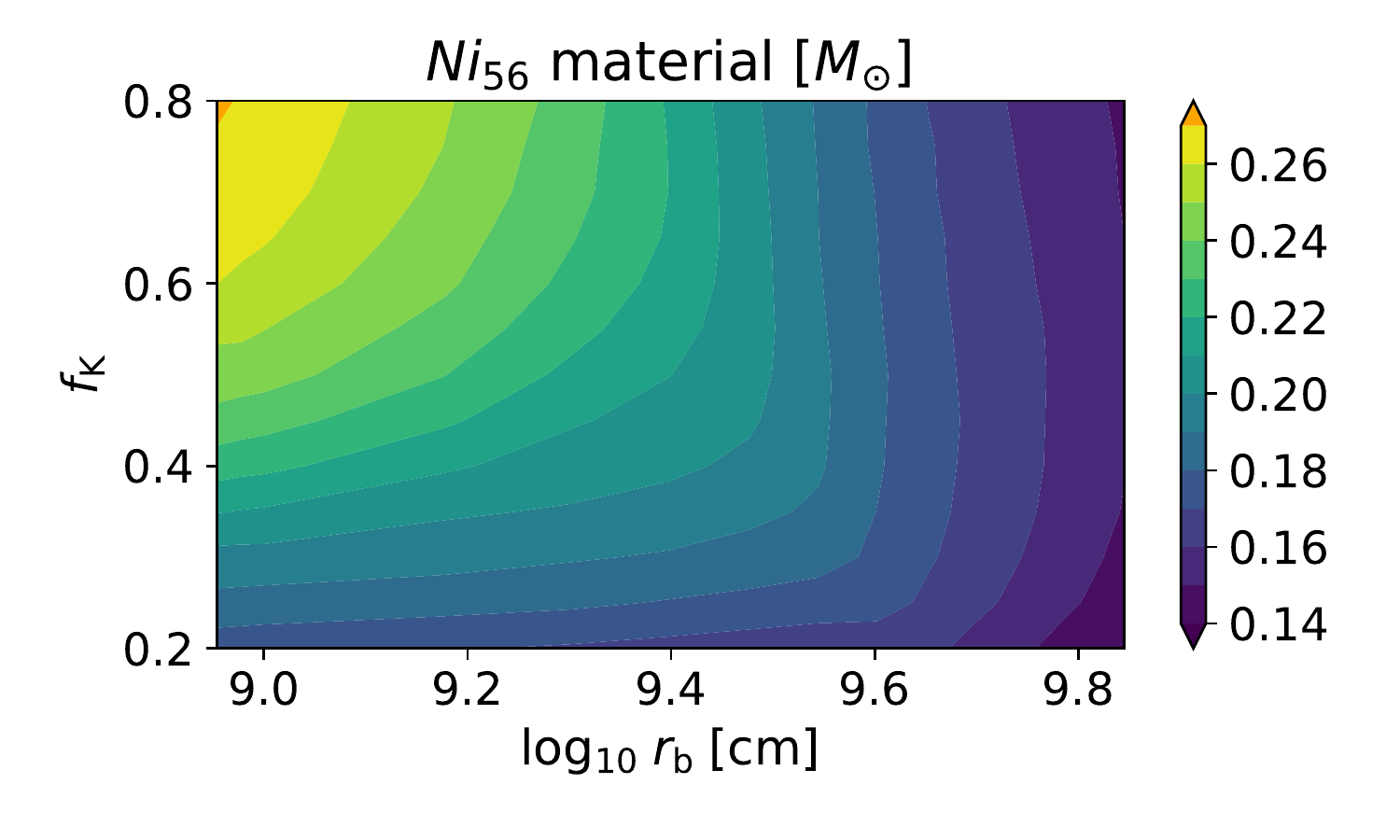}
\caption{Scan of the parameter space for model \texttt{E20} of \citet{heger_presupernova_2000}. Shown are the final BH mass (top), the total ejected mass [$M_\odot$] in heavy ($A>136$) $r$-process elements (center top), in light ($A<136$) $r$-process elements (center bottom), and $^{56}$Ni (bottom).}
\label{fig:Heger_ejecta}
\end{figure}

\begin{figure}
\centering
\includegraphics[width=0.98\linewidth]{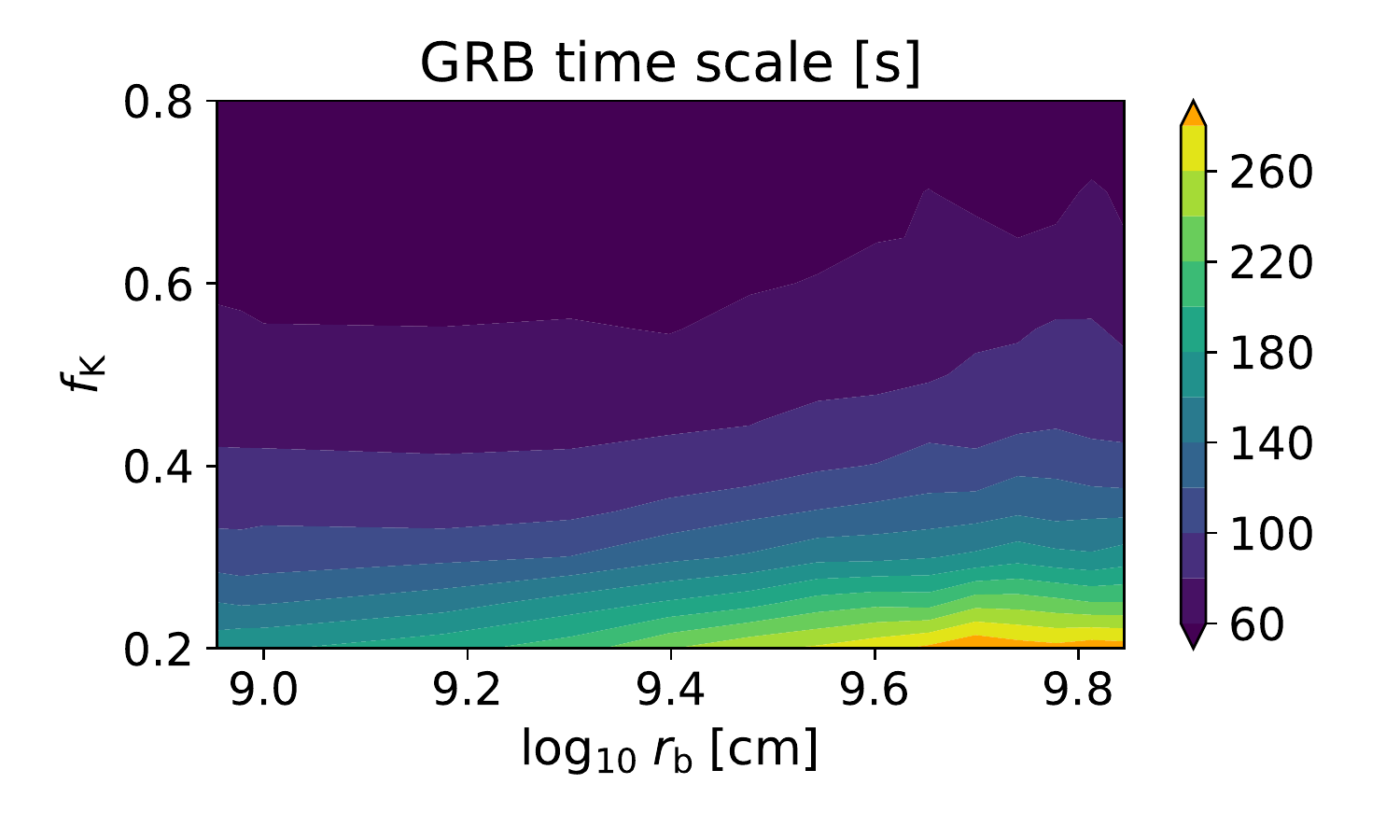}
\includegraphics[width=0.98\linewidth]{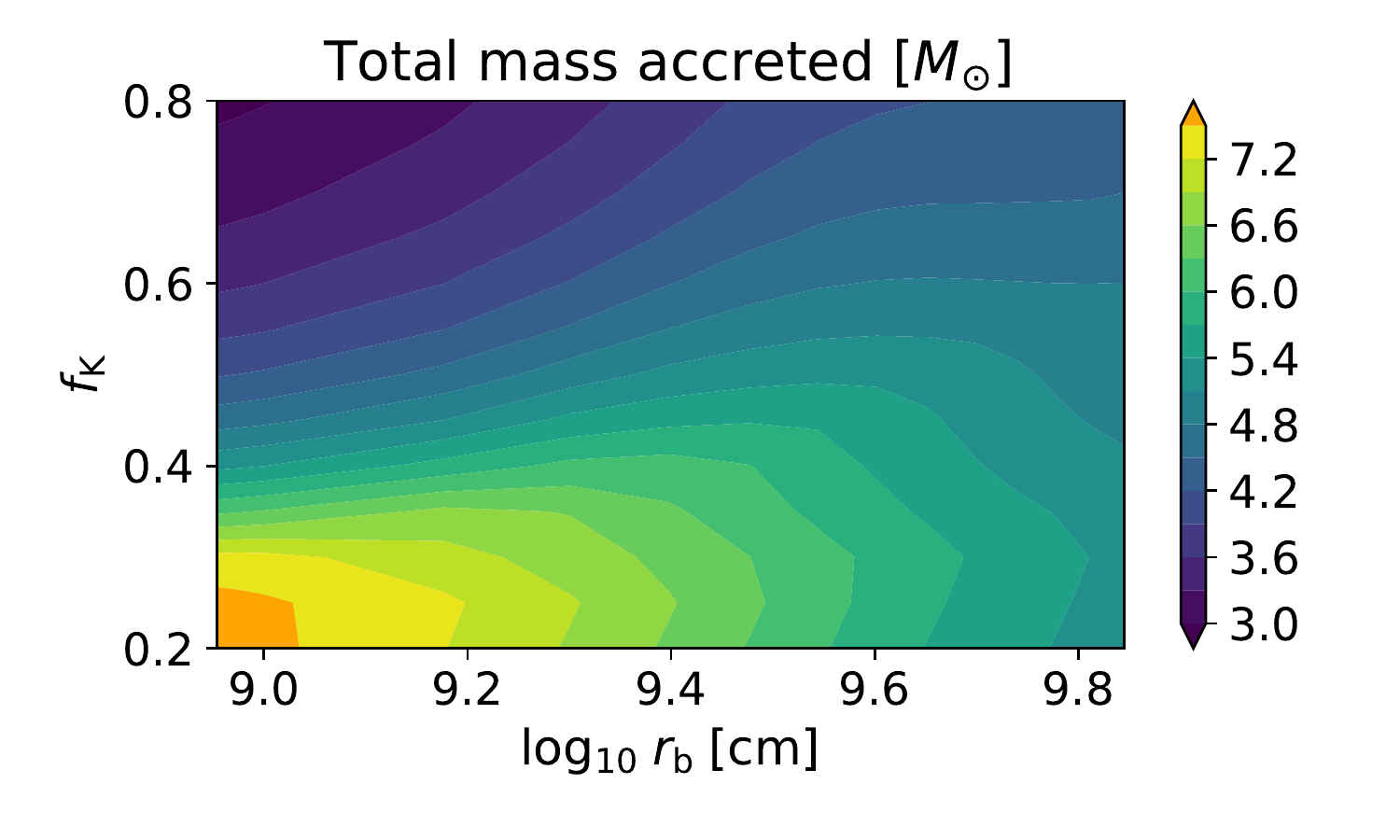}
\includegraphics[width=0.98\linewidth]{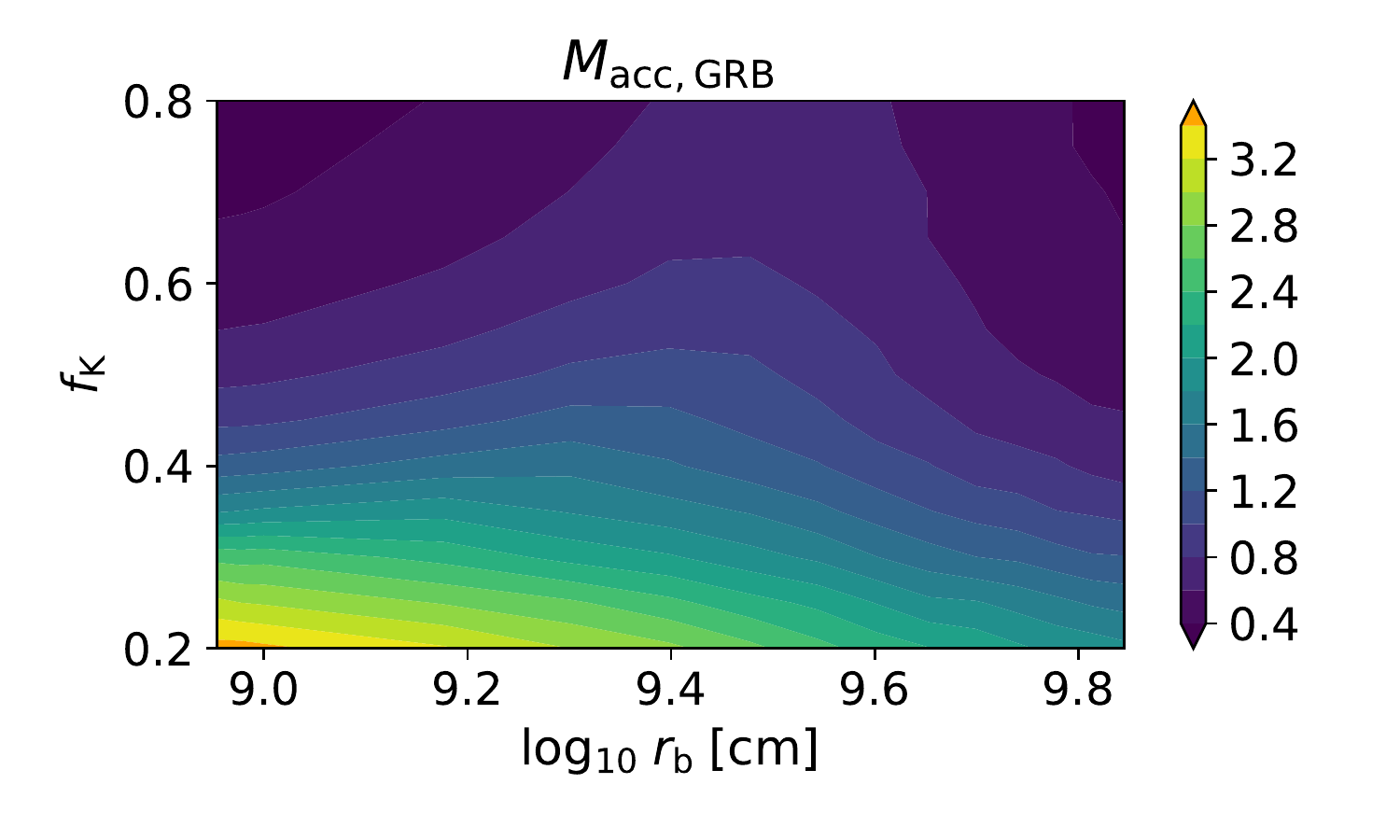}
\includegraphics[width=0.98\linewidth]{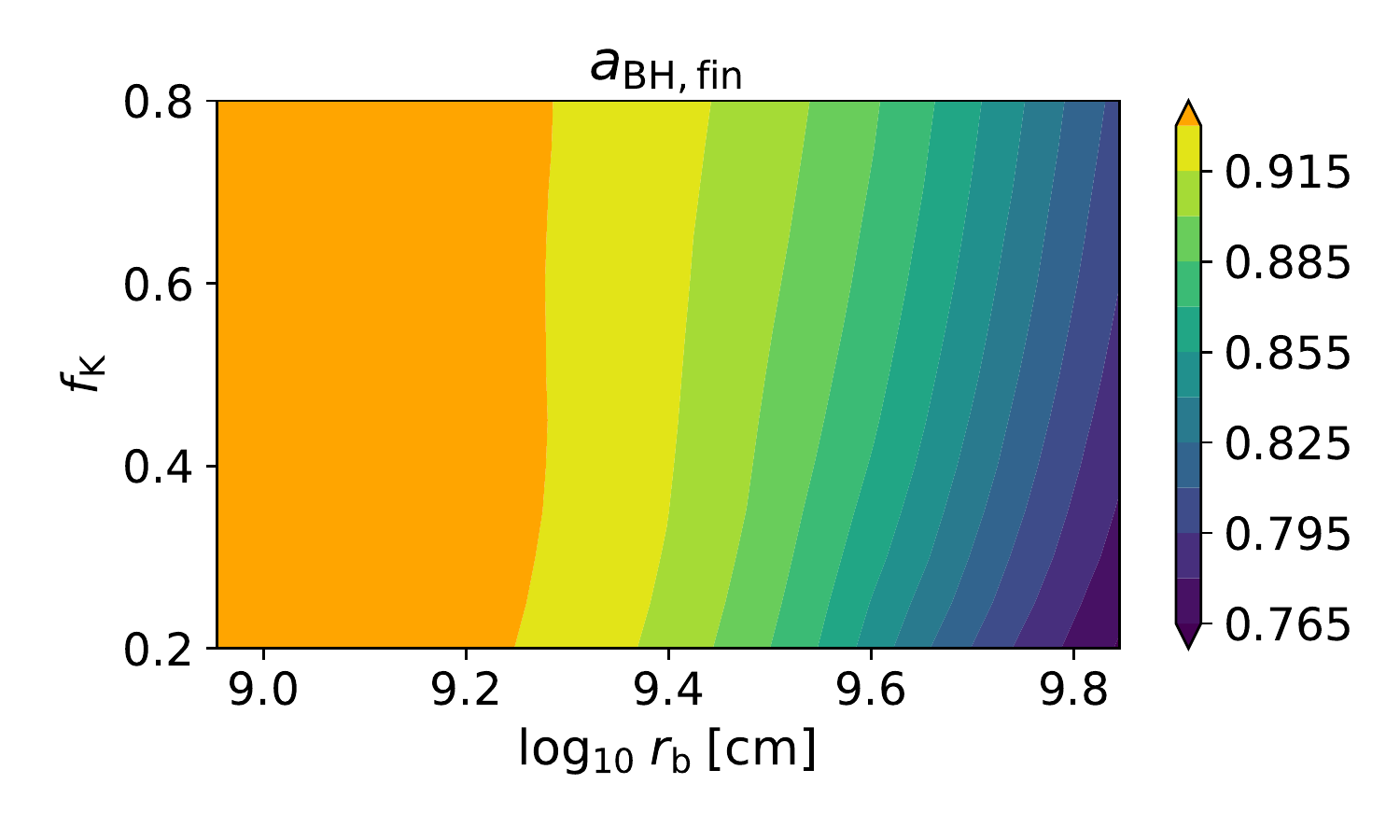}
\caption{Scan of the parameter space for model \texttt{E20} of \citet{heger_presupernova_2000}. Shown are the GRB timescale $t_{\rm GRB}$ (top), total accreted mass [$M_\odot$] (top center), and accreted mass $M_{\rm GRB}$ during the GRB phase [$M_\odot$] (bottom center), and final dimensionless BH spin $a_{\rm BH,fin}$ (bottom).}
\label{fig:Heger_GRB_accretion}
\end{figure}

\begin{figure}
\centering
\includegraphics[width=0.98\linewidth]{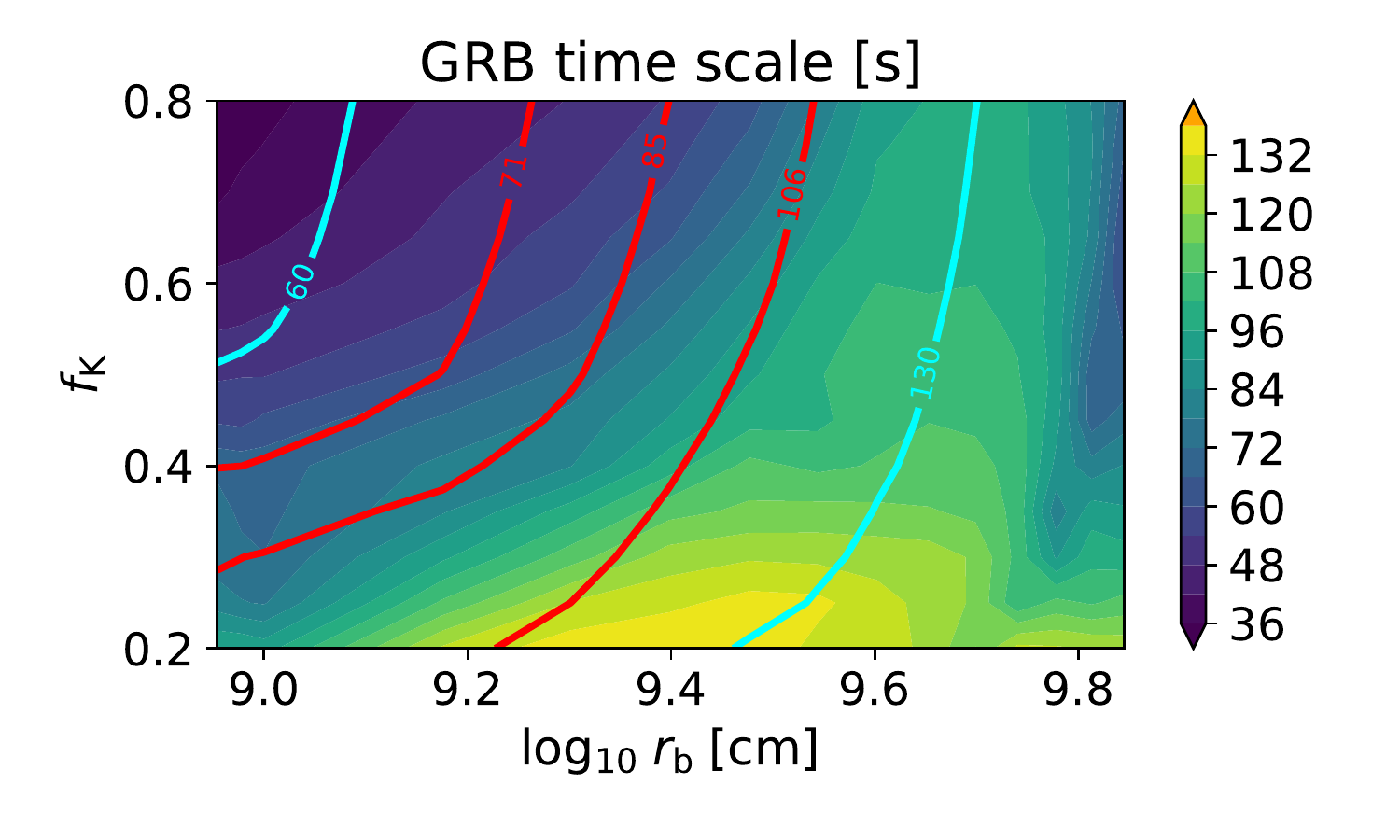}
\includegraphics[width=0.98\linewidth]{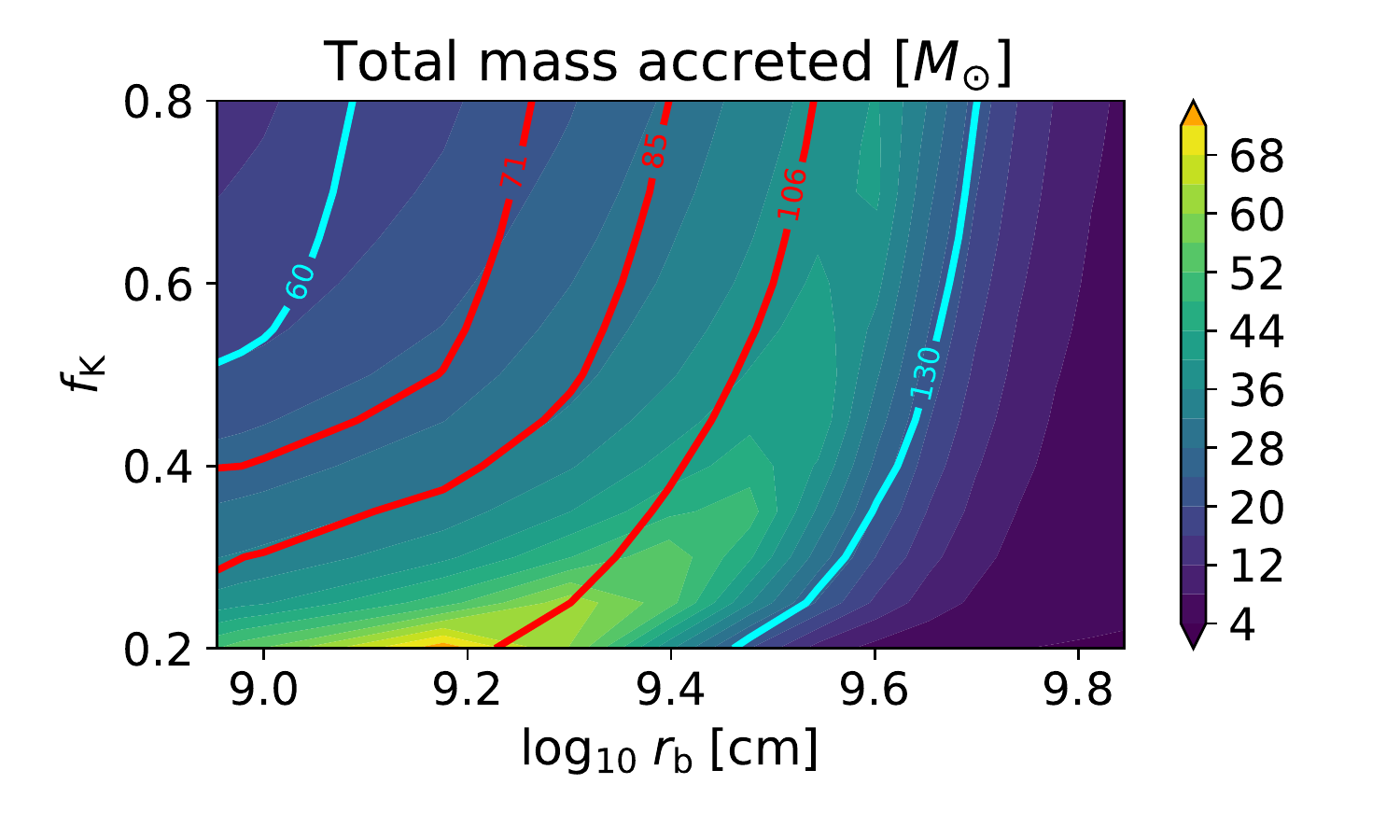}
\includegraphics[width=0.98\linewidth]{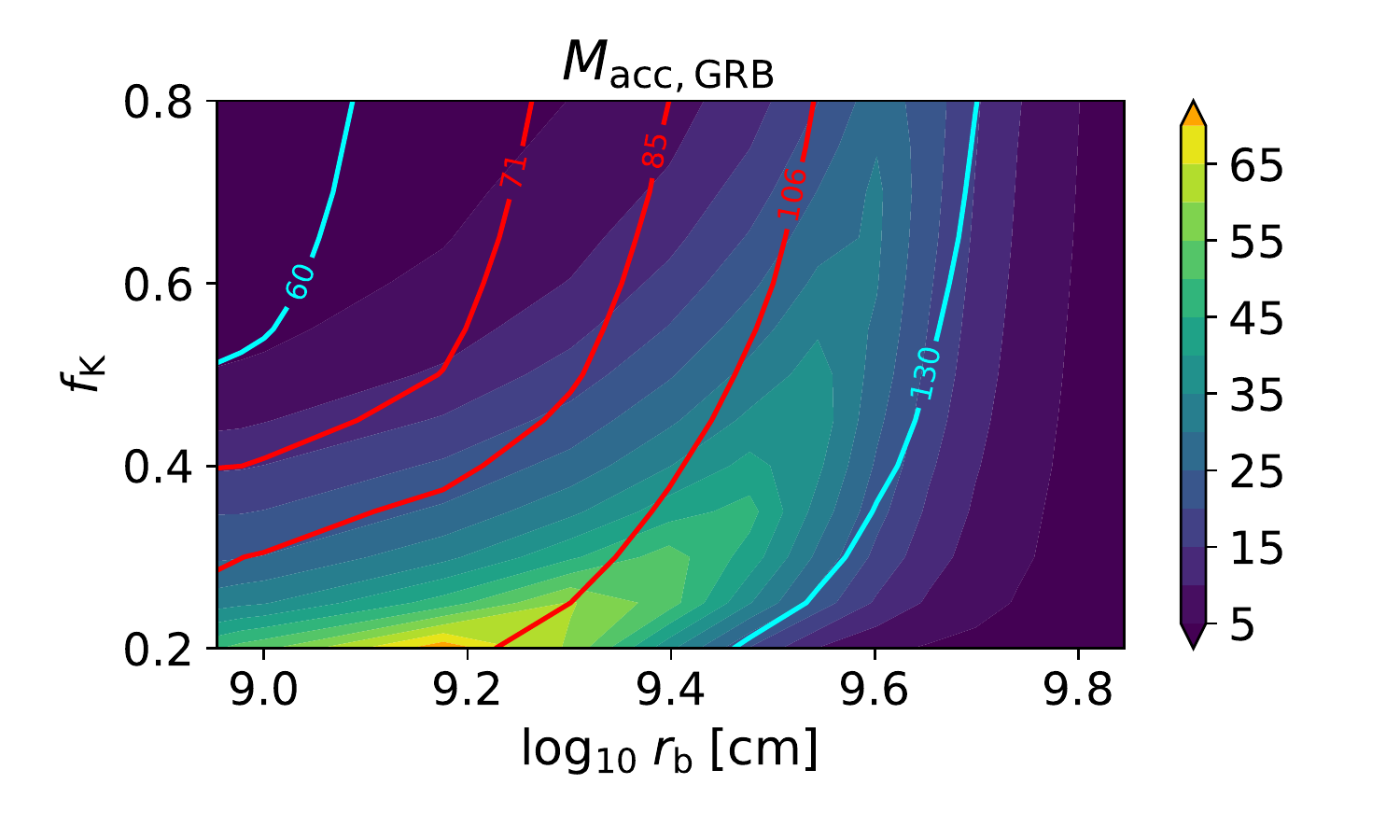}
\includegraphics[width=0.98\linewidth]{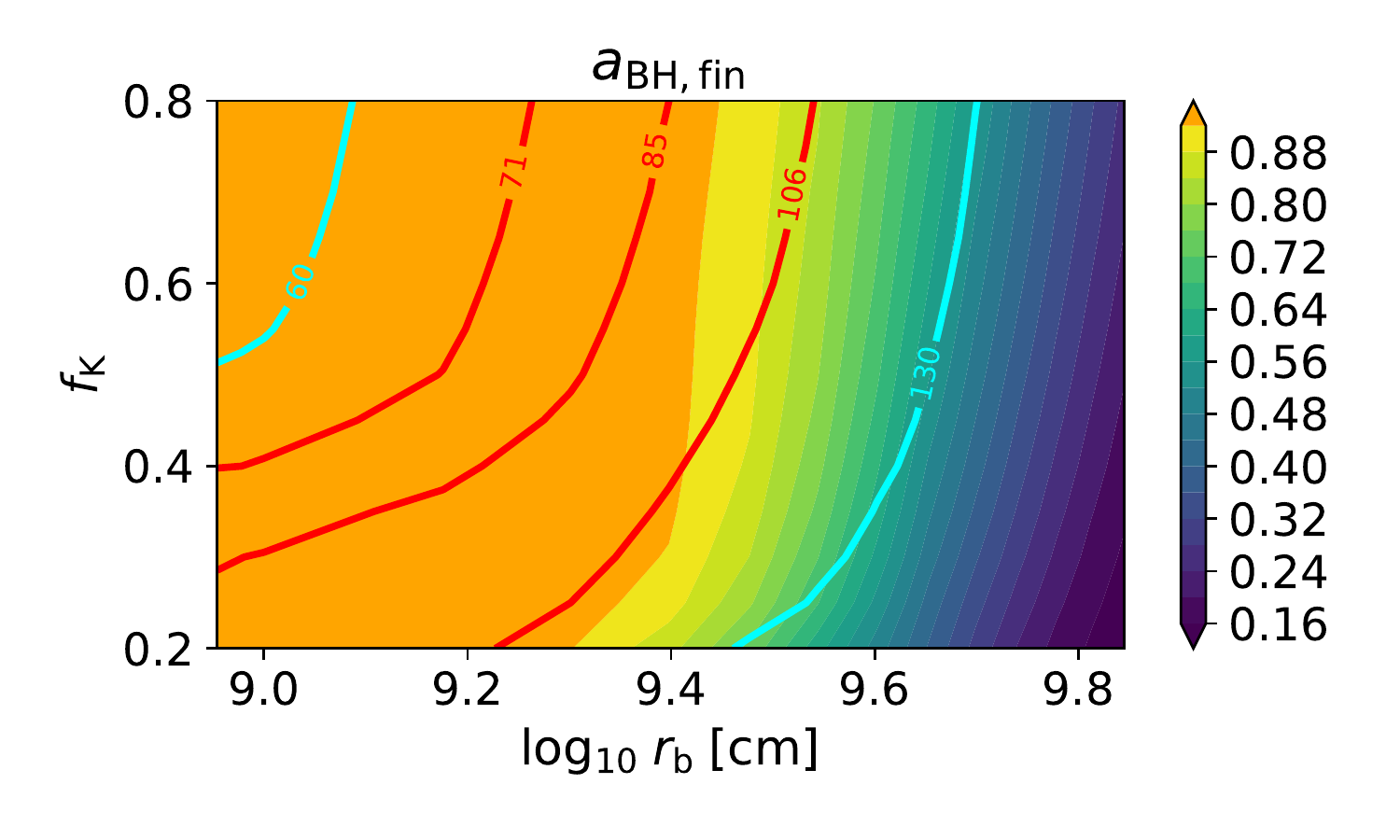}
\caption{Scan of the parameter space for model \texttt{250.25}. Shown are the GRB timescale $t_{\rm GRB}$ (top), total accreted mass [$M_\odot$] (center top), accreted mass $M_{\rm GRB}$ during the GRB phase [$M_\odot$] (center bottom), and final dimensionless BH spin $a_{\rm BH,fin}$ (bottom). Red contours indicate the inferred primary mass of \gw{} [$M_\odot$], together with its 90\% confidence limits. Cyan contours delineate final BH masses of 60\,$M_\odot$ and 130\,$M_\odot$, which approximately correspond to the lower and upper end of the PI mass gap.}
\label{fig:Renzo_GRB_accretion}
\end{figure}

Figure \ref{fig:Heger_ejecta} presents results for model \texttt{E20} of \citet{heger_presupernova_2000}, one representative case for ordinary collapsars below the PI BH mass gap. We vary the parameters of the adopted progenitor rotation profile (cf.~Eq.~\eqref{eq:j_profile}) within ranges motivated by the structure of the stellar evolution models (see Sec.~\ref{sec:fallback}), making nearly identical assumptions regarding the rotation profile as for mass-gap collapsars (Sec.~\ref{sec:results_massgap_collapsars}). Our results are almost insensitive to the exact value of the power-law index $p$, which we thus fix to $p=4.5$ for simplicity here.  For model \texttt{E20}, we find $\approx\!0.04-0.7\,M_\odot$ of $r$-process material, including $\approx\!0.01-0.04\,M_\odot$ of heavy ($A>136$) $r$-process material and $\approx\!0.03-0.65\,M_\odot$ of light ($A<136$) $r$-process material, and $\approx\!0.14-0.26\,M_\odot$ of $^{56}$Ni.

In comparison to \citet{siegel_collapsars_2019}, the updated model presented here tends to predict moderately less heavy $r$-process material, more light $r$-process material, and more $^{56}$Ni. This is the result of i) a more detailed treatment of the disk accretion rate onto the BH, ii) an additional nucleosynthesis regime of light $r$-process material only at high accretion rates $>\!\dot{M}_{\nu, \text{r}-\text{p}}$ (cf.~Eq.~\eqref{eq:accretion_regimes}), and iii) a detailed evolution of the nucleosynthesis regimes throughout the accretion process as a result of BH growth.  Overall, however, the mass ranges of all nucleosynthesis products found here broadly agree with \citet{siegel_collapsars_2019}. In particular, our refined analysis still predicts a sizable amount of lanthanide-bearing $r$-process ejecta of $\approx\!0.04-0.7\,M_\odot$ across various progenitor models of \citet{heger_presupernova_2000}. These results remain consistent with \citet{Miller+20}, insofar that the mass accretion and generation of disk winds occur over a wide range of accretion rates, which drift through nucleosynthesis regimes characterized by varying degrees of neutrino irradiation (cf.~Eq.~\eqref{eq:accretion_regimes}).  Interestingly, our new models result in disk-wind $^{56}$Ni yields approaching the values required to explain the light curves of observed GRB SNe (e.g., \citealt{cano_self-consistent_2016}), without a prompt shock-heated explosion (e.g., \citealt{barnes_grb_2018}) or explosive nucleosynthesis at larger radii in the disk \citep{zenati_nuclear_2020}.

Our collapsar model is also in good agreement with properties of observed GRBs. We check for consistency of our collapsar model with observed GRBs in terms of GRB timescales and energies. We assume that the accreted mass onto the BH is proportional to the radiated $\gamma$-ray energy, that is, $L_\gamma \propto \eta \dot{m}_{\rm acc} c^2$, where $L_\gamma$ is the observed gamma-ray luminosity and $\eta$ is an efficiency parameter.

A necessary requirement for collapsar accretion to explain observed GRBs is that the evolution time of the accretion rate be smaller or equal to the typical time required to generate a GRB in the engine frame, i.e., $\tau_{\dot{m}_\mathrm{acc}}\lesssim \tau_\mathrm{GRB}$. The timescale $\tau_{\dot{m}_\mathrm{acc}}$ increases with time, typically expected as a power-law $\tau_{\dot{m}_\mathrm{acc}} \propto t^\alpha$, where $\alpha\simeq 1$. Following \citet{siegel_collapsars_2019}, we define
\begin{equation}
  \tau_{\dot{m}_{\mathrm{acc}}} \equiv \left(\frac{\mathrm{d}\ln \dot{m}_\mathrm{acc}}{\mathrm{d}t}\right)^{-1}. \label{eq:tau_Mdot}
\end{equation}
Furthermore, let $t_\mathrm{GRB}$ denote the time relative to the onset of disk accretion at which the equality $\tau_{\dot{m}_\mathrm{acc}} = \tau_\mathrm{GRB}$ is reached. Consistency then requires
\begin{equation}
t_\mathrm{GRB} \sim \tau_\mathrm{GRB}. \label{eq:GRB_criterium}
\end{equation}
The GRB duration in the engine frame is determined by \citep{bromberg_observational_2012,sobacchi_common_2017}
\begin{equation}
  \tau_\mathrm{GRB} = \tau_\gamma + \tau_\mathrm{b},
\end{equation}
where $\tau_\gamma$ is the observed duration of a GRB in the engine rest frame and $\tau_\mathrm{b}$ is the time required for the jet to drill through the stellar envelope. Assuming a typical value of $\tau_\mathrm{b} = 57^{+13}_{-10}\,\mathrm{s}$ \citep{sobacchi_common_2017}, and a typical observed GRB duration of $\tau_\gamma = T_{90}/(1+z)=9\,\mathrm{s}$, with a characteristic $T_{90} \simeq 27\,\mathrm{s}$ and redshift $z\simeq 2$ \citep{narayana_bhat_third_2016}, one finds $\tau_\mathrm{GRB} \approx 66\,\mathrm{s}$. The top panel of Fig.~\ref{fig:Heger_GRB_accretion} shows that $t_{\rm GRB}\sim \tau_{\rm GRB}$ essentially throughout the parameter space. Similar results are found for the other models of \citet{heger_presupernova_2000}. Hence, consistency with observed GRB durations according to Eq.~\eqref{eq:GRB_criterium} holds.

Consistency with typical observed GRB energies requires that
\begin{eqnarray}
  E_{\gamma,{\rm iso}} f_{\rm b} \frac{\tau_{\rm GRB}}{\tau_\gamma}&\ll& \int_{t_{\rm GRB}} L_{\rm jet}\,{\rm d}t \\ 
   &<& \int_{t_{\rm GRB}} \dot{m}_{\rm acc}c^2\,{\rm d}t 
  \equiv M_{\rm acc,GRB}c^2,\nonumber
\end{eqnarray}
where $E_{\gamma,{\rm iso}}\sim 1\times 10^{53}$\,erg is the typical isotropic-equivalent gamma-ray energy of observed GRBs and $f_{\rm b}\simeq 0.006$ is the beaming fraction \citep{goldstein_estimating_2016}, $L_{\rm jet}$ is the luminosity of the accretion-powered jet, and $M_{\rm acc,GRB}$ is the accreted mass onto the BH through the disk during the GRB timescale $t_{\rm GRB}$. With these values, the condition translates into
\begin{equation}
  M_{\rm acc,GRB} \gg 2.5\times 10^{-3}\,M_\odot, \label{eq:GRB_mass_criterium}
\end{equation}
which we find is satisfied throughout the parameter space where the peak accretion rate reaches $\dot{m}_{\rm acc} > 10^{-4}\,M_\odot\,\text{s}^{-1}$ (cf.~Fig.~\ref{fig:Heger_GRB_accretion}, bottom panel), which we take as a threshold to postulate a successful jet (Sec.~\ref{sec:fallback}). Similar findings apply to other models of \citet{heger_presupernova_2000}. We therefore find that our ordinary collapsar models are consistent both with typical GRB duration times and energies, including drill time.

Figure \ref{fig:Renzo_GRB_accretion} shows a parameter-space scan for model \texttt{250.25}, a typical mass gap collapsar model. The GRB properties are in good agreement with observational constraints. While the GRB durations are typically similar to ordinary collapsars, the accreted mass during the GRB phase may be up to a factor $\sim\!10$ higher. One may thus speculate that these models give rise to GRBs that may be a factor $\sim\!10$ more luminous or energetic, if the gamma-ray luminosity tracks accreted mass.

\section{Gravitational-wave emission}
\label{app:GW_emission}

\begin{figure}
\centering
\includegraphics[width=0.98\linewidth]{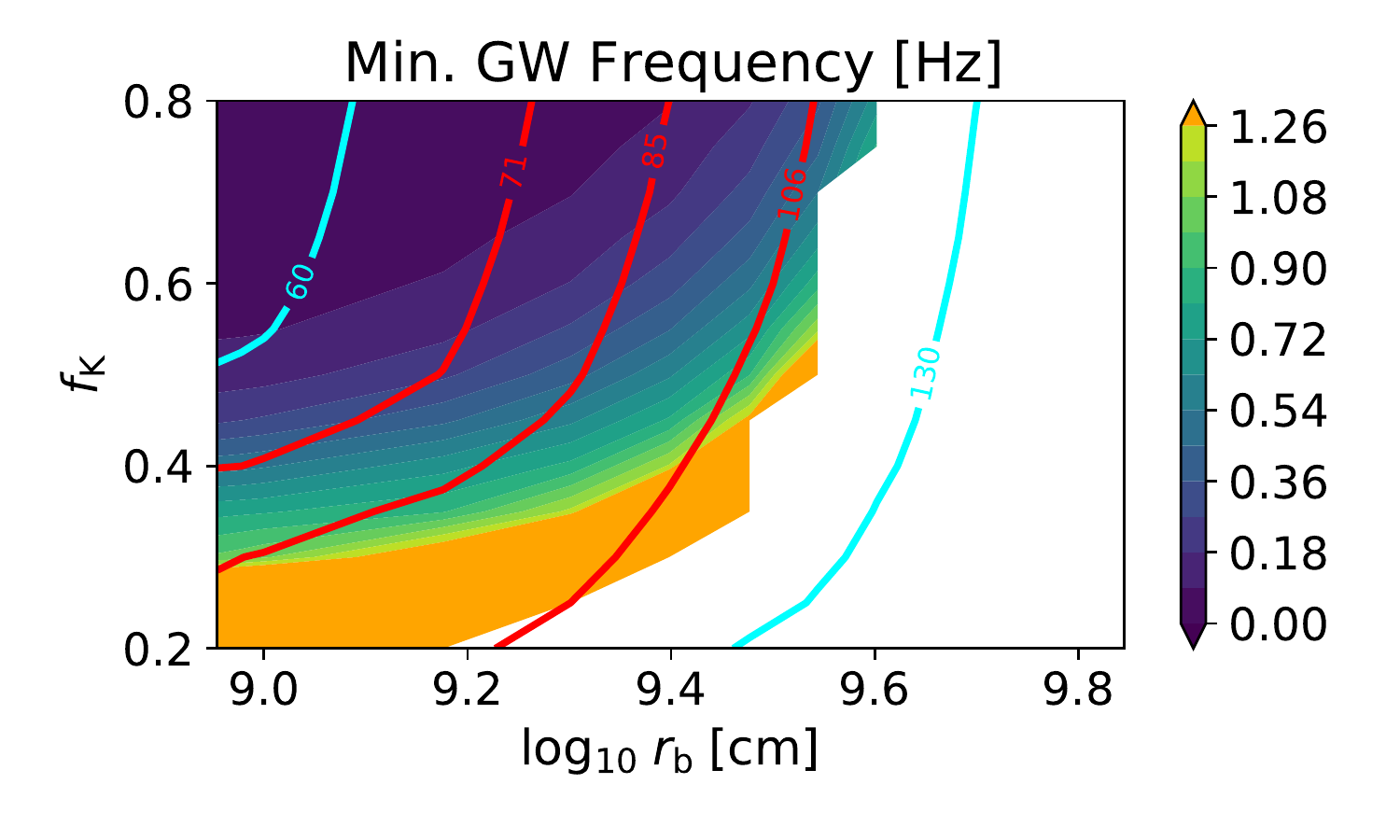}
\includegraphics[width=0.98\linewidth]{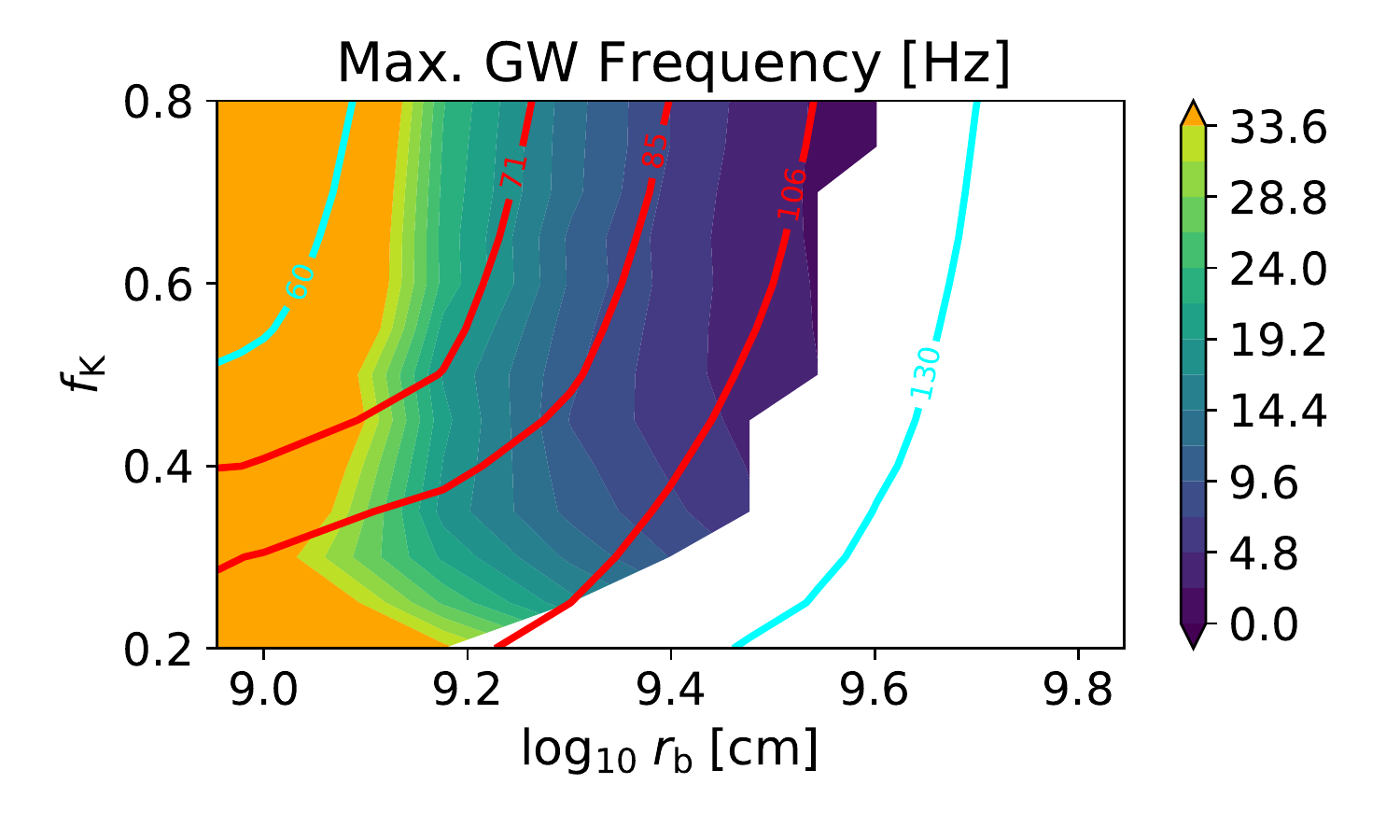}
\caption{Mininum (top) and maximum (bottom) frequency of the predicted $l=m=2$ gravitational wave emission during the gravitationally unstable phase of collapsar accretion, for the same model shown in Fig.~\ref{fig:gravitational_instabilities}. Final BH mass contours are drawn as in previous figures.}
\label{fig:gravitational_wave_frequency}
\end{figure}

We calculate the gravitational-wave strain of emitted gravitational waves by approximating `the lump' of the unstable disk (assumed to correspond to an over-density of $\delta \rho/\rho \gtrsim 0.1$; \citealt{shibata_alternative_2021,wessel_gravitational_2021}) and the central BH as two orbiting point masses. The frequencies of gravitational-wave emission can thus be directly predicted from the evolution of the disk angular velocity according to the collapsar model in Sec.~\ref{sec:fallback}. The maximum and minimum gravitational-wave frequencies vary considerably across the $\{f_{\rm K},r_{\rm b}\}$ parameter space (see Fig.~\ref{fig:gravitational_wave_frequency} for our fiducial model). 

According to the quadrupole formula, assuming that the orbital radius only slowly changes with respect to the orbital frequency, the plus ($h_{+}$) and cross ($h_{\times}$) polarizations of the gravitational waves at distance $r$ and inclination $\iota$ of the disk with respect to the observer can be written as
\begin{eqnarray}
  \mskip-20mu h_+(t) &=& \frac{4G}{rc^4}\mu r_{\rm disk}^2\Omega_{\rm K, disk}^2\frac{1+\cos^2\iota}{2}\cos[\Phi(t)], \label{eq:GW_strain_plus}\\
  \mskip-20mu h_\times(t) &=& \frac{4G}{rc^4}\mu r_{\rm disk}^2\Omega_{\rm K, disk}^2\cos\iota\sin[\Phi(t)], \label{eq:GW_strain_cross}
\end{eqnarray}
where $\Phi(t) \equiv \int_{t_0}^t 2\Omega_{\rm K, disk}(t')\,\rm{d}t'$, with $t=t_0$ denoting the start time of the gravitational instability. These expressions apply to the $l=m=2$ mode, while $\Omega_{\rm K, disk}$ is replaced by $0.5\Omega_{\rm K, disk}$ for the $m=1$ mode. Furthermore, $\mu=M_{\rm lump} M_\bullet/(M_{\rm lump} + M_\bullet)$ is the reduced mass of the lump--BH system, and we set $M_{\rm lump} = f_{\rm lump} M_{\rm disk}$ with $f_{\rm lump}=0.2$. Uncertainties in the value of $f_{\rm lump}$ can be absorbed into uncertainties of the scale height of the disk and the threshold mass fraction $f_{\rm disk, thr}\equiv M_{\bullet}/M_{\rm disk}$ at which gravitational instabilities set in (cf.~Eq.~\eqref{eq:gravitational_instability}). We neglect corrections $\propto \dot{\mu}$ to Eqs.~\eqref{eq:GW_strain_plus} and \eqref{eq:GW_strain_cross} due to a time-dependent reduced mass $\mu$ as a result of accretion onto the black hole, assuming that $M_\bullet$ changes only weakly over the course of gravitational-wave emission.

\begin{figure}
\centering
\includegraphics[width=0.98\linewidth]{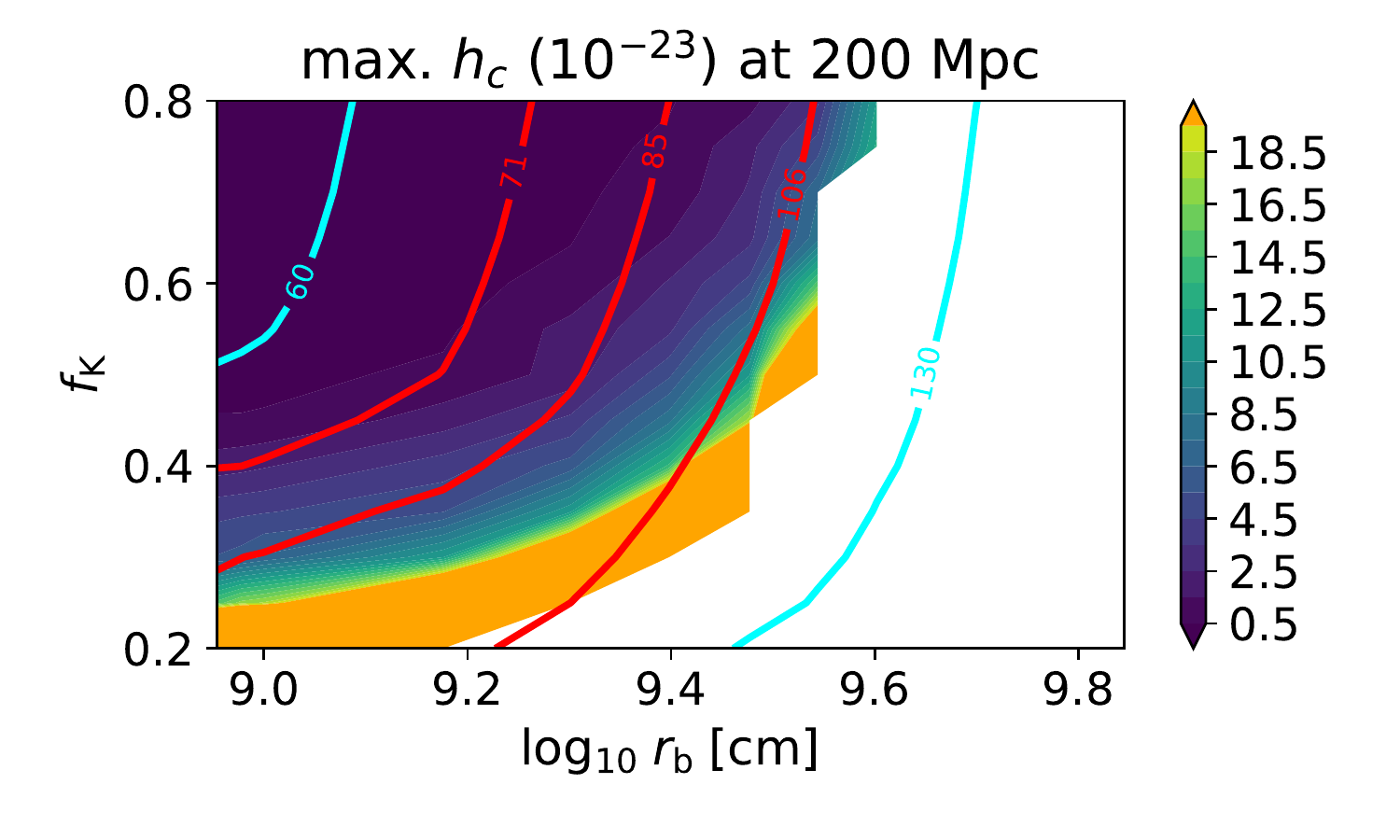}
\caption{Maximum strain amplitude of the characteristic strain $h_c$ of gravitational waves from non-axisymmetric instabilities in self-gravitating collapsar disks across the parameter space of $\{r_{\rm b}, f_{\rm K}\}$ for our fiducial model shown in Fig.~\ref{fig:Renzo_evolution} with $p=4.5$, $f_{\rm K}=0.3$ and $r_{\rm b}=1.5\times 10^{9}$\,cm. Characteristic strains range from $\sim 10^{-24} - 10^{-22}$ depending on the rotation profile of the progenitor. Final BH mass contours are drawn as in previous figures.}
\label{fig:gravitational_wave_strain_contour}
\end{figure}

\begin{figure}
\centering
\includegraphics[width=0.98\linewidth]{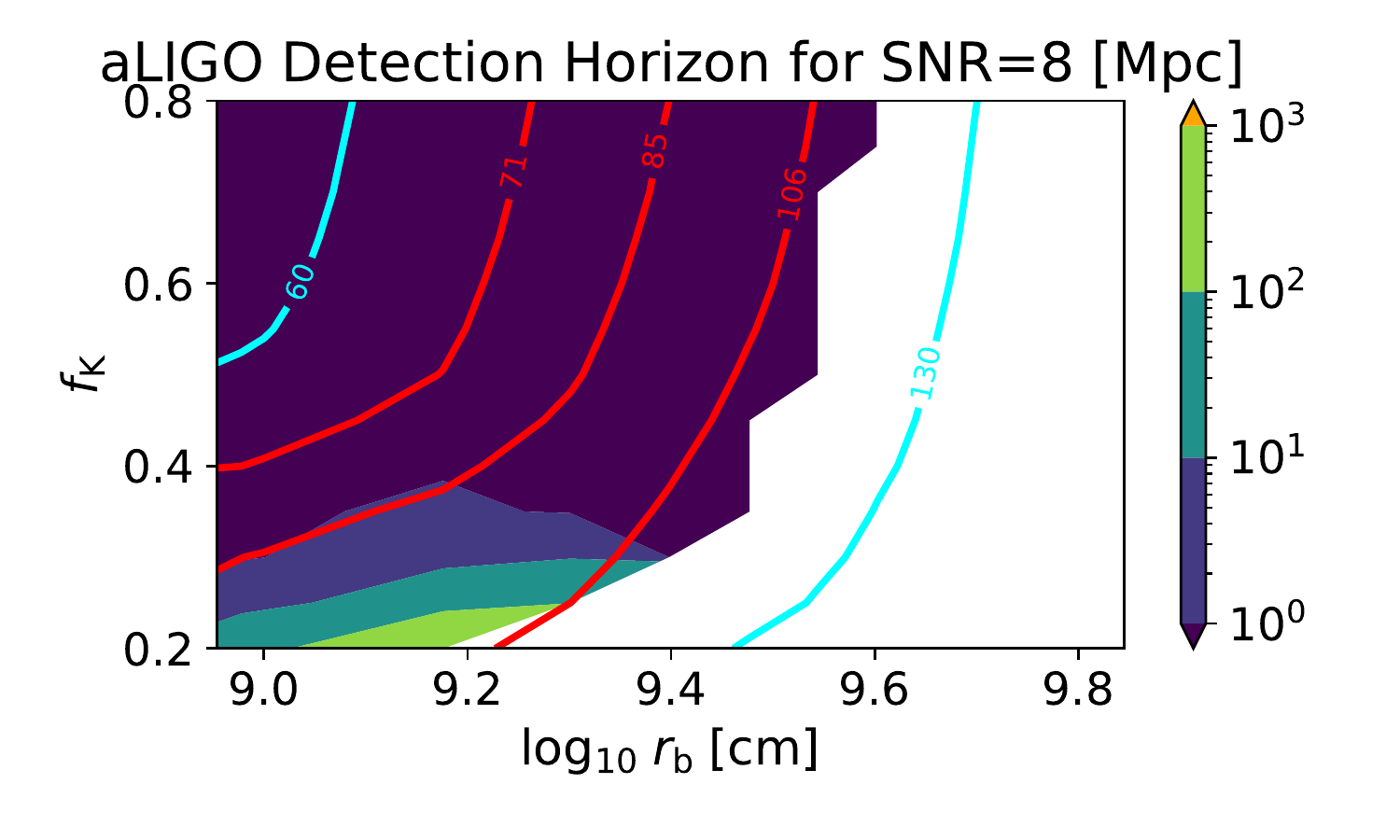}
\includegraphics[width=0.98\linewidth]{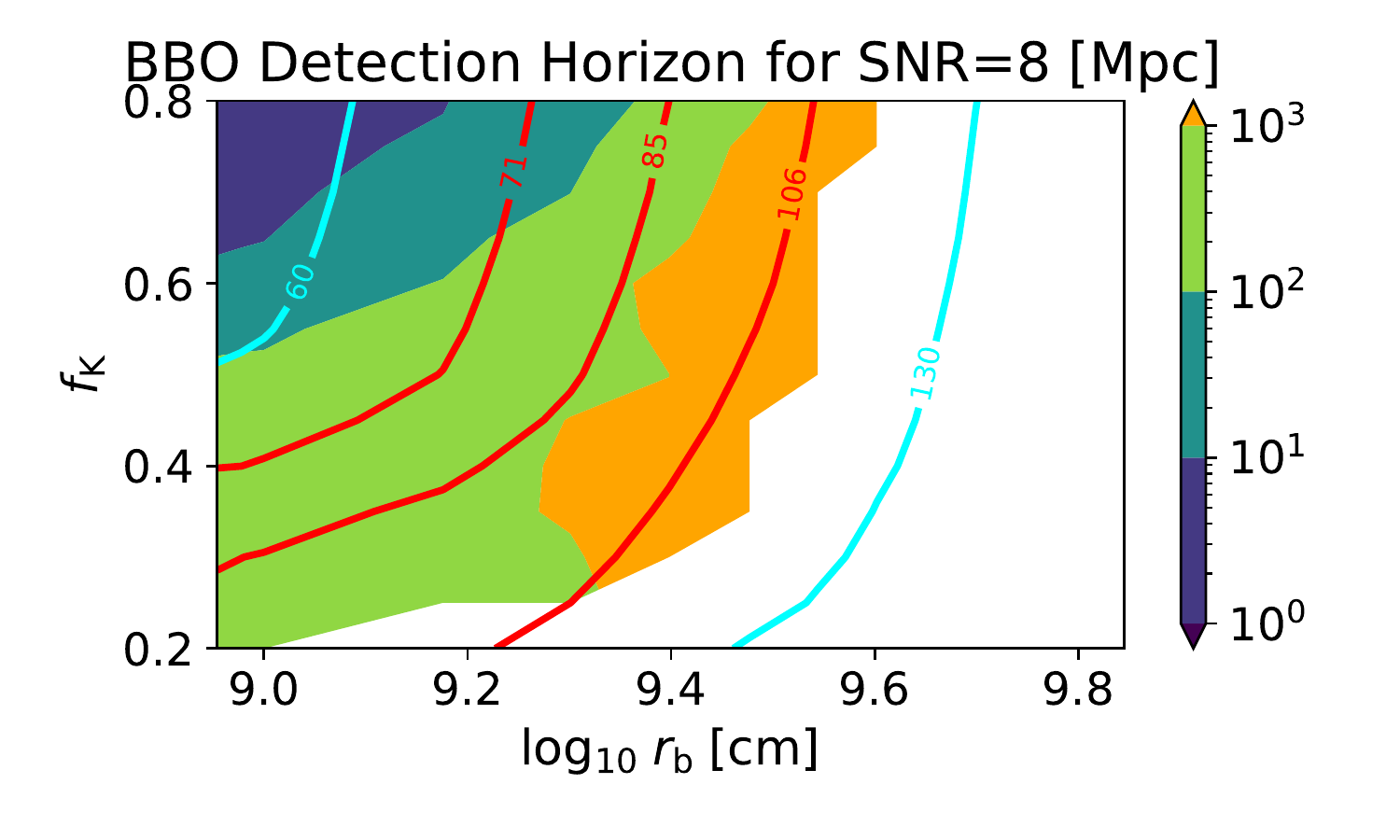}
\caption{Detection horizons for gravitational waves from our fiducial model shown in Fig.~\ref{fig:Renzo_evolution} with $p=4.5$, $f_{\rm K}=0.3$ and $r_{\rm b}=1.5\times 10^{9}$\,cm for advanced LIGO at design sensitivity (top) and Big Bang Observer (bottom), assuming optimal matched filtering and a signal-to-noise ratio of 8. While for aLIGO the detection horizon is typically limited to $\lesssim 100$\,Mpc, BBO will be able to detect such sources up to typically several hundred Mpc to several Gpc, with particular sensitivity for progenitors with low-angular momentum `cores' (medium to large values of $r_{\rm b}$). Contours delineate final BH masses as in previous figures.}
\label{fig:gravitational_wave_detection_horizon_2}
\end{figure}

We characterize gravitational-wave emission in the frequency domain (positive frequencies $f$ only) by computing the characteristic strain, defined as
\begin{equation}
    h_c=2f|\tilde{h}^{\rm res}(f)|.
    \label{eq:GW_char}
\end{equation}
For an estimate of the horizon distance we assume that the detector receives a signal from a directly overhead source and hence the optimal strain response at the detector can be written as
\begin{equation}
    \tilde{h}^{\rm res}=\sqrt{|\tilde{h}_{+}|^2 + |\tilde{h}_{\times}|^2},
    \label{eq:GW_response_strain}
\end{equation}
where $\tilde{h}_{+}$ and $\tilde{h}_{\times}$ are the Fourier transforms of the respective polarization strain amplitudes that we compute employing a Tukey window function limited to the physical frequencies between the maximum and minimum frequency expected from disk evolution (Fig.~\ref{fig:gravitational_wave_frequency}). We compare gravitational wave signals to detector sensitivity in terms of the amplitude spectral density $\sqrt{S_h}$ \citep{moore_gravitational-wave_2015},
\begin{equation}
   \sqrt{S_h}=2\sqrt{f}|\tilde{h}^{\rm res}|,
   \label{GW:ASD}
\end{equation}
where $S_h$ denotes the power spectral density, and calculate the signal-to-noise ratio (SNR) using an optimal filter \citep{moore_gravitational-wave_2015},
\begin{equation}
   \mathrm{SNR}=\left(\int df \frac{h_c^2}{S_n f^2}\right)^{\frac{1}{2}}.
   \label{GW:SNR}
\end{equation}

Characteristic strains of superKN collapsars range from $\sim 10^{-24} - 10^{-22}$ depending on the rotation profile of the progenitor. A representative example is shown in Fig.~\ref{fig:gravitational_wave_strain_contour}. Detection horizons for advanced LIGO and BBO assuming $\text{SNR}=8$ are shown in Fig.~\ref{fig:gravitational_wave_detection_horizon_2}, while those for CE, ET, and DECIGO are presented in Fig.~\ref{fig:gravitational_wave_detection_horizon}.

\section*{Data Availability}

The pre-collapse stellar models from \cite{renzo:20csm} and
\cite{heger_presupernova_2000}, are available at \url{https://zenodo.org/record/3406357} and
\url{https://2sn.org/stellarevolution/rotation/},
respectively. Data of our model runs together with visualization scripts will be made available at \url{https://doi.org/10.5281/zenodo.5639697}.

\bibliography{DanielBib,example}
\bibliographystyle{aasjournal}



\end{document}